\documentclass{IEEEoj}
\usepackage[table,xcdraw]{xcolor}
\usepackage{cite}
\usepackage{amsmath,amssymb,amsfonts}
\usepackage{algorithmic}
\usepackage{graphicx,color}
\usepackage{textcomp}
\usepackage{glossaries}
\usepackage{multirow}
\usepackage{url}
\usepackage{comment}
\usepackage[utf8]{inputenc}
\usepackage{pgfplots}
\usepackage{afterpage}
\usepackage{orcidlink}
\usepackage{ulem} 
\usepackage{soul}
\pgfplotsset{compat=newest}
\usepackage{tikz}
\usetikzlibrary{external,patterns,shapes.arrows}
\DeclareUnicodeCharacter{2212}{−}
\usepgfplotslibrary{groupplots,dateplot}
\usepackage{adjustbox}

\newcommand{\rev}[1]{\textcolor{black}{#1}}

\newacronym{ietf}{IETF}{Internet Engineering Task Force}
\newacronym{ts}{TS}{Technical Specification}
\newacronym{tr}{TR}{Technical Report}
\newacronym{qos}{QoS}{Quality of Service}
\newacronym{ran}{RAN}{Radio Access Network}
\newacronym{cn}{CN}{Core Network}
\newacronym{urllc}{URLLC}{Ultra Reliable Low Latency Communications}
\newacronym{embb}{eMBB}{enhanced Mobile Broadband}
\newacronym{miot}{mIOT}{massive Internet of Things}
\newacronym{3gpp}{3GPP}{3rd Generation Partnership Project}
\newacronym{htb}{HTB}{Hierarchical Token Bucket}
\newacronym{nest}{NEST}{Network Slice Type}
\newacronym{gst}{GST}{Generic Network Slice Template}
\newacronym{phb}{PHB}{Per-Hop-Behaviour}
\newacronym{drr}{DRR}{Deficit Round Robin}
\newacronym{gsma}{GSMA}{GSM Association}
\newacronym{5qi}{5QI}{5G Quality Indicator}
\newacronym{vnx}{VNX}{Virtual Networks over LinuX}
\newacronym{lxc}{LXC}{LinuX Containers}
\newacronym{tn}{TN}{Transport Network}
\newacronym{ifb}{IFB}{Intermediate Functional Block}
\newacronym{gbr}{GBR}{Guaranteed Bit Rate}
\newacronym{pir}{PIR}{Peak Information Rate}
\newacronym{cir}{CIR}{Committed Information Rate}
\newacronym{cbs}{CBS}{Committed Burst Size}
\newacronym{ebs}{EBS}{Exceeded Burst Size}
\newacronym{pbs}{PBS}{Peak Burst Size}
\newacronym{sdn}{SDN}{Software Defined Networking}
\newacronym{dscp}{DSCP}{Differentiated Services Code Point}
\newacronym{qfi}{QFI}{QoS flow Identifier}
\newacronym{upf}{UPF}{User Plane Function}
\newacronym{wan}{WAN}{Wide Area Network}
\newacronym{wfq}{WFQ}{Weighted Fair Queuing}
\newacronym{sla}{SLA}{Service Level Agreement}
\newacronym{nf}{NF}{Network Function}
\newacronym{sdp}{SDP}{Service Demarcation Point}
\newacronym{pe}{PE}{Provider Edge}
\newacronym{ce}{CE}{Customer Edge}
\newacronym{l2vpn}{L2VPN}{Layer 2 Virtual Private Network}
\newacronym{l3vpn}{L3VPN}{Layer 3 Virtual Private Network}
\newacronym{snssai}{S-NSSAI}{Single - Network Slice Selection Assistance Information}
\newacronym{pdb}{PDB}{Packet Delay Budget}
\newacronym{gtp}{GTP}{GPRS Tunnelling Protocol}
\newacronym{ac}{AC}{Attachment Circuit}
\newacronym{spf}{SPF}{Shortest Path Forwarding}
\newacronym{te}{TE}{Traffic Engineering}
\newacronym{tc}{TC}{Traffic Control}
\newacronym{rsvpte}{RSVP-TE}{Resource Reservation Protocol - Traffic Engineering}
\newacronym{srte}{SR-TE}{Segment Routing - Traffic Engineering}
\newacronym{vlan}{VLAN}{Virtual Local Area Network}
\newacronym{srh}{SRH}{Segment Router Header}
\newacronym{sr}{SR}{Segment Routing}
\newacronym{srv6}{SRv6}{Segment Routing IPv6}
\newacronym{mpls}{MPLS}{Multiprotocol Label Switching}
\newacronym{udp}{UDP}{User Datagram Protocol}
\newacronym{pdu}{PDU}{Packet Data Unit}
\newacronym{tod}{ToD}{Teleoperated Driving}
\newacronym{hctns}{HCTNS}{Hierarchically Controlled Transport Network Slicing}
\newacronym{trtcm}{trTCM}{two rate Three Color Marker}
\def\BibTeX{{\rm B\kern-.05em{\sc i\kern-.025em b}\kern-.08em
    T\kern-.1667em\lower.7ex\hbox{E}\kern-.125emX}}
\AtBeginDocument{\definecolor{ojcolor}{cmyk}{0.93,0.59,0.15,0.02}}
\def\OJlogo{\vspace{-4pt}\hskip-4pt}
\begin{document}
\title{A Slicing Model for Transport Networks with Traffic Burst Control and QoS Compliance for Traffic Flows}

\author{Aitor Encinas-Alonso\IEEEauthorrefmark{1}\orcidlink{0009-0000-1290-8815}, Carlos M. Lentisco\IEEEauthorrefmark{1}\orcidlink{0000-0002-7444-0872}, Ignacio Soto\IEEEauthorrefmark{1}\orcidlink{0000-0002-7421-3733 }, Luis Bellido\IEEEauthorrefmark{1}\orcidlink{0000-0001-9591-0928}, David Fernandez\IEEEauthorrefmark{1}\orcidlink{0000-0002-2172-9162}
}
\affil{Departamento de Ingeniería de Sistemas Telemáticos, ETSI Telecomunicación, Universidad Politécnica de Madrid, Spain}
\corresp{CORRESPONDING AUTHOR: Aitor Encinas-Alonso (e-mail: aitor.encinas.alonso@alumnos.upm.es).}
\authornote{This work was partially supported by the Remote Driver project (TSI-065100-2022-003) funded by the Ministerio de Asuntos Económicos y Transformación Digital (Gobierno de España) and the European Union's Horizon Europe research and innovation program under Grant Agreement No. 101097122 (ACROSS).}
\markboth{Preparation of Papers for IEEE OPEN JOURNALS}{Author \textit{et al.}}

\begin{abstract}
Network slicing has emerged as a key network technology, providing network operators with the means to offer virtual networks to vertical users over a single physical network infrastructure. Recent research has resulted mainly in techniques for managing and deploying network slices, but the implementation of network slices on a real physical transport network infrastructure has received much less attention. Standardization bodies, such as the \gls{ietf}, have provided some implementation recommendations. Still, there is a lack of mechanisms to implement network slices capable of handling traffic bursts while simultaneously meeting the \gls{qos} requirements of the traffic flows associated with the slices. In this paper, we propose a novel fine-grained resource control mechanism to implement transport network slices that meet traffic \gls{qos} requirements while both accepting limited traffic bursts, and enabling efficient bandwidth sharing within and across slices. The mechanism is executed at the edge of the transport network. The proposed model aligns with current standards on network slicing and has been tested on an experimental platform. Using this platform, we have conducted an extensive experimental campaign that demonstrates that our proposal can effectively control traffic bursts generated within the network slices while maximizing bandwidth utilization across the network.
\end{abstract}

\begin{IEEEkeywords}
Network Slicing, Quality of Service, Communication System Traffic Control,  Mobile Communication, Transport Network.
\end{IEEEkeywords}

\maketitle

\section{Introduction}
\label{sec:intro}
Although the deployment of commercial 5G networks is a reality in many countries worldwide, the most advanced features proposed in the \gls{3gpp} 5G standards are still under investigation. Among them, network slicing promises to revolutionize the way in which network infrastructures are exploited. With network slicing, a network infrastructure can be shared to provide different and isolated network services~\cite{3GPP2024_1}. Network slicing allows network operators to provide different communication services with tailored quality levels. Examples are \gls{urllc}, \gls{embb}, or \gls{miot}. Network slicing techniques can be applied to the different segments that compose a mobile network: the \gls{ran}, the \gls{tn}, and \gls{cn}. While the \gls{ran} and \gls{cn} fall under the domain of \gls{3gpp}, the \gls{tn} belongs to the \gls{ietf}'s domain. A mobile network operator may predefine a set of network slices, allowing external customers to subscribe to the communication services that best suit their applications. Additionally, operators can enable customers to define their own network slices based on specific needs. Vertical industries are expected to be one of the main beneficiaries of network slicing, obtaining customized connectivity services through a shared network infrastructure. 

Up to this date, the state of the art has focused on the problem of allocating network resources to different network slices efficiently, especially in the \gls{ran} segment of the network. Regarding the transport network, the \gls{ietf} has defined models for controlling the allocation of resources to different network slices, but these models do not provide the ability to limit the impact on the \gls{qos} of the traffic associated with services delivered over such slices. 

Our proposal is applied to a scenario where standard network slices such as \gls{urllc} and \gls{embb} 
are sharing the physical network infrastructure with a network slice for \gls{tod}. \gls{tod} is an advanced automotive industry use case that enables drivers to operate vehicles remotely~\cite{Amador2022}. 


In our target scenario, a \gls{tod} service provider signs a contract with an operator of a public mobile network to hire a network slice for the \gls{tod} service. Within the \gls{tod} service, various traffic flows with different \gls{qos} requirements (video, telemetry, and commands) are generated. Consequently, the network slice for \gls{tod} must be capable of providing differentiated traffic treatment for each type of traffic flow within the slice.

The main contributions of this work are:
\begin{itemize}

\item The design of a flexible slicing model for transport networks, named \gls{hctns}, aligned with current \gls{ietf} standards, enhancing bandwidth sharing between traffic classes and slices while providing explicit control over how bandwidth is allocated. The model also ensures consistent \gls{qos} behavior for all traffic accepted in the \gls{tn} within the same traffic class, or, in its absence, within the same slice.

\item \gls{hctns} incorporates a novel traffic policer that optimizes the utilization of available bandwidth both across slices and within individual slices. 

\item A new traffic burst control mechanism that enables the network to accommodate traffic bursts while managing their impact on the \gls{qos} experienced by the different traffic flows in the \gls{tn} through configurable parameters.

\item A performance evaluation, based on an implemented prototype, showing that by applying our proposal it is possible to meet the strict \gls{qos} requirements of services such as \gls{tod}. The results confirm the advantages of \gls{hctns} in ensuring that all traffic flows within any network slice adhere to the defined \gls{qos} requirements. \rev{Moreover, the experiments show that our proposal outperforms the 5G transport network slicing model that is being standardized by the IETF and a reference proposal in the literature.}
\end{itemize}

The remainder of the paper is organized as follows. Section~\ref{sec:soa} reviews the related work and highlights the contributions of this paper to the state of the art.
Section~\ref{sec:background} describes the IETF 5G transport network slicing model, which serves as the reference model for our proposed work. 
Section~\ref{sec:proposal} describes the \gls{hctns} network slicing and QoS model.
Section~\ref{sec:eval} presents our testbed with a simplified version of our proposed network slicing model for 5G transport networks, configured to support teleoperated driving services sharing the same physical network with other typical and constrained slices. This section is also dedicated to the evaluation of the proposal. Finally,  Section~\ref{sec:conclusion} summarizes the main conclusions of our work. 

\rev{A summary of key acronyms used in the article is presented in Table~\ref{table:acronyms}.}

\begin{table}[tbh]
\centering
\begin{tabular}{ |l l| } 
\hline
\rowcolor{lightgray} Acronym & Definition \\
\hline
\hline
5QI & 5G Quality Indicator \\
BE & Best Effort \\
CBS & Committed Burst Size \\
CIR & Committed Information Rate \\
DRR & Deficit Round Robin \\
DSCP & Differentiated Services Code Point \\
eMBB & enhanced Mobile Broadband \\
HCTNS & Hierarchically Controlled Transport Network Slicing \\
HTB & Hierarchical Token Bucket \\
PBS & Peak Burst Size \\
PDB & Packet Delay Budget \\
PE & Provider Edge \\
PHB & Per-Hop-Behaviour \\
PIR & Peak Information Rate \\
PQ & Priority Queue \\
S-NSSAI & Single - Network Slice Selection Assistance Information \\
SDP & Service Demarcation Point \\
SLA & Service Level Agreement \\
SR & Segment Routing \\
SRH & Segment Router Header \\
TC & Traffic Control \\
TE & Traffic Engineering \\
TN & Transport Network \\
ToD & Teleoperated Driving \\
trTCM & two rate Three Color Marker \\
TS & Technical Specification \\
UPF & User Plane Function \\
URLLC & Ultra Reliable Low Latency Communications \\
VLAN & Virtual Local Area Network \\
VNX & Virtual Networks over LinuX \\
WFQ & Weighted Fair Queuing \\
\hline
\end{tabular}
\vspace*{0.25cm}
\caption{\rev{Summary of important acronyms.}}
\label{table:acronyms}
\end{table}

\section{Related Work}
\label{sec:soa}
\subsection{Network Slicing Standards}
\label{subsec:soa-standards}
The \gls{3gpp} has done a great deal of standardization work on network slicing and \gls{qos} in 5G. \gls{ts} 23.501 \cite{3GPP2023_1} defines the overall 5G system architecture, including a framework providing \gls{qos} to network flows and an identification mechanism for managing network slices. 

\gls{ts} 28.530~\cite{3GPP2024_1} defines the general concepts and definitions for network slicing, as well as the phases that compose the lifecyle of a network slice instance. \gls{ts} 23.502~\cite{3GPP2023_2} describes the procedures to define the \gls{qos} policies in the 5G network functions and to connect mobile terminals to slices. \gls{tr} 28.801~\cite{3GPP2018} defines network slice management functions. Management operations and procedures for provisioning network slices are defined in \gls{ts} 28.533~\cite{3GPP2024_2}.

The \protect\gls{gsma}~\cite{GSMA} has defined generic network slices templates (GST) that contain a set of attributes that can characterize a type of network slice or service. A \gls{nest} is obtained by assigning specific values to the fields in the GST. 3GPP network slice management functions build a network slice based on a provided \gls{nest}.

The above standards define control and management plane mechanisms required for the administration and orchestration of network slices. However, they do not specify how network slices can be implemented in the data plane. A topic neither addressed by the above standards is how to define an end-to-end network slice across the \gls{ran}, \gls{tn}, and core networks. 

The \gls{ietf} has made progress in this area by defining: (1) an orchestration and management framework that enables the inter-operation between 5G network slices in non-\gls{3gpp} \glspl{tn} that connect with network slices defined within a \gls{3gpp} domain~\cite{draft-ietf-teas-5g-network-slice-application}; and (2) how to identify a \gls{3gpp} network slice for associating it to a \gls{tn} slice and how the \glspl{5qi}, which define levels of \gls{qos} for the traffic flows, are translated to the \gls{tn} classes in the non-\gls{3gpp} domain~\cite{draft-ietf-teas-5g-ns-ip-mpls}~\cite{draft-cbs-teas-5qi-to-dscp-mapping}. By means of this identification, the \gls{tn} is capable of treating these traffic flows in the \gls{tn} with the expected level of \gls{qos}. This makes it possible to create end-to-end slices across all the network segments.

Network slice management, coordination and signaling between \gls{3gpp} and \gls{ietf} control elements are also necessary and partially defined in \cite{rfc9543, draft-ietf-teas-5g-network-slice-application}, but they are out of the scope of this paper.

\subsection{Network Slicing Literature}
\label{subsec:soa-literature}
There is extensive literature covering the application of network slicing techniques across the different segments of the network. Wang et al.~\cite{Wang2019} followed an approach based on queuing disciplines implemented over a software-defined network based on P4. They proposed a hierarchical queuing structure that has two levels. The first one is composed of eight round robin queues that ensure a proper load balance between the processed traffic. In the second level, each round robin queue is connected to four priority queues. In this way, it is possible to limit the delay of different traffic classes in each round robin queue. The problem with this approach is that it is not aligned with the model proposed by \gls{ietf} for \glspl{tn}, for instance, because it does not consider a policer controlling the traffic entering the network.

Other works~\cite{Bosk2021},~\cite{Gajic2022} proposed using traffic shaping mechanisms based on \gls{htb} for providing each network slice with a bandwidth guarantee and allowing a network slice to consume unused bandwidth by other slices. The proposed traffic shaping mechanism provides traffic isolation between network slices that are sharing the network infrastructure. Raussi \textit{et al.}~\cite{Raussi2023} also proposed the use of traffic shaping mechanisms based on \glspl{htb} to improve the reliability of communications in smart grids scenarios with wired connections. Lin et al.~\cite{Lin2021} implemented \gls{qos} framework over a P4-based network composed of four functional blocks: a classifier, a marker, a policer and a packet scheduler based on priority queues. In this work, all traffic marked as \textit{``green"} from all of the slices is enqueued into the same priority queue, so, it is not possible to control the delay of the \textit{``green"} traffic of each slice. Additionally, traffic marked as \textit{``yellow"} is enqueued in a lower priority queue, so it does not meet the latency requirements. \rev{Chen \textit{et al.} proposal~\cite{Chen2022}  uses two priority queues, one for the \textit{``green"} traffic coming from all the slices, and a lower priority one for \textit{``yellow"} traffic from all the slices and for non-sliced or best-effort traffic (this is a similar arrangement to~\cite{Lin2021}). The lower priority queue is further divided into four queues using a \gls{drr} to separate different types of flows, which allows to control the sharing of the available bandwidth beyond the one guaranteed to the network slices. This proposal meets the bandwidth slice requirements and has a mechanism to control the sharing of the additional bandwidth available, but because all the slices share the same queue, bandwidth sharing among them is not controlled and delay constraints are not addressed. In general, proposals that use \textit{``yellow"} traffic to enable bandwidth sharing have the drawback that losses and delay for such traffic cannot be controlled.}

\rev{Huin \textit{et al.}~\cite{huin} and Martin \textit{et al.}~\cite{martin} have followed a different approach, where network resources are exclusively dedicated to each slice. While this approach provides a high level of isolation between slices, it reduces network utilization and efficiency by preventing the sharing of unused bandwidth from one slice with others, potentially leading to under-utilization of resources.}

None of the above works addressed the problem of control the delay of different types of traffic flows. On the other hand, \rev{Chang \textit{et al.}\rev{~\cite{chang-5growth, Chang2021}} proposed a network slicing technique at the data link layer (L2) that is able to meet the latency and bandwidth requirements of different network slices by using a queuing system based on priority queues and Active Queue Management (AQM). The proposed model is implemented over a network of P4-programmable data plane switches, an approach that is still not aligned with \gls{ietf} priorities. Besides, a worst-case scenario with bursty traffic, as proposed in this paper, has also not been analyzed either.}

Baba et al. ~\cite{Baba2019} analyzed the impact of micro-bursts on the \gls{qos} of the 5G network, comparing the case of using a priority queuing scheduler versus a \gls{wfq} scheduler. However, they did not propose a solution to limit this impact.

Regarding vehicular networks, several works~\cite{Cui2022,Ndikumana2023,Cui2023,Cui2024,Zamfirescu2024} focused on the management of resources on the radio interface to support network slicing. The work in~\cite{Khan2021}, in addition to studying slicing techniques in the radio interface, also explored how to apply network slicing in the core of the network, and proposed the use of priority queues to achieve a latency in mission critical traffic lower than in best effort traffic. For the specific case of \gls{tod}, the work in~\cite{Campolo2017,Campolo2018} identified the need to define a network slice capable of meeting the strict \gls{qos} requirements of the service. 

\rev{None of the existing works in the state of the art align with the slicing model defined by the \gls{ietf} for 5G \glspl{tn}. Additionally, they fail to account for multiple flows with different \gls{qos} requirements per slice or to address how to limit delays while guaranteeing bandwidth requirements in worst-case scenarios under bursty traffic conditions. The solution proposed in this paper makes it possible to satisfy both the bandwidth and the latency requirements of \glspl{tn} slices and incorporate traffic burst control for worst-case scenarios.} 

\section{Background: The IETF Network Slice Model}
\label{sec:background}
\rev{This section provides an overview of the key features of the slicing model being defined by the \gls{ietf} for transport networks.}


The slicing model defined by the \gls{ietf}~\cite{rfc9543, draft-ietf-teas-5g-ns-ip-mpls} is composed of three parts. First, the model defines how the \gls{3gpp} 5G network slice identifier, namely the \gls{snssai}, which cannot be used in the \gls{ietf} transport network domain, can be identified in this domain by using L2/L3 header fields such as the VLAN Identifier or the MPLS label. The second part of the model focuses on how the \gls{qos} indicators used in the 5G network are mapped to values of the DSCP field of the IP header to mark the traffic in the transport network, with concrete recommendations for this mapping process given in~\cite{draft-cbs-teas-5qi-to-dscp-mapping}. Finally, the model defines how the data plane traffic is treated in the transport network so that the \gls{qos} requirements of the flows are correspondingly satisfied. \rev{Our proposal, described in Section~\ref{sec:proposal}, focuses on network realization, but we  also describe the other features considered to be part of our model}.

An \gls{ietf} network slice is defined between a set of \glspl{sdp}, that is, points of attachment for customers connecting their corresponding network slices. An \gls{sdp} has a unique identifier in the provider network scope, for instance, an IP address, a VLAN tag, an MPLS label, or an interface/port number. The network slice provides connectivity between the \glspl{sdp} and satisfies the latency and bandwidth requirements agreed between the network slice provider and the network slice consumer in the \gls{sla}. 

Below, we explain the components of the \gls{ietf} network slice model in more detail.

\subsection{Network Slicing Identification}
\label{subsec:background-nsi}
End-to-end network slices in the \gls{3gpp} domain are identified by \glspl{snssai} that are not visible in the \gls{ietf} domain. \mbox{\glspl{snssai}} are managed in the \gls{ran} and core segments of the 5G network, for example, to apply differentiated treatment in terms of \gls{qos} to the slices defined in the network. But, since the \gls{snssai} is not visible in the \gls{ietf} domain, different mechanisms have been proposed for the \gls{tn} to be able to identify the network slice that is associated to the incoming data traffic. These mechanisms, known as hand-off methods~\cite{draft-ietf-teas-5g-ns-ip-mpls, draft-ietf-teas-5g-network-slice-application}, are based on the use of L2, L3, or L4 identifiers. The main hand-off methods are detailed below:

\begin{itemize}
    \item \gls{vlan} hand-off. A \gls{vlan} tag is added to all the traffic exchanged between the transport network and the \gls{ran} or core networks. The VLAN ID is used by the \gls{pe} transport network nodes to identify the 5G network slice. Each \gls{sdp} is represented by a \gls{vlan} ID (or double \gls{vlan} with QinQ) and each \gls{vlan} represents a separated logical interface on the \glspl{pe}. The VLAN ID is only used to identify the 5G network slice, so the VLAN tag is removed by the \gls{pe} routers when forwarding the data traffic within the transport network.

        
    \item IP hand-off (IPv4 or IPv6). It consists on establishing associations between \glspl{snssai} and source/destination IP addresses. The association can be done in three different ways: (1) the \gls{snssai} is associated to the IP address allocated to the gNB or the \gls{upf}; (2) the \gls{snssai} is associated to a range of IP addresses allocated to a set of gNBs and \glspl{upf}; (3) the \gls{snssai} is associated to a subset of bits of an IP address, or (4) the \gls{snssai} is associated to a \gls{dscp} value. 
    \item \gls{mpls} Label hand-off. \gls{mpls} labels are used by the \gls{pe} nodes to infer the identification of the 5G network slice.
    \item \gls{udp} source port hand-off. \gls{udp} source ports carried over \gls{gtp}-U tunnels are used to identify the 5G network slice. 
\end{itemize}

Regardless of the identification method used, it is also necessary to map 5G network slices to \gls{tn} slices. In \cite{draft-ietf-teas-5g-ns-ip-mpls}, the \gls{ietf} proposes three alternatives for making this mapping

\subsection{QoS Flow Identification}
\label{subsec:background-qfi}

As in the previous case, the \gls{3gpp} \gls{qos} identifiers are only visible in the \gls{3gpp} domain, which makes it necessary to translate them in the \gls{ietf} domain for providing different \gls{qos} treatments in the \glspl{tn} for each \gls{3gpp} 5G slice or for each 5G \gls{qos} flow within each \gls{3gpp} 5G slice. 

Each \gls{qos} flow in the \gls{3gpp} domain, either sent from the \gls{upf} to the \gls{ran} or vice versa, traverses the \gls{tn} through a \gls{gtp} tunnel, and it is identified by a \gls{qfi} that is carried in the \gls{gtp} header. Using these \glspl{qfi}, gNBs and \glspl{upf} apply the \gls{qos} policies defined in the \gls{3gpp} domain. A \gls{qfi} is associated with a \gls{5qi} that indicates the \gls{qos} characteristics of the traffic flow, such as the \gls{pdb} or the priority level. Packets from \gls{qos} flows are marked by using the \gls{dscp} field of the IP header. The \gls{ietf} describes an example of how the mapping between a \gls{5qi} and a \gls{dscp} can be done~\cite{draft-cbs-teas-5qi-to-dscp-mapping}. A \gls{5qi} can take multiple values, as has been defined in TS.23.501~\cite{3GPP2023_1}, and, in practice, \glspl{5qi} with similar characteristics are mapped to the same \gls{dscp} value. Packets received on the ingress ports of the \gls{pe} routers of the transport network are marked with these \gls{dscp} values. These marks enable the \gls{pe} routers to classify the incoming traffic and associate it with a \gls{tn} \gls{qos} class, so traffic can be treated according to the characteristics defined by the corresponding \glspl{5qi}. This association is defined by a label in the headers added to the packets when transported through a tunnel in the transport network, commonly referred to as the \gls{tn} \gls{qos} mark. These tunnels can be implemented over MPLS networks (and in this case, the Traffic Class field is used) or over IPv6 (\gls{dscp} values are used in this case). In the \gls{ietf} model, eight transport network classes have been considered, as typical router hardware supports up to eight traffic queues per port. These \gls{tn} \gls{qos} classes determine a \gls{phb} enforced to packets in each \gls{tn} node. From the eight available transport network classes, according to \cite{draft-cbs-teas-5qi-to-dscp-mapping}, four are reserved for the data plane traffic and the remaining four for control plane traffic. Consequently, data traffic is grouped into four \gls{5qi} categories: 

\begin{itemize}
    \item Group 1: delay-critical traffic with \gls{gbr}, allowing packet losses between $10^{-6}$ and $10^{-4}$).
    \item Group 2: traffic with moderated delay and varying packet loss levels.
    \item Group 3: other \gls{gbr} traffic not included in Groups 1 or 2.  
    \item Group 4: \gls{5qi} values assigned for non-\gls{gbr} traffic.
\end{itemize}

 For describing our proposal, the above groups will be referred as \gls{tn} \gls{qos} class A, \gls{tn} \gls{qos} class B, \gls{tn} \gls{qos} class C, and \gls{tn} \gls{qos} class D, respectively.

\subsection{IETF Network Slice Realization}
\label{subsec:background-realization}
An important part of the \gls{ietf} model described in~\cite{draft-ietf-teas-5g-ns-ip-mpls} is the realization of the transport network slices. Implementing a network slice in the \gls{ietf} domain requires combining different network mechanisms. The mechanisms identified in the \gls{ietf} draft are the following: L2VPN and/or L3VPN service instances for logical separation of the slices; fine-grained resource control for enforcing the bandwidth contract at the edges of the provider network for each \gls{qos} flow and slice; coarse-grained resource control for applying \gls{qos} mechanisms to flows aggregated in traffic classes within the transport network; and capacity planning/management mechanisms ensuring that enough capacity is available across the transport network for the network slices. Below, a more \rev{detailed} description of these mechanisms is provided.

\subsubsection{L2VPN and/or L3VPN for slices isolation}
L2VPN and/or L3VPN service instances might be deployed for achieving logical separation of slices. This results in an additional outer header due to the packet encapsulation carried out in the border nodes hosting the services instances, i.e., in the \glspl{pe}. This also provides a clear separation between the \gls{dscp} value that is used to identify the \gls{qos} in traffic received from the \gls{3gpp} domain and the \gls{tn} \gls{qos} mark used to determine the \gls{qos} in the \gls{tn}. These mechanisms might also be used to deploy different underlay transport paths optimized according to the \rev{latency and bandwidth requirements specified in the} \glspl{sla} of different 5G \gls{qos} flows. For example, for traffic with low latency requirements, a path with fewer hops would be preferable.

It is important to note that the \gls{tn} \gls{qos} classes and the underlay transport path have different targets. The \gls{tn} \gls{qos} class defines the \gls{phb} enforced to packets that transit the \gls{tn}, whereas the underlay transport determines the overall path taken by packets based on the operator’s requirements. These underlay transports can be realized through various mechanisms, such as \gls{rsvpte} or \gls{srte} tunnels.

\subsubsection{Fine-grained resource control at the edge of the TN}

This mechanism works as an admission control that enforces the bandwidth contract at the edge of the provider network by using traffic policers. A \gls{pe} receiving traffic from the 5G network, i.e., an ingress \gls{pe}, enforces a rate limitation policy to guarantee bandwidth per slice and per traffic class within each slice. Since resources are controlled for each traffic class and/or slice, this approach is referred to as fine-grained resource control. The \gls{ietf} describes a model (known as \textit{5QI-aware model}) in which the traffic policer is implemented by means of a hierarchical model. On the one hand, the \gls{pe} includes a traffic policer per slice (\textit{slice policer}). This traffic policer follows a two-rate three-color rate limiter approach. Traffic exceeding the \gls{pir} (maximum bitrate) is dropped. Traffic under the \gls{cir} is marked as ``green", and traffic between the \gls{cir} and \gls{pir} is marked as ``yellow''. In this way ``yellow'' traffic will be dropped in case of congestion. On the other hand, the \gls{pe} includes a traffic policer per \gls{qos} class. This traffic policer follows a single-rate two-color rate limiter approach, where only a \gls{cir} is defined and all traffic exceeding this \gls{cir} is dropped. The class policer does not apply any rate limitation to the best effort traffic, allowing this type of traffic to use the bandwidth in the slice that is not used by the \gls{qos} classes.

A \gls{pe} forwarding traffic to the 5G network, i.e., an egress \gls{pe}, may optionally incorporate a hierarchical scheduler or shaper that ensures the guaranteed bandwidth for each \gls{cir} slice and for each \gls{cir} \gls{qos} class defined.

\rev{In this paper, we propose a novel fine-grained resource control based on a \gls{htb} scheduler composed of three levels (Section~\ref{sec:proposal}.\ref{subsec:proposal-finegraine}).}

\subsubsection{Coarse-grained resource control in the transit links}

For the transit nodes of the transport network, the \gls{ietf} proposes a simplified traffic treatment mechanism based on the combination of priority queues and \gls{wfq} or \gls{drr} queues. As a result, there is no fine-grained resource control at the transit nodes as they do not perform traffic treatments per traffic class. Instead, they handle data traffic following a DiffServ model that supports up to eight transport network \gls{qos} classes (as explained in Section~\ref{sec:background}.\ref{subsec:background-qfi}). All the slices defined in the \gls{3gpp} domain are mapped to a \gls{tn} \gls{qos} class in the \gls{ietf} domain. In scenarios where the number of classes or slices is greater than the number of \gls{tn} \gls{qos} classes, it is possible to aggregate multiple slices or classes into a single \gls{tn} \gls{qos} class.  

Typically, traffic classes with strict latency requirements are associated with priority queues, while others are linked to \gls{wfq} or \gls{drr} non-priority queues. Non-priority queues are configured so that their weights are adjusted according to the \gls{qos} requirements of the classes sharing a \gls{tn} \gls{qos} class. With this coarse-grained resource control, traffic from best effort classes or traffic marked as ``yellow'' can be discarded during network congestion events while maintaining the \gls{qos} of the traffic flows in other classes.

This flat \gls{qos} model, based on traffic scheduling/prioritization, entails a different \gls{qos} treatment which is enforced in all \gls{tn} nodes at transit links to apply the same \gls{phb} for all packets that belong to each \gls{tn} \gls{qos} class.

\subsubsection{Capacity planning/management}
\gls{te} mechanisms, based on technologies such as \gls{rsvpte} or \gls{srte} tunnels, are employed to ensure that sufficient capacity is available along the transport network for all the accepted slices. This is a topic that is not covered in this paper. 

\section{HCTNS Model}
\label{sec:proposal}
\subsection{General View}
\label{subsec:general-view}

\begin{figure*}[th]
\vspace*{-0.12cm}
\centerline{\includegraphics[width=1.58\columnwidth]{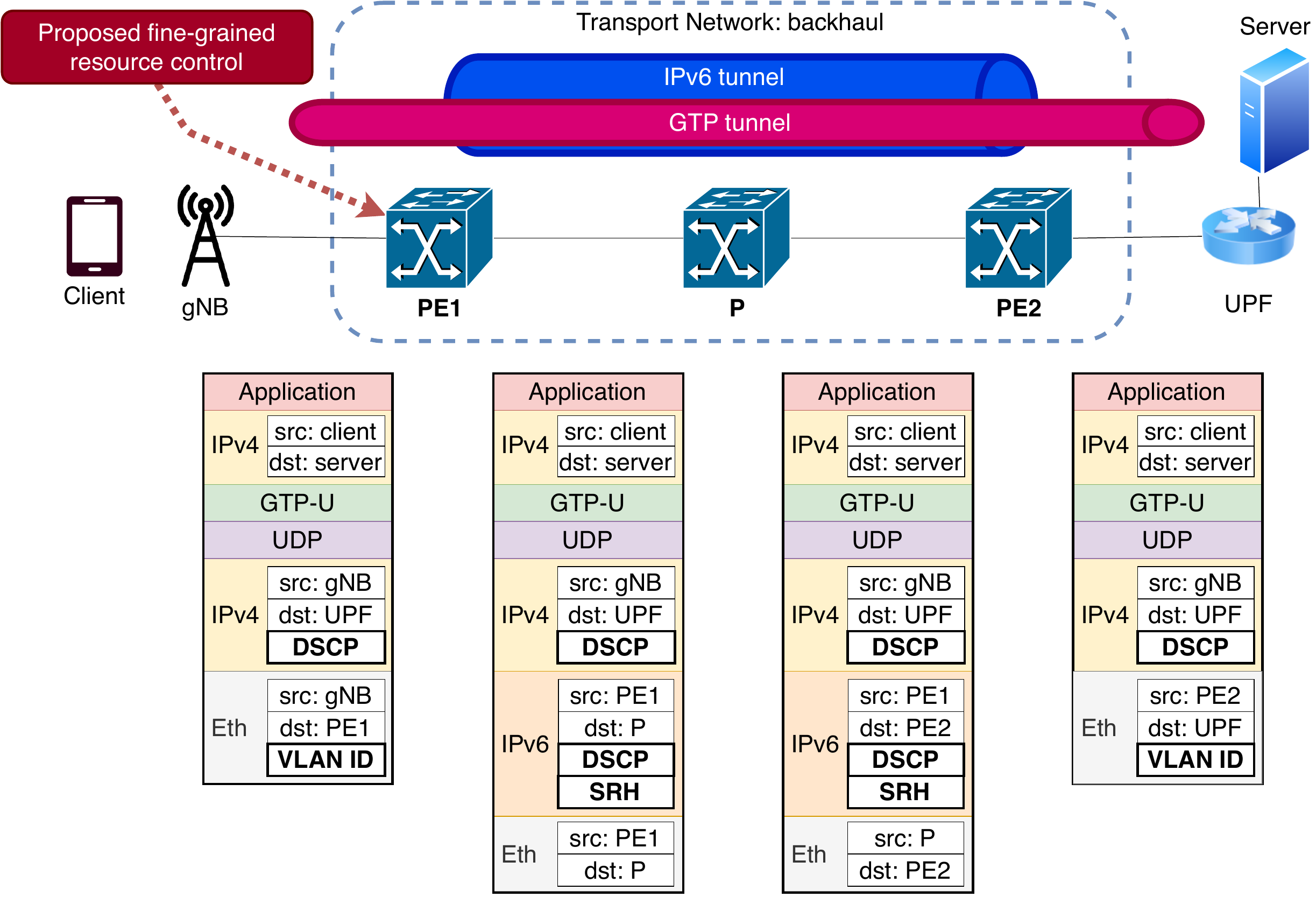}}
\caption{General view of the \gls{hctns} model.} 
\label{general-view}
\end{figure*}

A general view of the network slicing model proposed in this paper is illustrated in Fig.~\ref{general-view}, which also shows where our proposal for the ingress \textit{fine-grained resource control} is placed. The top of the figure shows a simplified view of a 5G transport network, illustrating how data traffic is double-encapsulated into two tunnels: GTP and IPv6. The bottom of the figure shows the protocol stack for packets transmitted in the uplink direction.

Our design for the fine-grained ingress resource control addresses some limitations we have identified in the \gls{ietf} model (described in Section~\ref{sec:background}.\ref{subsec:background-realization}): (1) Inflexible and inefficient bandwidth sharing between different 5G \gls{qos} classes and slices. The \gls{ietf} model allows unused bandwidth in a transport network class to be used only by the best-effort traffic. In contrast, \gls{hctns} improves resource utilization between 5G \gls{qos} classes and between slices because the bandwidth not consumed by a traffic class is made available for use by other traffic classes in the same slice or by other network slices. To address this, we propose a three-level hierarchical ingress policing mechanism, including a new top-level policer, referred to as \textit{global policer} (1st level), followed by \textit{slice policers} (2nd level), and \textit{class policers} (3rd level). Furthermore, we propose two mechanisms to determine how the bandwidth is shared between the 5G \gls{qos} classes of a slice and between the slices. (2) Lack of a traffic burst control mechanism. Some services may need a traffic burst control for meeting their applications demands. However, traffic bursts, as it will be demonstrated in Section~\ref{sec:eval}, can increase packet delays and compromise the \gls{qos} levels agreed in the \glspl{sla} associated with the classes and slices. For instance, packet delay in classes associated to \gls{tn} \gls{qos} class B may be increased when there are traffic bursts in the traffic classes that are assigned to \gls{tn} \gls{qos} class A. To address this, \gls{hctns} incorporates a mechanism to control traffic bursts. This mechanism is based on the definition of two new parameters at the class policers. The three-level structure of the \gls{htb} that we propose controls the maximum accepted rate, i.e., global \gls{cir}, while also guaranteeing the \glspl{cir} per class and slice even in worst-case networks scenarios with bursty traffic. (3) In the \gls{ietf} model, ``yellow'' traffic (the traffic of a class that arrives at the transport network's ingress at a rate between the \gls{cir} and the \gls{pir}) can cause network congestion, and may be discarded at transit nodes of the transport network. In contrast, our approach ensures that all traffic admitted by the policers and forwarded through the transport network is treated with the same \gls{qos} guarantees, without causing congestion, and with no packet losses. 

\rev{\gls{hctns}, in addition to our proposed \textit{fine-grained ingress resource control} component, incorporates some of the features of the slicing model defined by the \gls{ietf} (Section~\ref{sec:background}). These features are the following:}

\begin{itemize}

    \item Network Slice Identification (Section~\ref{sec:background}.\ref{subsec:background-nsi}): \gls{hctns} assumes the use of \gls{vlan} tags to identify network slices in the border between the \gls{3gpp} and \gls{ietf} domains. As shown in Fig.~\ref{general-view},  the Ethernet frame transmitted by the gNB to the ingress \gls{pe} router includes the 802.1Q extension header, which carries the VLAN identifier of the slice. Any other type of the hand-off mechanisms \rev{ described in Section~\ref{sec:background}.\ref{subsec:background-nsi}}, could also be applied. 
    
    \item \gls{qos} Flow Identification (Section~\ref{sec:background}.\ref{subsec:background-qfi}): \gls{hctns} adopts the \gls{ietf} \gls{5qi}-aware model, where the \glspl{5qi} values used in the \gls{3gpp} network are mapped to \gls{dscp} values in the transport network. As shown in Fig.~\ref{general-view}, the packet transmitted by the gNB to the UPF,  whether IPv4 or IPv6 (IPv4 in this case) carries the \gls{dscp} value corresponding to the \gls{5qi} of the traffic flow of a slice. The ingress \gls{pe} router uses this \gls{dscp} value to perform an additional mapping to a \gls{tn} \gls{qos} class. This mapping is reflected in the IPv6 header that encapsulates the IPv4 packet transmitted by the gNB. Flows with similar \gls{qos} requirements (i.e., \gls{5qi} values that ensure a comparable level of \gls{qos}) can be assigned to the same \gls{tn} \gls{qos} class.  
    
    \item Network Slice Realization (Section~\ref{sec:background}.\ref{subsec:background-realization}): \gls{hctns} considers the deployment of \glspl{l3vpn} based on IPv6 tunnels for: (1) the traffic isolation among slices, and (2) the transport of the \gls{tn} \gls{qos} class identifier, which is included in the \gls{dscp} field of the IPv6 packet header that is encapsulating the IPv4 packet. IPv6 tunnels are assumed to be implemented by using \gls{sr}. As shown in Fig.~\ref{general-view}, IPv6 packets exiting the ingress \gls{pe} router carry the \gls{dscp} identifying the \gls{tn} \gls{qos} class and a \gls{srh}, which is used to configure the path followed by packets in the underlying network. The implementation of IPv6 tunnels may also take into account capacity and planning/methods, something that falls outside the scope of this paper and that may be addressed in future steps of the investigation. Regarding coarse-grained resource control, our model is closely aligned with the model proposed by the \gls{ietf}, which is based on priority and \gls{drr} queues.     
    
\end{itemize}

\rev{Although the paper focuses on 5G transport networks, HCTNS has been designed based on the IETF framework, allowing the proposed network slicing model to be applied to any IP or MPLS network. In the case of MPLS networks, instead of using IPv6 tunnels and the DSCP field (which identifies the TN QoS Class for Coarse-grained Resource Control), MPLS tunnels (e.g., VPLS) and the MPLS Traffic Class field would be employed for the same purpose.}

A detailed description of the fine-grained resource control mechanism proposed is explained below.  

\subsection{A New Model for Fine-Grained Resource Control}
\label{subsec:proposal-finegraine}

\begin{figure}[t]
\centerline{\includegraphics[width=0.85\columnwidth]{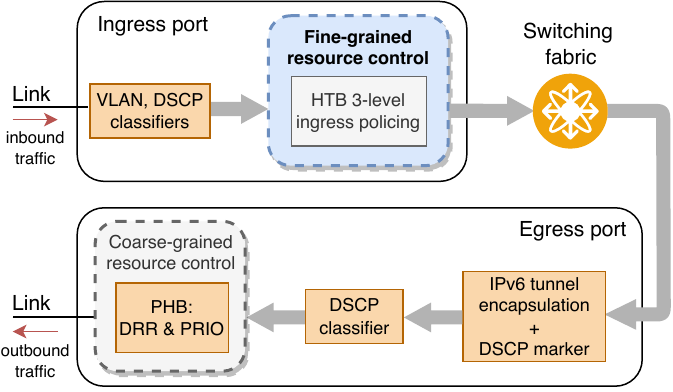}}
\caption{Input and output port processing at the \gls{pe} ingress node.}
\label{ingress_operations}
\end{figure}

\textit{Fine-grained resource control} is implemented in the ingress and egress \gls{pe} routers of the transport network, which handle, respectively,  various data plane functions in their input and output ports. As Fig.~\ref{ingress_operations} illustrates, the input port of a \gls{pe} is responsible for classifying the incoming packets according to the values of the VLAN tag and the \gls{dscp} header fields. This classification identifies the network slice and its associated \gls{qos} requirements.  Based on these identifiers, packets are enqueued into a \gls{htb} queuing system that implements our proposed three-level hierarchical ingress policing mechanism. Additionally, the identifiers are then used to map the packet to the corresponding \gls{tn} \gls{qos} class. As Fig.~\ref{ingress_operations} shows, the output port is responsible for sending the traffic over a IPv6 tunnel that carries its own \gls{dscp} mark. This marking is used by the coarse-grained resource control mechanism to identify the \gls{tn} \gls{qos} class and apply the appropiate \gls{phb} treatment. Packets are then enqueued into a queuing system based on priority and \gls{drr} queues, ensuring they are processed according to the \gls{qos} level defined for their \gls{tn} \gls{qos} class. 

\begin{figure}[b]
\centerline{\includegraphics[width=0.5\textwidth]{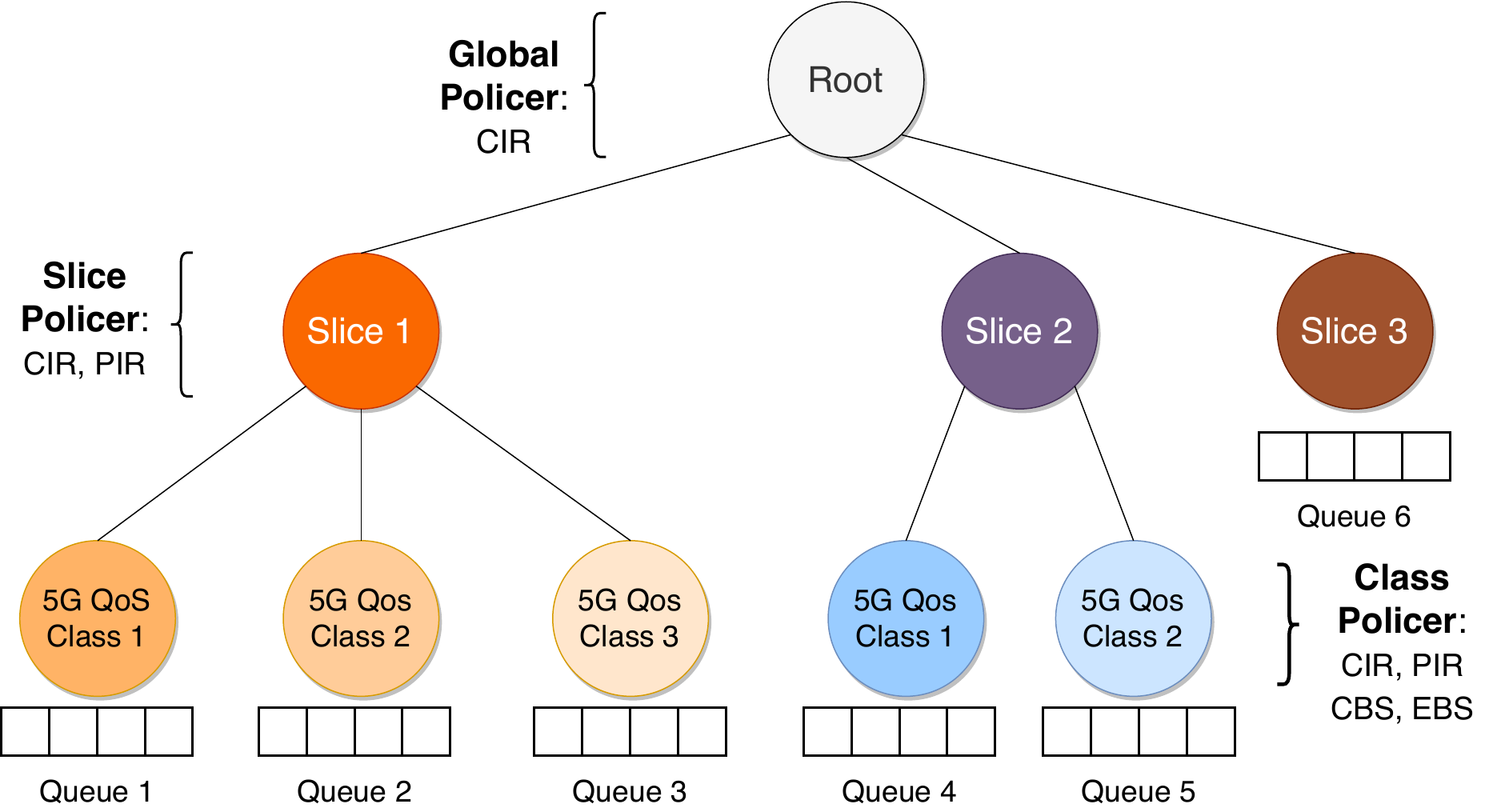}}
\caption{Three-level hierarchical ingress policing mechanism.}
\label{htb_logical}
\end{figure}

The three-level hierarchical ingress policing mechanism proposed in this paper, implemented using the \gls{htb} rate-limiting algorithm, is illustrated in Fig.~\ref{htb_logical}.  The \gls{htb} allows organizing traffic into a hierarchy of classes (or nodes), with each class ensuring its bandwidth guarantees. As shown in Fig.~\ref{htb_logical}, only the leaf nodes of the \gls{htb} tree have an associated queue, where packets can wait to be transmitted to an output port of the \gls{pe} router. The tree-based hierarchical structure of the \gls{htb} allows classes under the same parent node to share bandwidth not consumed by the other classes. This is a key advantage of our proposal. The \gls{ietf} in~\cite{draft-ietf-teas-5g-ns-ip-mpls} employs a two-level hierarchy (slice and class) for its policing mechanism, which makes bandwidth sharing between slices and 5G \gls{qos} classes inefficient and without \gls{qos} guarantees. In the \gls{ietf} model, unused bandwidth can only be used by the best-effort class within the same slice. 

Fig.~\ref{htb_logical} shows the global policer at the root of the \gls{htb} tree. The global policer limits the total amount of traffic allowed to pass through the ingress \gls{pe} router, defined by the \gls{cir} parameter. In our proposal, packets exceeding the \gls{cir} limit are discarded, making \gls{cir} and \gls{pir} equivalent in this case. All traffic admitted to traverse the transport network will be treated as ``green" by the transit nodes, and will not be dropped, as we will explain later. The global policer is also important to handle the bursty traffic, since it controls that, on average, the total maximum capacity is not exceeded. 

\begin{figure*}[hbtp]
\vspace*{-0.1cm}
\centerline{\includegraphics[width=0.9\textwidth]{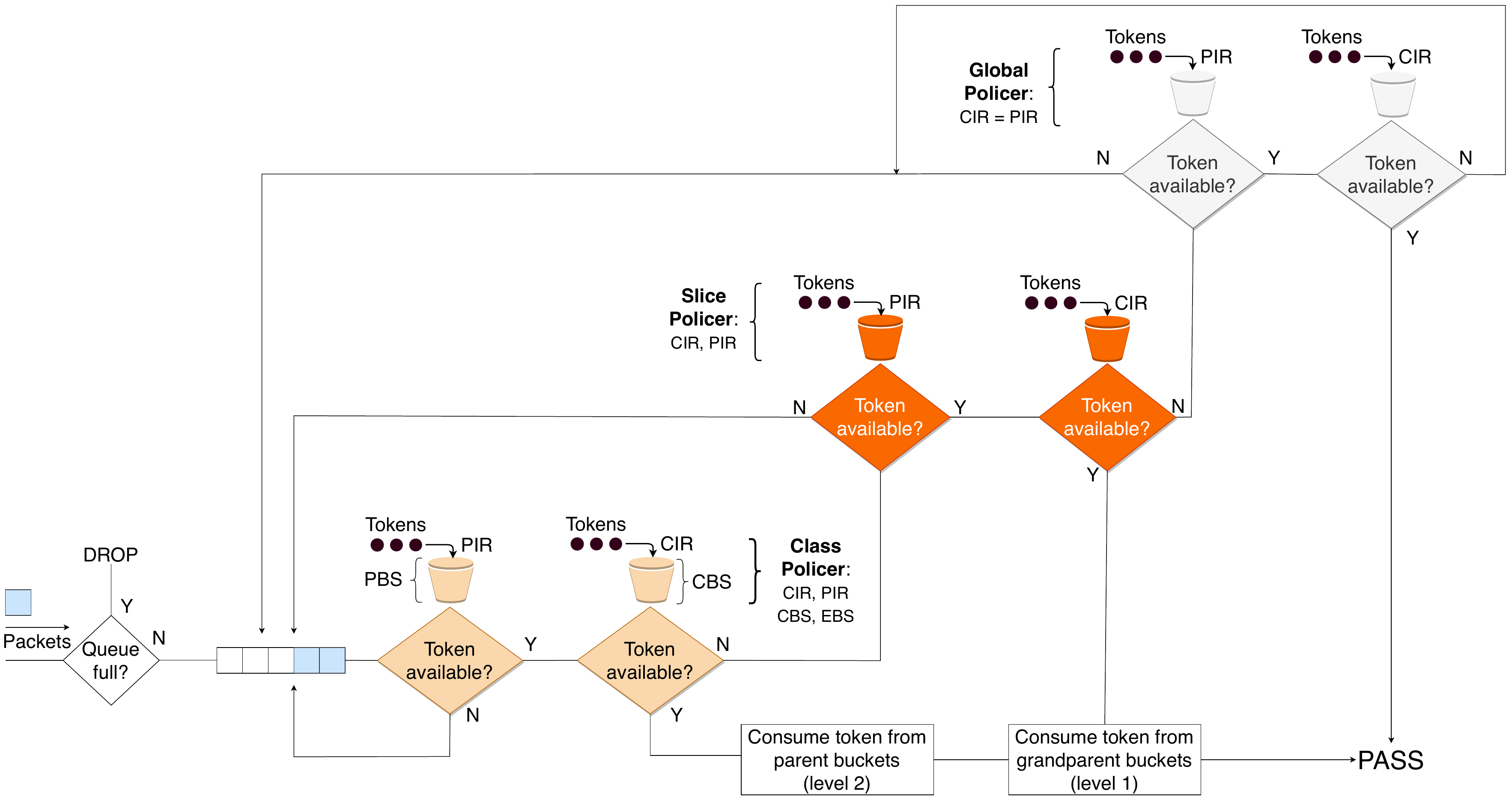}}
\caption{\gls{htb} operation diagram.}
\label{htb_physical}
\end{figure*}

The second level of the \gls{htb} hierarchy is occupied by slice policers, which enforce rate limiting at the slice level. These policers are defined by two parameters: \gls{cir} and \gls{pir}, which, in this case, may not be equivalent. The \gls{cir} represents the guaranteed bitrate of the slice, while the \gls{pir} defines its maximum. The sum of the \gls{cir} for all the slices must not exceed the \gls{cir} of the global policer. Regarding bandwidth sharing, any slice can make use of the bandwidth not consumed by other slices. The maximum bandwidth a slice can borrow from the root node is equal to its \gls{pir}-\gls{cir}. Slices where the \gls{cir} is equal to the \gls{pir} are not allowed to borrow bandwidth, making this an interesting method for controlling bandwidth consumption.  

The third level of the \gls{htb} hierarchy is occupied by class policers, which are used by slices transporting various \gls{qos} flows with different \gls{qos} requirements (e.g., slice 1 and 2 in Fig.~\ref{htb_logical}). In network slices where all traffic flows receive the same \gls{qos} treatment, only the global and slice policers are applied. Class policers are defined by four parameters: \gls{cir}, \gls{pir}, \gls{cbs}, and \gls{pbs}. As explained later, the \gls{cbs} and the \gls{pbs} parameters play a key role in controlling the bursty traffic of the slices. On the other hand, the sum of the \gls{cir} values for all classes must not exceed the \gls{cir} of the corresponding slice policer. Besides, any class can make use of the bandwidth not consumed by other classes within the same slice (i.e., under the same parent node).

Our bandwidth sharing mechanism allows all slices and classes to borrow excess available bandwidth equally. However, \gls{hctns} also introduces two more advanced methods to specify how excess or unused bandwidth can be shared between slices and, within a slice, among different \gls{qos} classes. These methods define how bandwidth is distributed when several nodes under the same parent node in the \gls{htb} try to obtain bandwidth from that parent node at the same time. The methods are implemented by controlling the distribution of tokens that is used to assign bandwidth in the \gls{htb} as we will explain later.  

\begin{itemize}
    \item Weight/Quantum-based bandwidth sharing. The available bandwidth is shared among slices and \gls{qos} classes based on weights (relative) or quantums (fix amount of bytes) assigned to the slice and/or class policers. When defined at the slice level, the average bandwidth a slice $\theta$ can borrow from the global available bandwidth (assuming that all the slices have packets to transmit), is calculated as: $B_\theta = B_r \cdot \frac{w_\theta}{\sum_{i=1}^{N} w_i}$, where $N$ is the total number of slices, $B_r$ is the global available bandwidth, $w_\theta$ is the weight/quantum assigned to slice $\theta$ and $w_i$ the weight/quantum assigned to the \textit{i}-th slice. When defined at both the slice and the class levels, the average bandwidth a class $\phi$ within a slice $\theta$ can borrow from the global available bandwidth  (assuming that all classes within the same slice have packets to transmit) is calculated as: $B_{\theta, \phi} = B_r \cdot \frac{w_\theta}{\sum_{i=1}^{N} w_i} \cdot \frac{w_\phi}{\sum_{j=1}^{C} w_j}$, where C is the total number of classes within the slice $\theta$, $w_\phi$ the weigh/quantum assigned to slice $\phi$ and $w_j$ the weight/quantum assigned to the \textit{j}-th class within the slice $\theta$. 
    \item Priority-based bandwidth sharing. In this method, each slice and/or \gls{qos} class is assigned a priority value, with lower values indicating  higher priority. The available global bandwidth is shared based on these priority levels. When defined at the slice level, slices with higher priority are granted bandwidth first, either until the bandwidth is exhausted or the slices have reached their \gls{pir} limit. When defined at both the slice and class levels, the bandwidth each class within a slice can borrow from the global available bandwidth depends on both the priority of the slice and the priority of the class within that slice. 
\end{itemize}

These methods can be combined in various ways. For instance, the weight-based method may be applied at the slice level while the priority-based method may be applied at the class level. Another option is to use both methods at both levels, as they are not mutually exclusive. Nevertheless, regardless of the bandwidth sharing method employed, it is important to note that the \gls{cir} of each slice and \gls{qos} class is always guaranteed because we are sharing only the available excess bandwidth.

The use of a \gls{htb} on the ingress port of the \gls{pe} router does not impact packet delay, despite the typical behavior of \glspl{htb}. The packet scheduler of the \gls{htb} first dequeues all traffic under the \gls{cir} limit for each leaf node, and after that, traffic between \gls{cir} and \gls{pir}. This causes traffic between \gls{cir} and \gls{pir} to wait for transmission, potentially violating the service delay requirements. However, in our proposal, the \gls{htb} operates on the \gls{pe} ingress interface, where packets traverse the switching fabric at a very high speed. As a result, packet delay is negligible when processed by our \gls{htb}-based fine-grained resource control mechanism.

The control of traffic bursts, i.e., traffic received at an unlimited rate for a bounded time, is a key feature of our proposal. Each class policer is defined by two parameters that regulate traffic bursts for the \gls{qos} flows: the \gls{cbs} and the \gls{pbs}. The \gls{cbs} indicates the maximum number of bits that can be transmitted above the \gls{cir} during a burst period, while the \gls{pbs} indicates the maximum number of bits allowed to be transmitted above the \gls{pir} during the same period. To better illustrate how bandwidth sharing and traffic burst control works in our proposal, Fig.~\ref{htb_physical} presents the internal structure of the proposed three-level \gls{htb}. 

Fig.~\ref{htb_physical} shows that each policer is associated with two token buckets. The bucket on the right, hereinafter referred to as \textit{\gls{cir}-bucket} is filled with tokens at the \gls{cir} rate, ensuring a guaranteed bandwidth. The bucket on the left, hereinafter referred to as \textit{\gls{pir}-bucket}, is filled with tokens at the \gls{pir} rate, ensuring that traffic does not exceeds a maximum bandwidth. The size of each bucket is determined by the \gls{cbs} and \gls{pbs} parameters. Buckets cannot store more tokens than their capacity (\gls{cbs} or \gls{pbs}) allows, so the tokens received when the buckets are full are discarded. Having this in mind, an \gls{htb} works as follows. If a packet arrives in the queue but there are no tokens available in the \textit{\gls{pir}-bucket}, the maximum bandwidth has been reached and the packet must wait in the queue until new tokens become available. If the \textit{\gls{pir}-bucket} is not empty, the \textit{\gls{cir}-bucket} is checked. If there are tokens available in the \textit{\gls{cir}-bucket}, the packet is within the \gls{cir} limit, and can be transmitted to the switching fabric, consuming a token from both buckets. Tokens consumed in the buckets of a child node must also be deducted from the parent nodes. For example, when a packet consumes tokens from the buckets associated with a class policer, tokens are also consumed from the buckets associated with the slice and global policers. In case there are no tokens available in the class policer \textit{\gls{cir}-bucket}, the child node requests tokens to the parent (slice) node, which first checks its \textit{\gls{pir}-bucket}. If no tokens are available in the \textit{\gls{pir}-bucket} the slice's maximum bandwidth has been reached and the packet must wait in the queue. If the \textit{\gls{pir}-bucket} is not empty, and there are tokens available in the slice's \textit{\gls{cir}-bucket}, the child node can borrow bandwidth from the parent (slice) node and the packet can be transmitted to the switching fabric (tokens are consumed in the slice and in the global level).  If no tokens are found at the slice level (the slice \textit{\gls{cir}-bucket} is empty), the slice policer requests tokens to the global policer. In summary, if the child node and its ancestors have no tokens for the packet transmission, the packet waits in the queue. However, if tokens are available in the \gls{htb} tree, tokens are consumed at different levels to maintain bandwidth control at both the slice and the global levels. The \gls{htb} queues have a limited size, and when they are full, packets begin to be dropped. This behavior enables the enforcement of the policying function.

In the following, we discuss how the \gls{htb} can be configured to control traffic bursts. Traffic burst control is effectively disabled in a bucket when it is sized to store just the tokens needed to transmit a single packet. If the traffic burst control is set in the \textit{\gls{cir}-bucket}, but not in the \textit{\gls{pir}-bucket}, packets can be dequeued at most at the \gls{pir} rate. This is because packet transmissions require the simultaneous consumption of tokens from both the \gls{cir} and the \gls{pir} buckets, and the \textit{\gls{pir}-bucket} is only storing tokens for transmitting one packet. In contrast, if the traffic burst control is set in the \textit{\gls{pir}-bucket} but not in the \textit{\gls{cir}-bucket}, the child node may request additional tokens to its parent node, as Fig.~\ref{htb_physical} shows. However, this does not guarantee the transmission of the traffic burst either. Finally, if the traffic burst control is set in both the \gls{cir} and \gls{pir} buckets, packets can be dequeued immediately as long as there are tokens available in both buckets, regardless of the availability of tokens in the buckets of the parent nodes. This is the configuration we have selected to control traffic bursts in our proposal, with the \gls{pbs} and \gls{cbs} set to the same value. Using this configuration, the \gls{pe} router can accept traffic bursts during a period that depends on the size of the buckets. After that burst period, a new burst will only be accepted if packets arrive at the ingress \gls{pe} port at a lower rate than the \gls{cir} rate, so that the \textit{\gls{cir}-bucket} can be refilled, and at a lower rate than the \gls{pir} rate, so that the \textit{\gls{pir}-bucket} can also be refilled. 

Our approach consists on enabling traffic burst control exclusively in leaf nodes, which implies specifying values for \gls{cbs} and \gls{pbs} only in the buckets associated with classes. To be able to accept traffic bursts, defining traffic burst control in the class policers is mandatory. Otherwise, even if the buckets associated with the slice and the global policer have tokens, a packet from a \gls{qos} class received at a rate higher than the class policer \gls{pir} would not be dequeued until a new token is received in the buckets associated with the class policer. The values of \gls{cbs} and \gls{pbs} will control the burst size allowed per \gls{qos} class. When the burst size is smaller than the \gls{pbs}/\gls{cbs}, there are sufficient tokens to forward all packets in the burst to the switching fabric at the rate they are received. However, if the burst size exceeds the \gls{pbs}/\gls{cbs}, only part of the burst is forwarded at the rate packets are received. This forwarding ends when both buckets are depleted. The remainder of the burst will be enqueued in the \gls{htb} queue until it reaches its capacity and will be transmitted at the \gls{cir} rate defined by the class policer. Any additional packets beyond the queue’s limit will be discarded.

As shown in Fig.~\ref{htb_physical}, packet transmission is allowed if there are tokens in the class policer, regardless the availability of tokens in the parent nodes. However, when packet transmission is allowed, tokens are consumed from the policers at all levels. This may result, when sending traffic bursts, in the parent nodes not having enough tokens in their buckets. In such cases, the token counters in the \gls{htb} parent nodes can take negative values, which can be considered as ''tokens to be paid''. Nodes with ''negative'' tokens cannot share tokens with their child classes, until the counters return to positive values. During this period, traffic will only flow at the \gls{cir} rate of the leaf nodes. This ensures that while the global maximum rate of the parent node may be temporally exceeded, it is eventually compensated to maintain the average maximum rate of the parent node within the defined limit. 

\begin{figure}[b]
\centerline{\includegraphics[width=0.9\columnwidth]{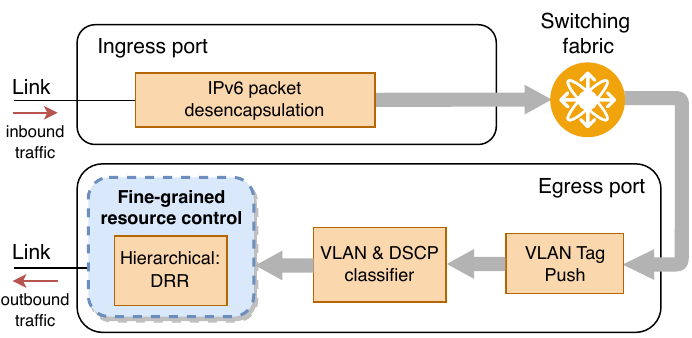}}
\caption{Input and output port processing at the \gls{pe} egress node.}
\label{egress_operation_flow}
\end{figure}

\begin{figure}[t]
\centerline{\includegraphics[width=0.8\columnwidth]{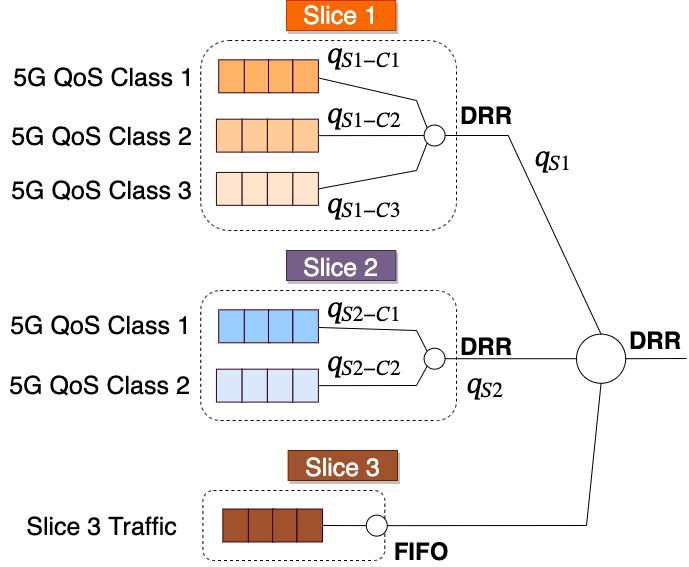}}
\caption{\rev{Hierarchical} \gls{drr} queuing system, as proposed by the \gls{ietf}.}
\label{egress_queue_scheme}
\end{figure}

As Fig.~\ref{egress_operation_flow} shows, a fine-grained resource control is also implemented in the \gls{pe} egress routers. The input port of the \gls{pe} router terminates the IPv6 tunnel, forwarding packets to the output ports of the \gls{pe} based on their IP destination address. On the physical output ports, various logical interfaces are defined with the VLAN-tagged mode enabled to carry tags identifying the network slices in the \gls{3gpp} domain. The output port applies a fine-grained resource control using a hierarchical two-level \gls{drr} queuing system. We have adopted the same model proposed by the \gls{ietf} in~\cite{draft-ietf-teas-5g-ns-ip-mpls}, and shown in Fig.~\ref{egress_queue_scheme}. The first \gls{drr} level is used to differentiate the 5G \gls{qos} classes within the same network slice, while the second \gls{drr} level differentiates the slices. 

The \gls{ietf} resource control mechanism can maintain the \gls{cir} of 5G slices and their associated traffic classes even in scenarios with network congestion. However, thanks to the global policer proposed in \gls{hctns}, if a slice sends traffic above its \gls{cir} it is mainly because there is unused bandwidth available from other slices. So, by applying \gls{hctns}, the 5G slices are allowed to consume more bandwidth without causing network congestion. 

Although the model has been described considering uplink communications, the concepts described here are also valid for downlink traffic. 

\subsection{Configuring the Coarse-Grained Resource Control}
\label{subsec:proposal-coarsegrained}  

\begin{figure}[tb]
\centerline{\includegraphics[width=1\columnwidth]{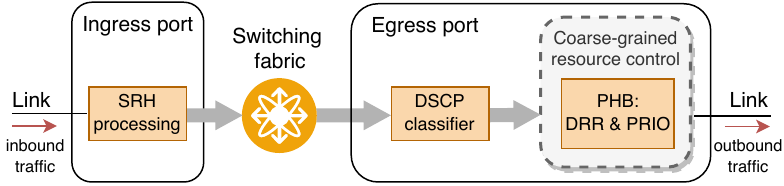}}
\caption{Input and output port processing at the P transit node.}
\label{P_operation_flow}
\end{figure}

The coarse-grained resource control is applied to the output ports of the \gls{pe} and P routers, as shown in Figs.~\ref{ingress_operations} and~\ref{P_operation_flow}. The output port of the \gls{pe} router first encapsulates packets in IPv6 tunnels based on \gls{sr}, adding a mark to the \gls{dscp} value in the IPv6 header that identifies the \gls{tn} class associated with the packet. A classifier uses this \gls{dscp} mark to enqueue the packets of each \gls{tn} class into the corresponding priority/\gls{drr} queue for the coarse-grained resource control. This allows traffic flows in the \gls{tn} to be treated according to their delay and bandwidth requirements. The output port of the P router works similarly but does not tunnel the packets, as P routers only forward traffic within the underlay network. Forwarding is done according to the \gls{srh} of the packet (referred to as \gls{srh} processing in Fig.~\ref{P_operation_flow}), which determines the next IPv6 hop in the underlay network. 

\begin{figure*}[bth]
\centerline{\includegraphics[width=2\columnwidth]{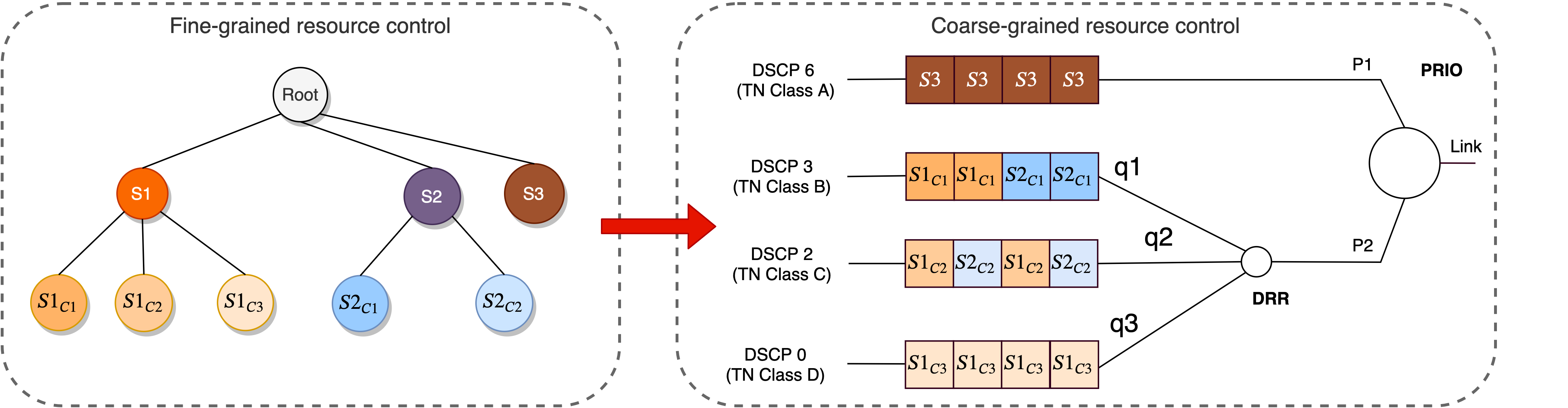}}
\caption{Queuing scheme at transit links for data plane traffic, based on the \gls{ietf} proposal \cite{draft-cbs-teas-5qi-to-dscp-mapping}.}
\label{transit_queues}
\end{figure*}

Fig.~\ref{transit_queues} shows the coarse-grained resource control (on the right) and its relationship with the fine-grained resource control (on the left). As an example, Fig.~\ref{transit_queues} shows the slice $S3$ as the highest priority slice. This slice has no associated classes, and therefore, does not have any child nodes in the \gls{htb} tree. Traffic from slice $S3$, marked with \gls{dscp} 6, enters the priority queue in the coarse-grained resource control. Slice $S1$ has three classes associated ($S1_{C1}$, $S1_{C2}$, and $S1_{C3})$, with traffic class $S1_{C1}$ marked with \gls{dscp} 3, $S1_{C2}$ marked with \gls{dscp} 2, and $S1_{C3}$ marked with \gls{dscp} 0. Similarly, slice $S2$ has two associated classes ($S2_{C1}$ and $S2_{C2}$), marked with \gls{dscp} 3 and 2 respectively. As shown in Fig.~\ref{transit_queues}, packets are associated to a queue in the \gls{drr} queuing system based on the \gls{dscp} value of the IP header, enabling the differentiation of traffic classes with different levels of \gls{qos}. Traffic from these classes must wait in the corresponding \gls{drr} queues until the priority queue is empty.

\section{Performance Evaluation}
\label{sec:eval}
\subsection{Experimental Platform}

\begin{figure}[t]
\centerline{\includegraphics[width=0.95\columnwidth]{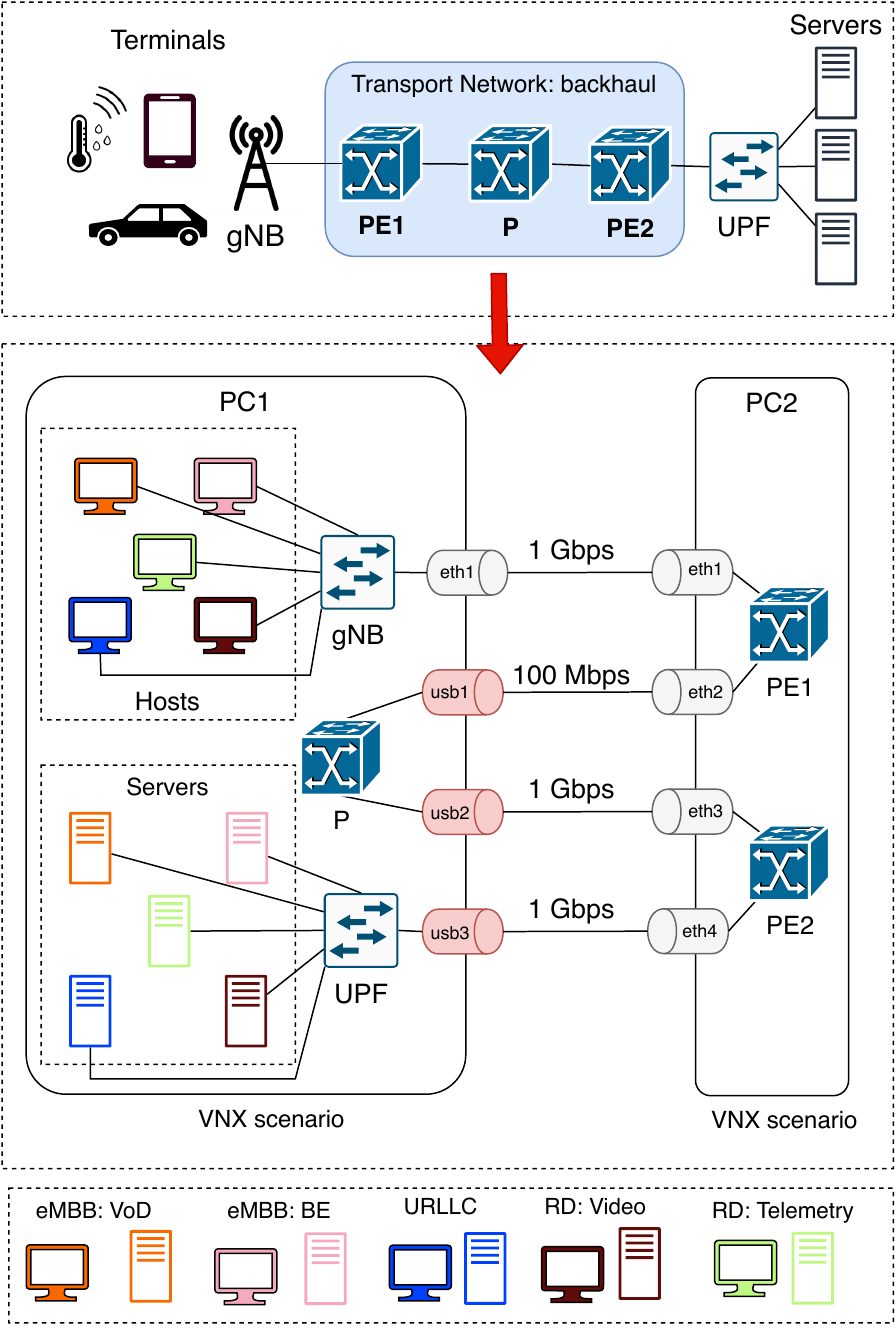}}
\caption{Experimental platform for transport network slicing.}
\label{platform}
\end{figure}

To evaluate \gls{hctns}, we have developed a platform that combines physical and virtual network equipment. The platform's software has been published in open access\footnote{\label{repo}https://github.com/giros-dit/net-slicing-emulator} for community use. The top of Fig.~\ref{platform} shows a conceptual view of the network scenario, consisting of a gNB and a \gls{upf} connected via a transport network. Terminals connect to standard 5G slices such as \gls{embb} or \gls{urllc}, as well as to more innovative slices, such as a \gls{tod} slice. The 5G network is interconnected via a simplified version of the transport network, consisting of three nodes: an ingress, a transit, and an egress \gls{pe} router. As shown in the middle part of Fig.~\ref{platform}, this network scenario has been deployed on two physical PCs connected via four communication links. 

PC1 is equipped with an Intel Core i7-8700 processor with 6 cores (clock speed 3.2\,GHz), 32\,GB of RAM, and a 512\,GB SSD. As Fig.~\ref{platform} shows, PC1 connects to PC2 by using a PCI Express Gigabit Ethernet network card with a single port and three USB-Ethernet adapters. PC2 is equipped with an Intel Core i7-4790 processor with 4 cores (clock speed 3.6\,GHz), 32\,GB of RAM, and a 256\,GB SSD. In this case, PC2 connects to PC1 using PCI Express Gigabit Ethernet network card with four Ethernet Ports. Both PCs run the Ubuntu 22.04 Linux distribution as their operating system. The network adapters can be configured flexibly to emulate links with capacities of 10\,Mbps, 100\,Mbps or 1\,Gbps, using Ethtool\footnote{https://linux.die.net/man/8/ethtool}. This approach was chosen instead of deploying the scenario in a fully virtualized environment, because setting virtual links speeds is more challenging and less realistic. To facilitate experiments, the communication links capacities in the testbed are scaled down compared to those in a real transport network~\cite{UC3M2018}. Additionally, to conduct  saturation tests on the input port of the \gls{pe} router, the link between PE1 and P has been set to 100\,Mbps, which is lower capacity than the rest of the links in the platform.

Within each PC, a virtual scenario has been deployed using the network virtualization tool \gls{vnx}\footnote{https://web.dit.upm.es/vnxwiki/index.php/Main\_Page}. The \gls{vnx} scenario on PC1, see Fig.~\ref{platform}, includes the P router and several hosts and servers, defined as LinuX Containers (LXC) running Ubuntu 22.04. These hosts and servers act as traffic sources and sinks for the traffic in our experiments. The \gls{vnx} scenario on PC1 also includes two Linux bridges emulating a gNB and a \gls{upf} 5G router. On the other hand, the \gls{vnx} scenario on PC2 includes the ingress and egress \gls{pe} routers. Using this setup, synthetic traffic was generated with the Iperf\footnote{https://iperf.fr/} tool to measure the \gls{qos} level of the traffic flows when the global available bandwidth is shared among different slices. 

Our experimental campaign focuses on analyzing the \gls{qos} levels achieved by traffic flows of these slices in the uplink direction, i.e., traffic transmitted from the gNB to the \gls{upf} router, although a similar analysis for downlink traffic would yield comparable results. This focus on the uplink is due to a particular interest in understanding how a \gls{tod} slice may be defined, as initially analyzed in~\cite{wimob}. The \gls{tod} slice is more bandwidth demanding in the uplink than in the downlink, as vehicles transmit video signals and telemetry data to a remote operations center. 

\begin{table}[t]
\vspace*{0.1cm}
\centering
\begin{tabular}{|c|c|c|c|c|}
\hline
Slice                 & Traffic Class & 5QI                                                   & DSCP & TN QoS Class \\ \hline
URLLC                 & URLLC         & 82                                                    & 46   & A            \\ \hline
\multirow{2}{*}{ToD}  & Video         & 130*                                                  & 38*  & B            \\ \cline{2-5} 
                      & Telemetry     & 131*                                                  & 28*  & C            \\ \hline
\multirow{2}{*}{eMBB} & VC            & 2                                                     & 28   & C            \\ \cline{2-5} 
                      & BE            & 9 (default)                                           & 0    & D            \\ \hline
\end{tabular}
\vspace*{0.3cm}
\caption{Slices and traffic classes in the experimental campaign.}
\label{tab:slices}
\end{table}

\begin{figure*}[t]
\vspace{-0.1cm}
\centering
\includegraphics[width=1.8\columnwidth]{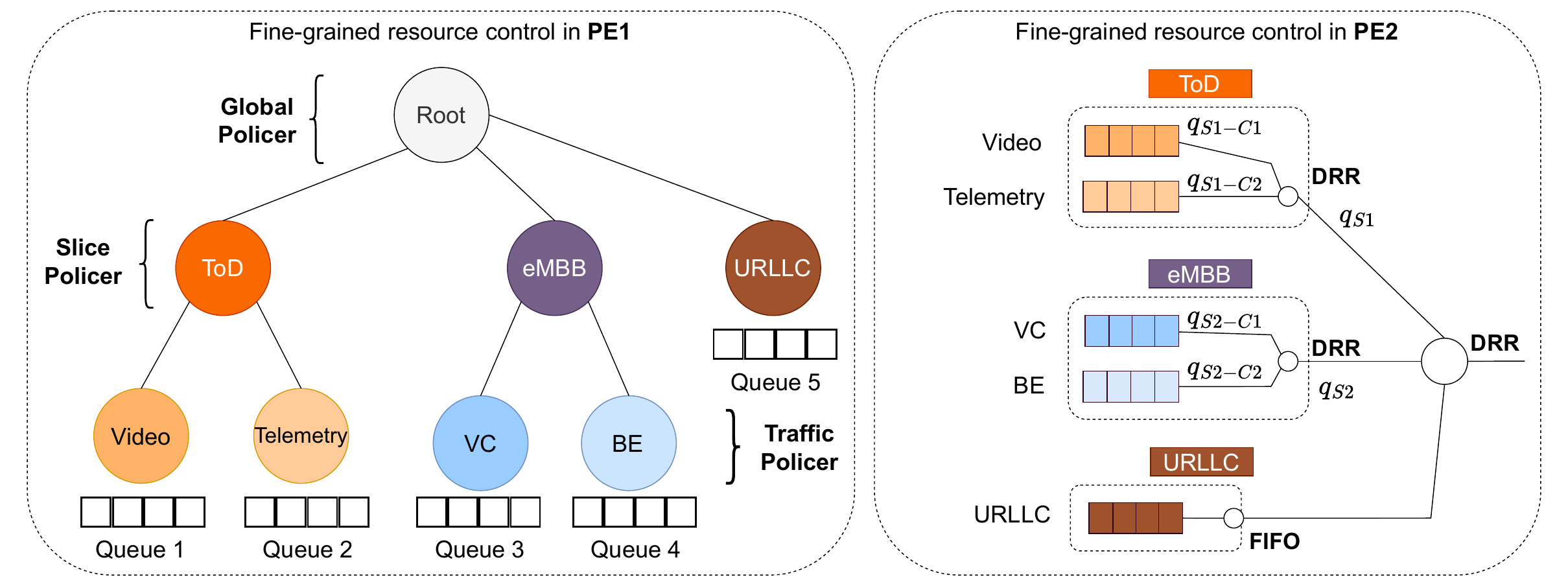}
\vspace*{0.5em}
\caption{Fine-grained resource control implementation in PE1 and PE2.}
\label{htb-lab}
\end{figure*}

At the logical level, we have three network slices in our experiments: the mentioned \gls{tod} slice, and an \gls{embb} and a \gls{urllc} slice. We define two traffic classes in the \gls{tod} slice: video and telemetry classes. The \gls{embb} slice has two associated traffic classes: Best Effort (BE) and Video Conferencing (VC). In contrast, the \gls{urllc} slice has no associated traffic classes. Table~\ref{tab:slices} summarizes this information and indicates a possible mapping between the \glspl{5qi} values that specify the \gls{qos} characteristics of the 5G network slices, the \gls{dscp} values, and their corresponding \gls{tn} \gls{qos} classes. This mapping is relevant for defining how the data traffic will be treated by the fine-and-coarse grained resource control mechanisms in our experiments. \glspl{5qi} values for the \gls{embb} and \gls{urllc} slices are obtained from~\cite{3GPP2023_1}, while their associated \gls{dscp} values and \gls{tn} \gls{qos} classes are derived from~\cite{draft-cbs-teas-5qi-to-dscp-mapping}. Since the \gls{tod} slice is not defined in the above documents, we propose new \glspl{5qi} values for the \gls{tod} video and telemetry traffic, along with a mapping of these \glspl{5qi} to \glspl{dscp} and \gls{tn} \gls{qos} classes.

Both the fine-grained and the coarse-grained mechanisms have been implemented using the Linux \gls{tc} tool\footnote{https://man7.org/linux/man-pages/man8/tc.8.html}. In the following, we provide implementation details of the \gls{hctns} model. For more details, please refer to our open-access code repository\footref{repo}. 

As described in Section~\ref{sec:proposal}, \gls{hctns} uses a three-level hierarchical \gls{htb} rate-limiting algorithm for fine-grained resource control at the ingress \gls{pe}1. In our platform, this \gls{htb} is implemented with the \gls{tc} \textit{htb qdisc} queuing discipline, which has been used to hierarchically organize the slices and traffic classes, as shown in Fig.~\ref{htb-lab}. \rev{\gls{tc} \textit{htb qdisc} was originally designed for shaping mechanisms applied to output interfaces,  and it cannot be directly applied to input interfaces, which is a limitation of the TC tool. To implement our proposed three-level hierarchical ingress policing mechanism, we have defined an Intermediate Functional Block (IFB) logical interface in the \gls{pe} router, enabling the redirection of incoming traffic to this logical interface and treating it as an output port. This logical interface is not the actual output port of the PE1 router, but an intermediary interface for handling ingress traffic according to the defined policing mechanism. Using TC, we have configured the different policers (class, slice, and global) for the three types of experiments conducted, as specified in Tables~\ref{tab:expa-params}, \ref{tab:expb-params} and \ref{tab:expc-params}, which define the CIR, PIR, CBS, and PBS values that determine the behavior of \gls{hctns} at the ingress interface of PE1.}

The token-sharing methods described in Section~\ref{sec:proposal}.\ref{subsec:proposal-finegraine} (Weight/Quantum-based and Priority-based) are configured in Linux \gls{htb} by using the \textit{quantum} and \textit{prio} parameters. \rev{However, the Linux \gls{htb} implementation has two differences with the proposed \gls{htb} model described in Subsection \ref{sec:proposal}.\ref{subsec:proposal-finegraine}. On the one hand, \gls{tc} allows those parameters to be defined only at the leaf nodes of the \gls{htb} tree. The description of the experiments will indicate how these parameters have been configured in our platform. On the other hand, the Linux \gls{htb} implementation does not distribute the excess bandwidth proportionally among the active classes. When quantum values are used to distribute the excess bandwidth, traffic classes that consume bandwidth 
not used by other classes within the same slice receive less proportion of the excess bandwidth.}

To ensure that the traffic generated by the hosts in our platform (see Fig.~\ref{platform}) is queued according to the corresponding traffic class in the \gls{htb}, \gls{tc} \textit{traffic filters} have been used. These filters emulate, to some extent, the classifiers of the transport network routers. In the current version of the platform, traffic classification is based on the source IP address of the packets. Similarly, we have defined \textit{traffic filters} to ensure that traffic exiting the logical IFB interface is correctly enqueued in the priority and \gls{drr} queues of the coarse-grained resource control associated with the output port of the \gls{pe} router.

The \gls{pe} egress router implementation differs slightly from \gls{pe}1, as the fine-grained resource control for outgoing traffic is based on a hierarchical \gls{drr} queuing discipline following the \gls{ietf} model (Section~\ref{sec:proposal}.\ref{subsec:proposal-finegraine}). For this, we use the \textit{\gls{drr} qdisc}, with three sub-classes, one per slice. The \gls{embb} and \gls{tod} slices  have a second level in the hierarchy for their corresponding traffic classes, as shown in Fig.~\ref{htb-lab} (on the right).

Coarse-grained resource control is implemented using the \gls{tc} \textit{PRIO qdisc} queuing discipline. With \gls{tc} \textit{PRIO qdisc} we have defined two classes with two priority levels, assigning higher priority to the \gls{urllc} slice over the \gls{embb} and \gls{tod} slices. As a result, \gls{embb} and \gls{tod} packets must wait for transmission if \gls{urllc} packets are present in the corresponding queue. The \gls{urllc} traffic is handled in the transport network using the \gls{tn} \gls{qos} class A. The lowest priority class is managed using a \textit{\gls{drr} qdisc}, with three sub-classes corresponding to \gls{tn} \gls{qos} classes B, C, and D. Table~\ref{tab:slices} shows the \gls{tn} \gls{qos} classes associated with each traffic class defined in our experiments. Each \gls{drr} queue is configured with a \textit{quantum}, which specifies the amount of bytes that can be dequeued from a class before the packet scheduler moves to the next class. As it will be described later, we have varied the values of the quantums associated with the \gls{drr} queues to analyze its impact on the network performance. 

Packets transmitted from the output port of the \gls{pe}1 router are sent to the P router at 100\,Mbps (see Fig.~\ref{platform}), the maximum global bandwidth allowed by the telco-operator in our emulated network. However, the link between the P router with the \gls{pe}2 router has a speed of 1\,Gbps, resulting in no queuing delay at either the P or \gls{pe}2 routers. 

\subsection{Test}
    \label{sec:test}

\rev{For validation and insight into the behavior of our proposed network slicing model, we conducted an experimental campaign comparing \gls{hctns} with the slicing models defined by the \gls{ietf} and the proposal in Lin \textit{et al.}~\cite{Lin2021}. Based on the related work (Section~\ref{sec:soa}.\ref{subsec:soa-literature}), we selected the model proposed by Lin \textit{et al.} because similar approaches are commonly used in the literature to implement network slices in transport networks. The campaign is composed of three experiments: \textit{Experiment A}, \textit{Experiment B} and \textit{Experiment C}}. 

\rev{\textit{Experiment A} compares the performance of the \gls{ietf} model~\cite{rfc9543, draft-ietf-teas-5g-ns-ip-mpls}, the state-of-the-art model \cite{Lin2021}, and \gls{hctns}} by analyzing the evolution of latency, bandwidth, packet losses, and number of packets waiting in the queuing system associated with the output port of the \gls{pe}1 router. \rev{This is done while generating UDP traffic at a constant bitrate of 100 Mbps for all defined traffic classes.} In this experiment, we also evaluate a combination of the proposed quantum and priority based bandwidth sharing mechanisms described in Section~\ref{sec:proposal}.\ref{subsec:proposal-finegraine}. Traffic burst control is disabled. 

\textit{Experiment B} also compares \gls{hctns} and the \gls{ietf} network slicing models but with a focus on analyzing the impact of coarse-grained resource control configuration on the bandwidth consumed by the telemetry and video traffic classes of the \gls{tod} slice. As in the previous experiment, traffic burst control is disabled and UDP traffic has been generated at a constant bitrate \rev{(100\,Mbps)}. \rev{In this experiment, the model in \cite{Lin2021} is excluded from the analysis because it cannot incorporate different coarse-grained resource control configurations.} 

Finally, \textit{Experiment C} demonstrates the operation of the traffic burst control mechanism in \gls{hctns}, highlighting the impact of traffic bursts on latency experienced by packets from different traffic classes and slices. Under these conditions, we analyze how the configuration of coarse-grained resource control affects the latency experienced by packets of the different traffic classes. \rev{In~\cite{Lin2021}, traffic bursts are not considered. However, the policer the~\cite{Lin2021} model uses to split \textit{``green"} and \textit{``yellow"} traffic has parameters to accept traffic bursts. Therefore, we have configured them to compare the performance in the presence of bursts with \gls{hctns}}. 

\rev{The packet size was set to 1538 bytes (including link and physical headers) in all experiments. Since the analyzed models, including our proposal, manage rate in byte/s (not in packet/s), packet size, apart from the overhead, should not affect the behavior of the models or their performance, although we plan to carry out more experiments in our future work to further corroborate this.}

\begin{table*}
\centering
\resizebox{\textwidth}{!}{%
\begin{tabular}{c|
>{\columncolor[HTML]{FFFFFF}}c 
>{\columncolor[HTML]{FFFFFF}}l 
>{\columncolor[HTML]{FFFFFF}}c 
>{\columncolor[HTML]{FFFFFF}}c 
>{\columncolor[HTML]{FFFFFF}}c 
>{\columncolor[HTML]{FFFFFF}}l 
>{\columncolor[HTML]{FFFFFF}}c 
>{\columncolor[HTML]{FFFFFF}}c 
>{\columncolor[HTML]{FFFFFF}}c 
>{\columncolor[HTML]{FFFFFF}}c 
>{\columncolor[HTML]{FFFFFF}}c 
>{\columncolor[HTML]{FFFFFF}}c |
>{\columncolor[HTML]{FFFFFF}}c 
>{\columncolor[HTML]{FFFFFF}}c |}
\cline{2-15}
& \multicolumn{12}{c|}{\cellcolor[HTML]{F5F5F5}\textbf{Fine-grained Ingress Policer Resource Control}}                                                                                                                                                                                                                                                                                                                                                                                                                                                                                                                                                                                                                               & \multicolumn{2}{c|}{\cellcolor[HTML]{DAE8FC}\textbf{\begin{tabular}[c]{@{}c@{}}Coarse-grained \\ Resource Control\end{tabular}}}                                                             \\ \hline
\multicolumn{1}{|c|}{\cellcolor[HTML]{FFFFFF}}                                         & \multicolumn{4}{c|}{\cellcolor[HTML]{FFFFFF}\textbf{Class Policer}}                                                                                                                         & \multicolumn{4}{c|}{\cellcolor[HTML]{FFFFFF}\textbf{Slice Policer}}                                                                                                                                                          & \multicolumn{2}{c|}{\cellcolor[HTML]{FFFFFF}}                                                                                                & \multicolumn{2}{c|}{\cellcolor[HTML]{FFFFFF}}                                                                                                           & \multicolumn{2}{c|}{\cellcolor[HTML]{FFFFFF}}                                                                                                                                                \\ \cline{2-9}
\multicolumn{1}{|c|}{\cellcolor[HTML]{FFFFFF}}                                         & \multicolumn{2}{c|}{\cellcolor[HTML]{FFFFFF}\textbf{CIR}}        & \multicolumn{2}{c|}{\cellcolor[HTML]{FFFFFF}\textbf{PIR}}                                                                & \multicolumn{2}{c|}{\cellcolor[HTML]{FFFFFF}\textbf{CIR}}                & \multicolumn{2}{c|}{\cellcolor[HTML]{FFFFFF}\textbf{PIR}}                                                                                         & \multicolumn{2}{c|}{\multirow{-2}{*}{\cellcolor[HTML]{FFFFFF}\textbf{Global Policer}}}                                                       & \multicolumn{2}{c|}{\multirow{-2}{*}{\cellcolor[HTML]{FFFFFF}\textbf{\begin{tabular}[c]{@{}c@{}}trTCM \cite{Lin2021}\end{tabular}}}} & \multicolumn{2}{c|}{\multirow{-2}{*}{\cellcolor[HTML]{FFFFFF}\textbf{\begin{tabular}[c]{@{}c@{}}TN QoS Class: \\ Parameters\end{tabular}}}}                                                  \\ \cline{2-15} 
\multicolumn{1}{|c|}{\multirow{-3}{*}{\cellcolor[HTML]{FFFFFF}\textbf{Traffic Class}}} & \multicolumn{2}{c|}{\cellcolor[HTML]{FFFFFF}\textbf{IETF/HCTNS}} & \multicolumn{1}{c|}{\cellcolor[HTML]{FFFFFF}\textbf{IETF}} & \multicolumn{1}{c|}{\cellcolor[HTML]{FFFFFF}\textbf{HCTNS}} & \multicolumn{2}{c|}{\cellcolor[HTML]{FFFFFF}\textbf{IETF/HCTNS}}         & \multicolumn{1}{c|}{\cellcolor[HTML]{FFFFFF}\textbf{IETF}}              & \multicolumn{1}{c|}{\cellcolor[HTML]{FFFFFF}\textbf{HCTNS}}             & \multicolumn{1}{c|}{\cellcolor[HTML]{FFFFFF}\textbf{IETF}}         & \multicolumn{1}{c|}{\cellcolor[HTML]{FFFFFF}\textbf{HCTNS}}             & \multicolumn{1}{c|}{\cellcolor[HTML]{FFFFFF}\textbf{CIR}}                             & \cellcolor[HTML]{FFFFFF}\textbf{PIR}                            & \multicolumn{1}{c|}{\cellcolor[HTML]{FFFFFF}\textbf{IETF/HCTNS}}        & \cellcolor[HTML]{FFFFFF}\cite{Lin2021}                                                            \\ \hline
\multicolumn{1}{|c|}{\cellcolor[HTML]{FFFFFF}URLLC}                                    & \multicolumn{2}{c|}{\cellcolor[HTML]{FFFFFF}N/A}                 & \multicolumn{1}{c|}{\cellcolor[HTML]{FFFFFF}N/A}           & \multicolumn{1}{c|}{\cellcolor[HTML]{FFFFFF}N/A}            & \multicolumn{2}{c|}{\cellcolor[HTML]{FFFFFF}1.2 Mbps}                    & \multicolumn{1}{c|}{\cellcolor[HTML]{FFFFFF}100 Mbps}                   & \multicolumn{1}{c|}{\cellcolor[HTML]{FFFFFF}100 Mbps}                   & \multicolumn{1}{c|}{\cellcolor[HTML]{FFFFFF}}                      & \multicolumn{1}{c|}{\cellcolor[HTML]{FFFFFF}}                           & \multicolumn{1}{c|}{\cellcolor[HTML]{FFFFFF}1.2 Mbps}                                 & 100 Mbps                                                        & \multicolumn{1}{c|}{\cellcolor[HTML]{FFFFFF}A: Priority Queue (PQ)}                      & \cellcolor[HTML]{FFFFFF}                                                                                           \\ \cline{1-9} \cline{12-14}
\multicolumn{1}{|c|}{\cellcolor[HTML]{FFFFFF}Video}                                    & \multicolumn{2}{c|}{\cellcolor[HTML]{FFFFFF}32 Mbps}             & \multicolumn{1}{c|}{\cellcolor[HTML]{FFFFFF}N/A}           & \multicolumn{1}{c|}{\cellcolor[HTML]{FFFFFF}100 Mbps}       & \multicolumn{2}{c|}{\cellcolor[HTML]{FFFFFF}}                            & \multicolumn{1}{c|}{\cellcolor[HTML]{FFFFFF}}                           & \multicolumn{1}{c|}{\cellcolor[HTML]{FFFFFF}}                           & \multicolumn{1}{c|}{\cellcolor[HTML]{FFFFFF}}                      & \multicolumn{1}{c|}{\cellcolor[HTML]{FFFFFF}}                           & \multicolumn{1}{c|}{\cellcolor[HTML]{FFFFFF}32 Mbps}                                  & 100 Mbps                                                        & \multicolumn{1}{c|}{\cellcolor[HTML]{FFFFFF}B: quantum = 1538 B}                   & \cellcolor[HTML]{FFFFFF}                                                                                           \\ \cline{1-5} \cline{12-14}
\multicolumn{1}{|c|}{\cellcolor[HTML]{FFFFFF}Telemetry}                                & \multicolumn{2}{c|}{\cellcolor[HTML]{FFFFFF}4 Mbps}              & \multicolumn{1}{c|}{\cellcolor[HTML]{FFFFFF}N/A}           & \multicolumn{1}{c|}{\cellcolor[HTML]{FFFFFF}100 Mbps}       & \multicolumn{2}{c|}{\multirow{-2}{*}{\cellcolor[HTML]{FFFFFF}36 Mbps}}   & \multicolumn{1}{c|}{\multirow{-2}{*}{\cellcolor[HTML]{FFFFFF}N/A}}      & \multicolumn{1}{c|}{\multirow{-2}{*}{\cellcolor[HTML]{FFFFFF}100 Mbps}} & \multicolumn{1}{c|}{\cellcolor[HTML]{FFFFFF}}                      & \multicolumn{1}{c|}{\cellcolor[HTML]{FFFFFF}}                           & \multicolumn{1}{c|}{\cellcolor[HTML]{FFFFFF}4 Mbps}                                   & 100 Mbps                                                        & \multicolumn{1}{c|}{\cellcolor[HTML]{FFFFFF}}                           & \cellcolor[HTML]{FFFFFF}                                                                                           \\ \cline{1-9} \cline{12-13}
\multicolumn{1}{|c|}{\cellcolor[HTML]{FFFFFF}VC}                                       & \multicolumn{2}{c|}{\cellcolor[HTML]{FFFFFF}52.8 Mbps}           & \multicolumn{1}{c|}{\cellcolor[HTML]{FFFFFF}N/A}           & \multicolumn{1}{c|}{\cellcolor[HTML]{FFFFFF}100 Mbps}       & \multicolumn{2}{c|}{\cellcolor[HTML]{FFFFFF}}                            & \multicolumn{1}{c|}{\cellcolor[HTML]{FFFFFF}}                           & \multicolumn{1}{c|}{\cellcolor[HTML]{FFFFFF}}                           & \multicolumn{1}{c|}{\cellcolor[HTML]{FFFFFF}}                      & \multicolumn{1}{c|}{\cellcolor[HTML]{FFFFFF}}                           & \multicolumn{1}{c|}{\cellcolor[HTML]{FFFFFF}52.8 Mbps}                                & 100 Mbps                                                        & \multicolumn{1}{c|}{\multirow{-2}{*}{\cellcolor[HTML]{FFFFFF}C: quantum = 1538 B}} & \cellcolor[HTML]{FFFFFF}                                                                                           \\ \cline{1-5} \cline{12-14}
\multicolumn{1}{|c|}{\cellcolor[HTML]{FFFFFF}BE}                                       & \multicolumn{2}{c|}{\cellcolor[HTML]{FFFFFF}N/A}                 & \multicolumn{1}{c|}{\cellcolor[HTML]{FFFFFF}N/A}           & \multicolumn{1}{c|}{\cellcolor[HTML]{FFFFFF}N/A}            & \multicolumn{2}{c|}{\multirow{-2}{*}{\cellcolor[HTML]{FFFFFF}52.8 Mbps}} & \multicolumn{1}{c|}{\multirow{-2}{*}{\cellcolor[HTML]{FFFFFF}100 Mbps}} & \multicolumn{1}{c|}{\multirow{-2}{*}{\cellcolor[HTML]{FFFFFF}100 Mbps}} & \multicolumn{1}{c|}{\multirow{-5}{*}{\cellcolor[HTML]{FFFFFF}N/A}} & \multicolumn{1}{c|}{\multirow{-5}{*}{\cellcolor[HTML]{FFFFFF}100 Mbps}} & \multicolumn{1}{c|}{\cellcolor[HTML]{FFFFFF}N/A}                                      & 100 Mbps                                                        & \multicolumn{1}{c|}{\cellcolor[HTML]{FFFFFF}D: quantum = 1538 B}                   & \multirow{-5}{*}{\cellcolor[HTML]{FFFFFF}\begin{tabular}[c]{@{}c@{}}A: High PQ (green)\\ B: Low PQ (yellow)\end{tabular}} \\ \hline
\end{tabular}%
}
\vspace*{0.25cm}
\caption{Experiment A System Parameter Configuration}
\label{tab:expa-params}
\end{table*}

\rev{\subsubsection{Experiment A: Comparison between IETF, \cite{Lin2021} and HCTNS models}}

For this experiment, the fine-grained and coarse-grained resource control mechanisms of the \gls{pe}1 router in our emulated \gls{tn} scenario were configured using the parameters and values listed in Table~\ref{tab:expa-params}. This table shows some differences between \rev{the \cite{Lin2021} model,} the \gls{ietf} model and our proposed model. The class policers defined by the \gls{ietf} only allow the configuration of the \gls{cir}. For slices that do not have best-effort traffic classes (such as the \gls{tod} slice), the \gls{pir} setting in slice policers is not applicable, as traffic classes can only transmit at their \gls{cir}. The \gls{ietf} model does not define a global policer, which is also reflected in Table~\ref{tab:expa-params}. \rev{Since the model in \cite{Lin2021} considers only slices and not different \gls{qos} flows within them, we conducted experiments by treating each flow as a separate slice, where a \gls{cir} and \gls{pir} can be defined for each slice. Moreover,  \cite{Lin2021} does not implement a hierarchical policing at the \gls{tn} ingress, as \gls{ietf} and \gls{hctns} do. Instead, it uses a single level mechanism based on a \gls{trtcm} per slice. Furthermore, \cite{Lin2021} implements a queuing system at the output ports of the \gls{tn} routers composed of two priority queues. In contrast, \gls{hctns} and the \gls{ietf} model use a queuing system composed of a priority queue and three \gls{drr} queues, each corresponding to a different \gls{tn} \gls{qos} class. Consequently, we assume that the highest priority queue (HPQ) of \cite{Lin2021} corresponds to \gls{tn} \gls{qos} class A, and the lowest priority queue (LPQ) to \gls{tn} \gls{qos} class B.}

This experiment lasts 100\,s and is divided into six time intervals. In each interval, different UDP traffic flows of 100\,Mbps, representing different traffic classes, are transmitted over the transport network. This configuration has allowed us to evaluate the performance of our solution under diverse conditions. The traffic flows active in each interval are as follows:
 
\begin{itemize}
    \item Interval 1 (from \( t = 0 \, \text{s} \) to \( t = 20 \, \text{s} \)): BE traffic.
    \item Interval 2 (from \( t = 20 \, \text{s} \) to \( t = 40 \, \text{s} \)): BE, video, and telemetry traffic.
    \item Interval 3 (from \( t = 40 \, \text{s} \) to \( t = 60 \, \text{s} \)): All traffic classes are active.
    \item Interval 4 (from \( t = 60 \, \text{s} \) to \( t = 70 \, \text{s} \)): All traffic classes except \gls{tod} video are active
    \item Interval 5 (from \( t = 70 \, \text{s} \) to \( t = 80 \, \text{s} \)): BE, telemetry, and \gls{urllc} traffic.
    \item Interval 6 (from \( t = 80 \, \text{s} \) to the end of the experiment): BE traffic, as in the initial state.
\end{itemize}

\afterpage{
\begin{figure*}[h!]
    \centering
        \vspace{-2em}
        \includegraphics[width=\textwidth, trim=50 10 50 10, clip]{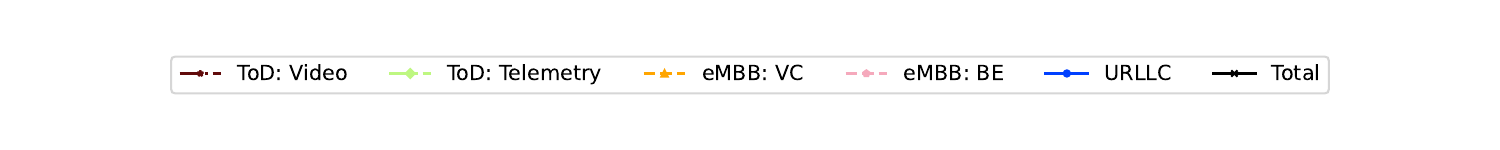}
    \begin{minipage}[t]{0.3\textwidth}
        \centering
\begin{tikzpicture}

\definecolor{blue064255}{RGB}{0,64,255}
\definecolor{darkgrey176}{RGB}{176,176,176}
\definecolor{lightpink245169188}{RGB}{245,169,188}
\definecolor{maroon971111}{RGB}{97,11,11}
\definecolor{orange}{RGB}{255,165,0}
\definecolor{palegreen190247129}{RGB}{190,247,129}

\begin{axis}[
height=0.8\textwidth,
tick align=outside,
tick pos=left,
width=\textwidth,
x grid style={darkgrey176},
xlabel={t (s)},
xmajorgrids,
xmin=0, xmax=100,
xtick style={color=black},
y grid style={darkgrey176},
ylabel={BW (Mbps)},
ymajorgrids,
ymin=-5.03872622282609, ymax=105.813250679348,
ytick style={color=black}
]
\addplot [semithick, maroon971111, dash pattern=on 1pt off 3pt on 3pt off 3pt, mark=asterisk, mark size=1, mark options={solid}]
table {%
0 0
1 0
2 0
3 0
4 0
5 0
6 0
7 0
8 0
9 0
10 0
11 0
12 0
13 0
14 0
15 0
16 0
17 0
18 0
19 0
20 31.9720108695652
21 31.9720108695652
22 31.9720108695652
23 31.9720108695652
24 31.9720108695652
25 31.9720108695652
26 31.9720108695652
27 31.9720108695652
28 31.9720108695652
29 31.9720108695652
30 31.9720108695652
31 31.9720108695652
32 31.9720108695652
33 31.9720108695652
34 31.9720108695652
35 31.9720108695652
36 31.9720108695652
37 31.9720108695652
38 31.867527173913
39 31.9720108695652
40 31.9720108695652
41 31.9720108695652
42 31.9720108695652
43 31.7630434782609
44 31.867527173913
45 31.9720108695652
46 31.9720108695652
47 31.9720108695652
48 31.9720108695652
49 31.9720108695652
50 31.9720108695652
51 31.9720108695652
52 31.9720108695652
53 24.2402173913043
54 31.9720108695652
55 31.9720108695652
56 31.867527173913
57 31.867527173913
58 31.9720108695652
59 29.6733695652174
60 0
61 0
62 0
63 0
64 0
65 0
66 0
67 0
68 0
69 0
70 0
71 0
72 0
73 0
74 0
75 0
76 0
77 0
78 0
79 0
80 0
81 0
82 0
83 0
84 0
85 0
86 0
87 0
88 0
89 0
90 0
91 0
92 0
93 0
94 0
95 0
96 0
97 0
98 0
99 0
};
\addplot [semithick, palegreen190247129, dash pattern=on 1pt off 3pt on 3pt off 3pt, mark=diamond*, mark size=1, mark options={solid}]
table {%
0 0
1 0
2 0
3 0
4 0
5 0
6 0
7 0
8 0
9 0
10 0
11 0
12 0
13 0
14 0
15 0
16 0
17 0
18 0
19 0
20 4.00172554347826
21 4.00172554347826
22 3.99127717391304
23 3.99127717391304
24 3.99127717391304
25 3.99127717391304
26 3.99127717391304
27 3.99127717391304
28 3.99127717391304
29 3.99127717391304
30 3.99127717391304
31 4.00172554347826
32 3.99127717391304
33 3.99127717391304
34 3.99127717391304
35 3.99127717391304
36 3.99127717391304
37 3.99127717391304
38 3.99127717391304
39 3.99127717391304
40 2.14191576086956
41 1.00722282608696
42 1.45232336956522
43 1.19111413043478
44 0.81079347826087
45 0.859900815217391
46 0.859900815217391
47 1.69263586956522
48 1.91205163043478
49 1.36873641304348
50 1.32694293478261
51 1.66129076086957
52 1.58815217391304
53 2.53895380434783
54 1.94339673913043
55 1.69263586956522
56 1.66129076086957
57 1.60904891304348
58 1.73442934782609
59 2.06877717391304
60 2.24639945652174
61 2.08967391304348
62 1.96429347826087
63 1.94339673913043
64 2.20460597826087
65 2.18370923913043
66 2.1001222826087
67 2.01653532608696
68 2.17326086956522
69 2.14191576086956
70 4.35697010869565
71 3.99127717391304
72 3.99127717391304
73 4.00172554347826
74 3.99127717391304
75 3.99127717391304
76 3.99127717391304
77 3.99127717391304
78 3.99127717391304
79 3.99127717391304
80 0
81 0
82 0
83 0
84 0
85 0
86 0
87 0
88 0
89 0
90 0
91 0
92 0
93 0
94 0
95 0
96 0
97 0
98 0
99 0
};
\addplot [semithick, orange, dashed, mark=triangle*, mark size=1, mark options={solid}]
table {%
0 0
1 0
2 0
3 0
4 0
5 0
6 0
7 0
8 0
9 0
10 0
11 0
12 0
13 0
14 0
15 0
16 0
17 0
18 0
19 0
20 0
21 0
22 0
23 0
24 0
25 0
26 0
27 0
28 0
29 0
30 0
31 0
32 0
33 0
34 0
35 0
36 0
37 0
38 0
39 0
40 31.240625
41 32.3899456521739
42 31.9720108695652
43 32.2854619565217
44 32.5989130434783
45 32.4944293478261
46 32.4944293478261
47 31.6585597826087
48 31.4495923913043
49 32.0764945652174
50 32.0764945652174
51 31.7630434782609
52 31.7630434782609
53 34.7930706521739
54 31.4495923913043
55 31.6585597826087
56 31.7630434782609
57 31.867527173913
58 31.6585597826087
59 32.4944293478261
60 46.9131793478261
61 47.3311141304348
62 47.4355978260869
63 47.4355978260869
64 47.1221467391304
65 47.1221467391304
66 47.2266304347826
67 47.3311141304348
68 47.2266304347826
69 47.2266304347826
70 0
71 0
72 0
73 0
74 0
75 0
76 0
77 0
78 0
79 0
80 0
81 0
82 0
83 0
84 0
85 0
86 0
87 0
88 0
89 0
90 0
91 0
92 0
93 0
94 0
95 0
96 0
97 0
98 0
99 0
};
\addplot [semithick, lightpink245169188, dashed, mark=pentagon*, mark size=1, mark options={solid}]
table {%
0 99.8864130434782
1 99.7819293478261
2 99.8864130434782
3 99.8864130434782
4 99.8864130434782
5 99.7819293478261
6 99.7819293478261
7 99.8864130434782
8 99.7819293478261
9 99.7819293478261
10 99.8864130434782
11 99.8864130434782
12 99.8864130434782
13 99.8864130434782
14 99.8864130434782
15 99.8864130434782
16 99.7819293478261
17 99.7819293478261
18 99.7819293478261
19 99.7819293478261
20 64.5709239130435
21 63.9440217391304
22 63.9440217391304
23 63.9440217391304
24 63.9440217391304
25 63.9440217391304
26 63.9440217391304
27 63.9440217391304
28 63.9440217391304
29 63.9440217391304
30 63.9440217391304
31 63.9440217391304
32 63.9440217391304
33 63.9440217391304
34 63.9440217391304
35 63.9440217391304
36 63.9440217391304
37 63.9440217391304
38 63.9440217391304
39 63.9440217391304
40 20.7922554347826
41 7.77358695652174
42 9.68563858695652
43 12.5380434782609
44 7.19892663043478
45 7.31385869565217
46 8.57811141304348
47 11.9111413043478
48 7.62730978260869
49 7.58551630434783
50 6.83323369565217
51 8.73483695652174
52 8.63035326086956
53 10.4483695652174
54 9.12142663043478
55 9.53936141304348
56 11.7021739130435
57 9.53936141304348
58 7.8676222826087
59 10.3334375
60 10.5528532608696
61 19.4339673913043
62 14.2097826086957
63 7.90941576086956
64 12.2245923913043
65 12.9559782608696
66 10.8663043478261
67 13.3739130434783
68 16.2994565217391
69 17.0308423913043
70 38.9724184782609
71 49.5252717391304
72 57.4660326086956
73 56.5256793478261
74 48.3759510869565
75 40.1217391304348
76 56.8391304347826
77 48.689402173913
78 36.673777173913
79 50.2566576086956
80 95.7070652173913
81 97.5877717391304
82 99.6774456521739
83 99.8864130434782
84 99.8864130434782
85 99.8864130434782
86 99.8864130434782
87 99.8864130434782
88 99.8864130434782
89 99.8864130434782
90 99.8864130434782
91 99.8864130434782
92 99.8864130434782
93 99.8864130434782
94 99.8864130434782
95 99.8864130434782
96 99.8864130434782
97 99.8864130434782
98 99.8864130434782
99 99.8864130434782
};
\addplot [semithick, blue064255, mark=*, mark size=1, mark options={solid}]
table {%
0 0
1 0
2 0
3 0
4 0
5 0
6 0
7 0
8 0
9 0
10 0
11 0
12 0
13 0
14 0
15 0
16 0
17 0
18 0
19 0
20 0
21 0
22 0
23 0
24 0
25 0
26 0
27 0
28 0
29 0
30 0
31 0
32 0
33 0
34 0
35 0
36 0
37 0
38 0
39 0
40 14.6277173913043
41 26.9567934782609
42 24.7626358695652
43 22.255027173913
44 27.3747282608696
45 27.2702445652174
46 25.7029891304348
47 22.5684782608696
48 26.9567934782609
49 27.1657608695652
50 27.5836956521739
51 25.9119565217391
52 25.807472826087
53 28.0016304347826
54 25.0760869565217
55 25.2850543478261
56 22.6729619565217
57 25.0760869565217
58 26.8523097826087
59 25.2850543478261
60 39.4948369565217
61 31.0316576086956
62 36.673777173913
63 42.5248641304348
64 38.3455163043478
65 37.3006793478261
66 39.5993206521739
67 37.1961956521739
68 34.270652173913
69 33.4347826086956
70 42.7338315217391
71 45.4504076086956
72 37.9275815217391
73 39.8082880434783
74 47.5400815217391
75 56.0032608695652
76 38.6589673913043
77 47.8535326086956
78 58.9288043478261
79 44.6145380434783
80 0
81 0
82 0
83 0
84 0
85 0
86 0
87 0
88 0
89 0
90 0
91 0
92 0
93 0
94 0
95 0
96 0
97 0
98 0
99 0
};
\addplot [semithick, black, mark=x, mark size=1, mark options={solid}]
table {%
0 99.8864130434782
1 99.7819293478261
2 99.8864130434782
3 99.8864130434782
4 99.8864130434782
5 99.7819293478261
6 99.7819293478261
7 99.8864130434782
8 99.7819293478261
9 99.7819293478261
10 99.8864130434782
11 99.8864130434782
12 99.8864130434782
13 99.8864130434782
14 99.8864130434782
15 99.8864130434782
16 99.7819293478261
17 99.7819293478261
18 99.7819293478261
19 99.7819293478261
20 100.544660326087
21 99.9177581521739
22 99.9073097826087
23 99.9073097826087
24 99.9073097826087
25 99.9073097826087
26 99.9073097826087
27 99.9073097826087
28 99.9073097826087
29 99.9073097826087
30 99.9073097826087
31 99.9177581521739
32 99.9073097826087
33 99.9073097826087
34 99.9073097826087
35 99.9073097826087
36 99.9073097826087
37 99.9073097826087
38 99.8028260869565
39 99.9073097826087
40 100.774524456522
41 100.099559782609
42 99.8446195652174
43 100.032690217391
44 99.8508885869565
45 99.9104442934782
46 99.607441576087
47 99.8028260869565
48 99.9177581521739
49 100.168519021739
50 99.7923777173913
51 100.043138586957
52 99.7610326086956
53 100.022241847826
54 99.5625135869565
55 100.147622282609
56 99.6669972826087
57 99.9595516304348
58 100.084932065217
59 99.8550679347826
60 99.2072690217391
61 99.8864130434783
62 100.283451086957
63 99.8132744565217
64 99.8968614130435
65 99.5625135869565
66 99.7923777173913
67 99.9177581521739
68 99.97
69 99.8341711956522
70 86.0632201086956
71 98.9669565217391
72 99.3848913043478
73 100.335692934783
74 99.9073097826087
75 100.116277173913
76 99.489375
77 100.534211956522
78 99.5938586956522
79 98.862472826087
80 95.7070652173913
81 97.5877717391304
82 99.6774456521739
83 99.8864130434782
84 99.8864130434782
85 99.8864130434782
86 99.8864130434782
87 99.8864130434782
88 99.8864130434782
89 99.8864130434782
90 99.8864130434782
91 99.8864130434782
92 99.8864130434782
93 99.8864130434782
94 99.8864130434782
95 99.8864130434782
96 99.8864130434782
97 99.8864130434782
98 99.8864130434782
99 99.8864130434782
};
\end{axis}

\end{tikzpicture}
        \hspace*{3.35em}
        \centering \small (a) IETF: Bandwidth Evolution.
    \end{minipage}
    \vspace*{0.25em}
    \hfill
    \begin{minipage}[t]{0.3\textwidth}
        \centering
\begin{tikzpicture}

\definecolor{blue064255}{RGB}{0,64,255}
\definecolor{darkgrey176}{RGB}{176,176,176}
\definecolor{lightpink245169188}{RGB}{245,169,188}
\definecolor{maroon971111}{RGB}{97,11,11}
\definecolor{orange}{RGB}{255,165,0}
\definecolor{palegreen190247129}{RGB}{190,247,129}

\begin{axis}[
height=0.8\textwidth,
tick align=outside,
tick pos=left,
width=\textwidth,
x grid style={darkgrey176},
xlabel={t (s)},
xmajorgrids,
xmin=0, xmax=100,
xtick style={color=black},
y grid style={darkgrey176},
ylabel={BW (Mbps)},
ymajorgrids,
ymin=-5.17716711956522, ymax=108.72050951087,
ytick style={color=black}
]
\addplot [semithick, maroon971111, dash pattern=on 1pt off 3pt on 3pt off 3pt, mark=asterisk, mark size=1, mark options={solid}]
table {%
0 0
1 0
2 0
3 0
4 0
5 0
6 0
7 0
8 0
9 0
10 0
11 0
12 0
13 0
14 0
15 0
16 0
17 0
18 0
19 0
20 55.7942934782609
21 58.9288043478261
22 61.854347826087
23 57.7794836956522
24 53.0777173913043
25 51.8239130434783
26 51.9283967391304
27 52.86875
28 51.1970108695652
29 52.0328804347826
30 57.2570652173913
31 57.5705163043478
32 53.3911684782609
33 54.9584239130435
34 58.301902173913
35 54.2270380434783
36 51.5104619565217
37 52.0328804347826
38 52.5552989130435
39 54.1225543478261
40 35.8379076086956
41 33.7482336956522
42 33.5392663043478
43 33.2258152173913
44 33.9572010869565
45 33.64375
46 33.4347826086956
47 33.3302989130435
48 33.3302989130435
49 33.5392663043478
50 33.3302989130435
51 33.64375
52 33.9572010869565
53 33.5392663043478
54 33.3302989130435
55 33.016847826087
56 33.8527173913043
57 33.4347826086956
58 33.4347826086956
59 33.4347826086956
60 0
61 0
62 0
63 0
64 0
65 0
66 0
67 0
68 0
69 0
70 0
71 0
72 0
73 0
74 0
75 0
76 0
77 0
78 0
79 0
80 0
81 0
82 0
83 0
84 0
85 0
86 0
87 0
88 0
89 0
90 0
91 0
92 0
93 0
94 0
95 0
96 0
97 0
98 0
99 0
};
\addplot [semithick, palegreen190247129, dash pattern=on 1pt off 3pt on 3pt off 3pt, mark=diamond*, mark size=1, mark options={solid}]
table {%
0 0
1 0
2 0
3 0
4 0
5 0
6 0
7 0
8 0
9 0
10 0
11 0
12 0
13 0
14 0
15 0
16 0
17 0
18 0
19 0
20 26.5388586956522
21 21.8370923913043
22 24.1357336956522
23 22.8819293478261
24 25.0760869565217
25 27.1657608695652
26 26.6433423913043
27 26.6433423913043
28 23.0908967391304
29 24.9716032608696
30 25.4940217391304
31 25.5985054347826
32 25.5985054347826
33 26.0164402173913
34 26.7478260869565
35 25.2850543478261
36 25.7029891304348
37 26.5388586956522
38 26.9567934782609
39 24.658152173913
40 7.35565217391304
41 5.18239130434783
42 4.93163043478261
43 4.83759510869565
44 4.79580163043478
45 4.91073369565217
46 4.83759510869565
47 5.06745923913043
48 5.06745923913043
49 4.85849184782609
50 4.88983695652174
51 4.83759510869565
52 4.59728260869565
53 4.6286277173913
54 4.95252717391304
55 4.96297554347826
56 4.70176630434783
57 4.69131793478261
58 4.79580163043478
59 4.83759510869565
60 8.09748641304348
61 8.45273097826087
62 8.32735054347826
63 9.38263586956522
64 9.47667119565217
65 9.16322010869565
66 9.79012228260869
67 9.17366847826087
68 9.65429347826087
69 9.79012228260869
70 38.6589673913043
71 33.5392663043478
72 34.270652173913
73 40.853125
74 36.7782608695652
75 34.1661684782609
76 36.7782608695652
77 37.5096467391304
78 36.8827445652174
79 36.3603260869565
80 0
81 0
82 0
83 0
84 0
85 0
86 0
87 0
88 0
89 0
90 0
91 0
92 0
93 0
94 0
95 0
96 0
97 0
98 0
99 0
};
\addplot [semithick, orange, dashed, mark=triangle*, mark size=1, mark options={solid}]
table {%
0 0
1 0
2 0
3 0
4 0
5 0
6 0
7 0
8 0
9 0
10 0
11 0
12 0
13 0
14 0
15 0
16 0
17 0
18 0
19 0
20 0
21 0
22 0
23 0
24 0
25 0
26 0
27 0
28 0
29 0
30 0
31 0
32 0
33 0
34 0
35 0
36 0
37 0
38 0
39 0
40 52.6597826086956
41 55.5853260869565
42 56.4211956521739
43 56.4211956521739
44 56.2122282608696
45 56.2122282608696
46 56.0032608695652
47 56.5256793478261
48 56.3167119565217
49 56.7346467391304
50 56.8391304347826
51 56.5256793478261
52 56.7346467391304
53 56.9436141304348
54 56.6301630434783
55 56.3167119565217
56 56.3167119565217
57 56.5256793478261
58 56.4211956521739
59 56.0032608695652
60 68.7502717391304
61 68.5413043478261
62 67.8099184782609
63 68.0188858695652
64 68.3323369565217
65 68.2278532608696
66 67.6009510869565
67 68.0188858695652
68 69.8995923913044
69 70.9444293478261
70 0
71 0
72 0
73 0
74 0
75 0
76 0
77 0
78 0
79 0
80 0
81 0
82 0
83 0
84 0
85 0
86 0
87 0
88 0
89 0
90 0
91 0
92 0
93 0
94 0
95 0
96 0
97 0
98 0
99 0
};
\addplot [semithick, lightpink245169188, dashed, mark=pentagon*, mark size=1, mark options={solid}]
table {%
0 99.8864130434782
1 99.8864130434782
2 99.7819293478261
3 99.7819293478261
4 99.8864130434782
5 99.8864130434782
6 99.8864130434782
7 99.8864130434782
8 99.8864130434782
9 99.7819293478261
10 99.7819293478261
11 99.8864130434782
12 99.8864130434782
13 99.7819293478261
14 99.8864130434782
15 99.7819293478261
16 99.8864130434782
17 99.8864130434782
18 99.8864130434782
19 99.8864130434782
20 21.2101902173913
21 19.0160326086956
22 14.1052989130435
23 18.8070652173913
24 22.0460597826087
25 20.5832880434783
26 21.3146739130435
27 20.8967391304348
28 25.1805706521739
29 23.2998641304348
30 17.0308423913043
31 16.7173913043478
32 20.4788043478261
33 19.6429347826087
34 14.8366847826087
35 19.6429347826087
36 23.2998641304348
37 21.1057065217391
38 20.1653532608696
39 20.3743206521739
40 2.72702445652174
41 2.21505434782609
42 1.51501358695652
43 1.441875
44 1.52546195652174
45 1.62994565217391
46 1.69263586956522
47 1.78667119565217
48 1.70308423913043
49 1.38963315217391
50 1.55680706521739
51 1.46277173913043
52 1.441875
53 1.45232336956522
54 1.32694293478261
55 1.6194972826087
56 1.41052989130435
57 1.54635869565217
58 1.73442934782609
59 1.82846467391304
60 6.90637228260869
61 10.1035733695652
62 10.1035733695652
63 9.28860054347826
64 7.53327445652174
65 8.77663043478261
66 9.46622282608696
67 10.187160326087
68 8.94380434782609
69 7.209375
70 25.807472826087
71 35.1065217391304
72 35.2110054347826
73 35.0020380434783
74 35.7334239130435
75 35.419972826087
76 35.8379076086956
77 35.2110054347826
78 32.7033967391304
79 35.1065217391304
80 89.8559782608696
81 99.8864130434782
82 99.8864130434782
83 99.8864130434782
84 99.8864130434782
85 99.8864130434782
86 99.8864130434782
87 99.8864130434782
88 99.8864130434782
89 99.8864130434782
90 99.8864130434782
91 99.8864130434782
92 99.8864130434782
93 99.8864130434782
94 99.8864130434782
95 99.8864130434782
96 99.8864130434782
97 99.8864130434782
98 99.8864130434782
99 99.8864130434782
};
\addplot [semithick, blue064255, mark=*, mark size=1, mark options={solid}]
table {%
0 0
1 0
2 0
3 0
4 0
5 0
6 0
7 0
8 0
9 0
10 0
11 0
12 0
13 0
14 0
15 0
16 0
17 0
18 0
19 0
20 0
21 0
22 0
23 0
24 0
25 0
26 0
27 0
28 0
29 0
30 0
31 0
32 0
33 0
34 0
35 0
36 0
37 0
38 0
39 0
40 1.23290760869565
41 3.0613722826087
42 3.57334239130435
43 3.85544836956522
44 3.364375
45 3.51065217391304
46 3.81365489130435
47 3.37482336956522
48 3.43751358695652
49 3.25989130434783
50 3.35392663043478
51 3.47930706521739
52 3.23899456521739
53 3.21809782608696
54 3.68827445652174
55 3.89724184782609
56 3.53154891304348
57 3.63603260869565
58 3.41661684782609
59 3.83455163043478
60 13.2694293478261
61 12.9559782608696
62 13.3739130434783
63 13.1649456521739
64 14.7322010869565
65 13.5828804347826
66 13.0604619565217
67 12.4335597826087
68 11.5976902173913
69 12.1201086956522
70 30.8226902173913
71 31.240625
72 29.7778532608696
73 24.658152173913
74 26.9567934782609
75 30.1957880434783
76 27.7926630434783
77 27.7926630434783
78 29.1509510869565
79 28.9419836956522
80 0
81 0
82 0
83 0
84 0
85 0
86 0
87 0
88 0
89 0
90 0
91 0
92 0
93 0
94 0
95 0
96 0
97 0
98 0
99 0
};
\addplot [semithick, black, mark=x, mark size=1, mark options={solid}]
table {%
0 99.8864130434782
1 99.8864130434782
2 99.7819293478261
3 99.7819293478261
4 99.8864130434782
5 99.8864130434782
6 99.8864130434782
7 99.8864130434782
8 99.8864130434782
9 99.7819293478261
10 99.7819293478261
11 99.8864130434782
12 99.8864130434782
13 99.7819293478261
14 99.8864130434782
15 99.7819293478261
16 99.8864130434782
17 99.8864130434782
18 99.8864130434782
19 99.8864130434782
20 103.543342391304
21 99.7819293478261
22 100.095380434783
23 99.4684782608695
24 100.199864130435
25 99.5729619565217
26 99.8864130434783
27 100.408831521739
28 99.4684782608696
29 100.304347826087
30 99.7819293478261
31 99.8864130434783
32 99.4684782608696
33 100.617798913043
34 99.8864130434783
35 99.155027173913
36 100.513315217391
37 99.6774456521739
38 99.6774456521739
39 99.155027173913
40 99.8132744565217
41 99.7923777173913
42 99.9804483695652
43 99.7819293478261
44 99.8550679347826
45 99.9073097826087
46 99.7819293478261
47 100.084932065217
48 99.8550679347826
49 99.7819293478261
50 99.97
51 99.9491032608695
52 99.97
53 99.7819293478261
54 99.9282065217391
55 99.8132744565217
56 99.8132744565217
57 99.8341711956522
58 99.8028260869565
59 99.9386548913044
60 97.0235597826087
61 100.053586956522
62 99.6147554347826
63 99.8550679347826
64 100.074483695652
65 99.7505842391304
66 99.9177581521739
67 99.8132744565217
68 100.095380434783
69 100.064035326087
70 95.2891304347826
71 99.8864130434783
72 99.2595108695652
73 100.513315217391
74 99.4684782608696
75 99.7819293478261
76 100.408831521739
77 100.513315217391
78 98.7370923913043
79 100.408831521739
80 89.8559782608696
81 99.8864130434782
82 99.8864130434782
83 99.8864130434782
84 99.8864130434782
85 99.8864130434782
86 99.8864130434782
87 99.8864130434782
88 99.8864130434782
89 99.8864130434782
90 99.8864130434782
91 99.8864130434782
92 99.8864130434782
93 99.8864130434782
94 99.8864130434782
95 99.8864130434782
96 99.8864130434782
97 99.8864130434782
98 99.8864130434782
99 99.8864130434782
};
\end{axis}

\end{tikzpicture}
        \hspace*{3.5em}
        \centering \small (b) \cite{Lin2021}: Bandwidth Evolution.
    \end{minipage}
    \vspace*{0.25em}
    \hfill
    \begin{minipage}[t]{0.3\textwidth}
        \centering
\begin{tikzpicture}

\definecolor{blue064255}{RGB}{0,64,255}
\definecolor{darkgrey176}{RGB}{176,176,176}
\definecolor{lightpink245169188}{RGB}{245,169,188}
\definecolor{maroon971111}{RGB}{97,11,11}
\definecolor{orange}{RGB}{255,165,0}
\definecolor{palegreen190247129}{RGB}{190,247,129}

\begin{axis}[
height=0.8\textwidth,
tick align=outside,
tick pos=left,
width=\textwidth,
x grid style={darkgrey176},
xlabel={t (s)},
xmajorgrids,
xmin=0, xmax=100,
xtick style={color=black},
y grid style={darkgrey176},
ylabel={BW (Mbps)},
ymajorgrids,
ymin=-5.10037160326087, ymax=107.107803668478,
ytick style={color=black}
]
\addplot [semithick, maroon971111, dash pattern=on 1pt off 3pt on 3pt off 3pt, mark=asterisk, mark size=1, mark options={solid}]
table {%
0 0
1 0
2 0
3 0
4 0
5 0
6 0
7 0
8 0
9 0
10 0
11 0
12 0
13 0
14 0
15 0
16 0
17 0
18 0
19 0
20 37.6141304347826
21 37.5096467391304
22 37.5096467391304
23 37.5096467391304
24 37.5096467391304
25 37.5096467391304
26 37.5096467391304
27 37.5096467391304
28 37.5096467391304
29 37.5096467391304
30 37.5096467391304
31 37.5096467391304
32 37.6141304347826
33 37.5096467391304
34 37.5096467391304
35 37.5096467391304
36 37.5096467391304
37 37.6141304347826
38 37.5096467391304
39 37.5096467391304
40 34.0616847826087
41 34.0616847826087
42 34.0616847826087
43 34.0616847826087
44 34.0616847826087
45 34.0616847826087
46 34.0616847826087
47 34.0616847826087
48 34.0616847826087
49 34.0616847826087
50 34.0616847826087
51 34.0616847826087
52 34.0616847826087
53 34.0616847826087
54 34.0616847826087
55 34.0616847826087
56 34.0616847826087
57 34.0616847826087
58 34.0616847826087
59 34.0616847826087
60 0
61 0
62 0
63 0
64 0
65 0
66 0
67 0
68 0
69 0
70 0
71 0
72 0
73 0
74 0
75 0
76 0
77 0
78 0
79 0
80 0
81 0
82 0
83 0
84 0
85 0
86 0
87 0
88 0
89 0
90 0
91 0
92 0
93 0
94 0
95 0
96 0
97 0
98 0
99 0
};
\addplot [semithick, palegreen190247129, dash pattern=on 1pt off 3pt on 3pt off 3pt, mark=diamond*, mark size=1, mark options={solid}]
table {%
0 0
1 0
2 0
3 0
4 0
5 0
6 0
7 0
8 0
9 0
10 0
11 0
12 0
13 0
14 0
15 0
16 0
17 0
18 0
19 0
20 9.59160326086956
21 9.58115489130435
22 9.58115489130435
23 9.58115489130435
24 9.57070652173913
25 9.58115489130435
26 9.58115489130435
27 9.59160326086956
28 9.58115489130435
29 9.58115489130435
30 9.58115489130435
31 9.59160326086956
32 9.58115489130435
33 9.58115489130435
34 9.58115489130435
35 9.58115489130435
36 9.59160326086956
37 9.58115489130435
38 9.58115489130435
39 9.59160326086956
40 6.08095108695652
41 6.0705027173913
42 6.08095108695652
43 6.0705027173913
44 6.08095108695652
45 6.06005434782609
46 6.10184782608696
47 6.09139945652174
48 6.08095108695652
49 6.08095108695652
50 6.12274456521739
51 6.08095108695652
52 6.09139945652174
53 6.09139945652174
54 6.14364130434783
55 6.12274456521739
56 6.08095108695652
57 6.09139945652174
58 6.12274456521739
59 6.10184782608696
60 35.8379076086956
61 35.9423913043478
62 35.9423913043478
63 35.9423913043478
64 35.9423913043478
65 35.9423913043478
66 35.9423913043478
67 35.9423913043478
68 35.9423913043478
69 35.9423913043478
70 35.9423913043478
71 35.9423913043478
72 35.9423913043478
73 35.9423913043478
74 35.9423913043478
75 35.9423913043478
76 35.9423913043478
77 35.9423913043478
78 35.9423913043478
79 35.9423913043478
80 0
81 0
82 0
83 0
84 0
85 0
86 0
87 0
88 0
89 0
90 0
91 0
92 0
93 0
94 0
95 0
96 0
97 0
98 0
99 0
};
\addplot [semithick, orange, dashed, mark=triangle*, mark size=1, mark options={solid}]
table {%
0 0
1 0
2 0
3 0
4 0
5 0
6 0
7 0
8 0
9 0
10 0
11 0
12 0
13 0
14 0
15 0
16 0
17 0
18 0
19 0
20 0
21 0
22 0
23 0
24 0
25 0
26 0
27 0
28 0
29 0
30 0
31 0
32 0
33 0
34 0
35 0
36 0
37 0
38 0
39 0
40 54.3315217391304
41 54.4360054347826
42 54.3315217391304
43 54.4360054347826
44 54.3315217391304
45 54.4360054347826
46 54.2270380434783
47 54.3315217391304
48 54.3315217391304
49 54.3315217391304
50 54.2270380434783
51 54.3315217391304
52 54.3315217391304
53 54.3315217391304
54 54.1225543478261
55 54.2270380434783
56 54.3315217391304
57 54.3315217391304
58 54.2270380434783
59 54.2270380434783
60 55.3763586956522
61 55.3763586956522
62 55.5853260869565
63 55.4808423913043
64 55.5853260869565
65 55.271875
66 55.271875
67 55.3763586956522
68 55.3763586956522
69 55.271875
70 0
71 0
72 0
73 0
74 0
75 0
76 0
77 0
78 0
79 0
80 0
81 0
82 0
83 0
84 0
85 0
86 0
87 0
88 0
89 0
90 0
91 0
92 0
93 0
94 0
95 0
96 0
97 0
98 0
99 0
};
\addplot [semithick, lightpink245169188, dashed, mark=pentagon*, mark size=1, mark options={solid}]
table {%
0 99.8864130434782
1 99.8864130434782
2 99.8864130434782
3 99.8864130434782
4 99.8864130434782
5 99.8864130434782
6 99.7819293478261
7 99.8864130434782
8 99.8864130434782
9 99.8864130434782
10 99.8864130434782
11 99.8864130434782
12 99.8864130434782
13 99.8864130434782
14 99.8864130434782
15 99.8864130434782
16 99.8864130434782
17 99.8864130434782
18 99.8864130434782
19 99.8864130434782
20 54.6449728260869
21 52.7642663043478
22 52.7642663043478
23 52.7642663043478
24 52.7642663043478
25 52.7642663043478
26 52.7642663043478
27 52.7642663043478
28 52.7642663043478
29 52.7642663043478
30 52.7642663043478
31 52.7642663043478
32 52.7642663043478
33 52.7642663043478
34 52.7642663043478
35 52.7642663043478
36 52.7642663043478
37 52.7642663043478
38 52.7642663043478
39 52.7642663043478
40 4.23158967391304
41 2.08967391304348
42 2.08967391304348
43 2.07922554347826
44 2.11057065217391
45 2.07922554347826
46 2.1001222826087
47 2.14191576086956
48 2.1001222826087
49 2.1001222826087
50 2.12101902173913
51 2.1001222826087
52 2.1001222826087
53 2.11057065217391
54 2.1628125
55 2.14191576086956
56 2.1001222826087
57 2.12101902173913
58 2.11057065217391
59 2.12101902173913
60 3.58379076086957
61 3.65692934782609
62 3.56289402173913
63 3.63603260869565
64 3.56289402173913
65 3.71961956521739
66 3.69872282608696
67 3.68827445652174
68 3.64648097826087
69 3.71961956521739
70 50.5701086956522
71 52.7642663043478
72 52.7642663043478
73 52.7642663043478
74 52.7642663043478
75 52.7642663043478
76 52.7642663043478
77 52.7642663043478
78 52.7642663043478
79 52.7642663043478
80 97.7967391304348
81 99.8864130434782
82 99.8864130434782
83 99.8864130434782
84 99.8864130434782
85 99.8864130434782
86 99.8864130434782
87 99.8864130434782
88 99.8864130434782
89 99.7819293478261
90 99.8864130434782
91 99.8864130434782
92 99.8864130434782
93 99.8864130434782
94 99.8864130434782
95 99.8864130434782
96 99.8864130434782
97 99.8864130434782
98 99.8864130434782
99 99.8864130434782
};
\addplot [semithick, blue064255, mark=*, mark size=1, mark options={solid}]
table {%
0 0
1 0
2 0
3 0
4 0
5 0
6 0
7 0
8 0
9 0
10 0
11 0
12 0
13 0
14 0
15 0
16 0
17 0
18 0
19 0
20 0
21 0
22 0
23 0
24 0
25 0
26 0
27 0
28 0
29 0
30 0
31 0
32 0
33 0
34 0
35 0
36 0
37 0
38 0
39 0
40 3.3016847826087
41 3.28078804347826
42 3.28078804347826
43 3.27033967391304
44 3.28078804347826
45 3.28078804347826
46 3.34347826086957
47 3.3016847826087
48 3.28078804347826
49 3.29123641304348
50 3.3016847826087
51 3.29123641304348
52 3.29123641304348
53 3.3016847826087
54 3.35392663043478
55 3.32258152173913
56 3.3016847826087
57 3.29123641304348
58 3.32258152173913
59 3.32258152173913
60 4.82714673913043
61 4.83759510869565
62 4.76445652173913
63 4.80625
64 4.76445652173913
65 4.90028532608696
66 4.90028532608696
67 4.85849184782609
68 4.85849184782609
69 4.95252717391304
70 11.1797554347826
71 11.1797554347826
72 11.1797554347826
73 11.1797554347826
74 11.1797554347826
75 11.1797554347826
76 11.1797554347826
77 11.1797554347826
78 11.1797554347826
79 11.1797554347826
80 0
81 0
82 0
83 0
84 0
85 0
86 0
87 0
88 0
89 0
90 0
91 0
92 0
93 0
94 0
95 0
96 0
97 0
98 0
99 0
};
\addplot [semithick, black, mark=x, mark size=1, mark options={solid}]
table {%
0 99.8864130434782
1 99.8864130434782
2 99.8864130434782
3 99.8864130434782
4 99.8864130434782
5 99.8864130434782
6 99.7819293478261
7 99.8864130434782
8 99.8864130434782
9 99.8864130434782
10 99.8864130434782
11 99.8864130434782
12 99.8864130434782
13 99.8864130434782
14 99.8864130434782
15 99.8864130434782
16 99.8864130434782
17 99.8864130434782
18 99.8864130434782
19 99.8864130434782
20 101.850706521739
21 99.8550679347826
22 99.8550679347826
23 99.8550679347826
24 99.8446195652174
25 99.8550679347826
26 99.8550679347826
27 99.8655163043478
28 99.8550679347826
29 99.8550679347826
30 99.8550679347826
31 99.8655163043478
32 99.9595516304348
33 99.8550679347826
34 99.8550679347826
35 99.8550679347826
36 99.8655163043478
37 99.9595516304348
38 99.8550679347826
39 99.8655163043478
40 102.007432065217
41 99.9386548913044
42 99.8446195652174
43 99.9177581521739
44 99.8655163043478
45 99.9177581521739
46 99.8341711956521
47 99.9282065217391
48 99.8550679347826
49 99.8655163043478
50 99.8341711956522
51 99.8655163043478
52 99.875964673913
53 99.8968614130435
54 99.8446195652174
55 99.875964673913
56 99.875964673913
57 99.8968614130435
58 99.8446195652174
59 99.8341711956521
60 99.6252038043478
61 99.8132744565217
62 99.8550679347826
63 99.8655163043478
64 99.8550679347826
65 99.8341711956522
66 99.8132744565217
67 99.8655163043478
68 99.823722826087
69 99.8864130434782
70 97.6922554347826
71 99.8864130434783
72 99.8864130434783
73 99.8864130434783
74 99.8864130434783
75 99.8864130434783
76 99.8864130434783
77 99.8864130434783
78 99.8864130434783
79 99.8864130434783
80 97.7967391304348
81 99.8864130434782
82 99.8864130434782
83 99.8864130434782
84 99.8864130434782
85 99.8864130434782
86 99.8864130434782
87 99.8864130434782
88 99.8864130434782
89 99.7819293478261
90 99.8864130434782
91 99.8864130434782
92 99.8864130434782
93 99.8864130434782
94 99.8864130434782
95 99.8864130434782
96 99.8864130434782
97 99.8864130434782
98 99.8864130434782
99 99.8864130434782
};
\end{axis}

\end{tikzpicture}
        \hspace*{3.35em}
        \centering \small (c) \gls{hctns}: Bandwidth Evolution.
    \end{minipage}
    \vspace*{0.25em}

    \begin{minipage}[t]{0.3\textwidth}
        \centering
\begin{tikzpicture}

\definecolor{blue064255}{RGB}{0,64,255}
\definecolor{darkgrey176}{RGB}{176,176,176}
\definecolor{lightpink245169188}{RGB}{245,169,188}
\definecolor{maroon971111}{RGB}{97,11,11}
\definecolor{orange}{RGB}{255,165,0}
\definecolor{palegreen190247129}{RGB}{190,247,129}

\begin{axis}[
height=0.8\textwidth,
tick align=outside,
tick pos=left,
width=\textwidth,
x grid style={darkgrey176},
xlabel={t (s)},
xmajorgrids,
xmin=0, xmax=100,
xtick style={color=black},
y grid style={darkgrey176},
ylabel={Latency (ms)},
ymajorgrids,
ymin=-19.09595, ymax=401.01495,
ytick style={color=black}
]
\addplot [semithick, maroon971111, dash pattern=on 1pt off 3pt on 3pt off 3pt, mark=asterisk, mark size=1, mark options={solid}]
table {%
1 0
2 0
3 0
4 0
5 0
6 0
7 0
8 0
9 0
10 0
11 0
12 0
13 0
14 0
15 0
16 0
17 0
18 0
19 0
20 0
21 2.301
22 2.343
23 2.245
24 2.26
25 2.355
26 2.656
27 2.536
28 2.485
29 2.266
30 2.219
31 2.24
32 2.468
33 2.49
34 2.577
35 2.418
36 2.399
37 2.412
38 2.34
39 2.347
40 2.377
41 4.248
42 2.587
43 2.613
44 8.09
45 9.978
46 9.722
47 5.67
48 4.054
49 3.476
50 5.149
51 6.644
52 6.039
53 6.307
54 10.211
55 11.149
56 11.366
57 2.865
58 3.289
59 8.492
60 16.293
61 0
62 0
63 0
64 0
65 0
66 0
67 0
68 0
69 0
70 0
71 0
72 0
73 0
74 0
75 0
76 0
77 0
78 0
79 0
80 0
81 0
82 0
83 0
84 0
85 0
86 0
87 0
88 0
89 0
90 0
91 0
92 0
93 0
94 0
95 0
96 0
97 0
98 0
99 0
100 0
};
\addplot [semithick, palegreen190247129, dash pattern=on 1pt off 3pt on 3pt off 3pt, mark=diamond*, mark size=1, mark options={solid}]
table {%
1 0
2 0
3 0
4 0
5 0
6 0
7 0
8 0
9 0
10 0
11 0
12 0
13 0
14 0
15 0
16 0
17 0
18 0
19 0
20 0
21 4.866
22 4.89
23 4.936
24 5.052
25 4.956
26 5.357
27 5.229
28 5.096
29 4.882
30 4.879
31 4.939
32 4.914
33 4.916
34 5.085
35 4.993
36 4.925
37 4.826
38 4.911
39 4.899
40 4.888
41 372.957
42 373.087
43 373.045
44 373.105
45 375.479
46 380.054
47 373.706
48 376.181
49 374.118
50 375.044
51 375.325
52 376.987
53 374.375
54 373.588
55 373.32
56 381.919
57 372.83
58 372.985
59 373.121
60 374.897
61 349.764
62 253.979
63 253.745
64 253.704
65 253.615
66 253.618
67 253.776
68 253.69
69 253.629
70 253.735
71 253.541
72 5.065
73 5.199
74 4.896
75 4.829
76 4.864
77 4.881
78 5.247
79 5.018
80 4.92
81 0
82 0
83 0
84 0
85 0
86 0
87 0
88 0
89 0
90 0
91 0
92 0
93 0
94 0
95 0
96 0
97 0
98 0
99 0
100 0
};
\addplot [semithick, orange, dashed, mark=triangle*, mark size=1, mark options={solid}]
table {%
1 0
2 0
3 0
4 0
5 0
6 0
7 0
8 0
9 0
10 0
11 0
12 0
13 0
14 0
15 0
16 0
17 0
18 0
19 0
20 0
21 0
22 0
23 0
24 0
25 0
26 0
27 0
28 0
29 0
30 0
31 0
32 0
33 0
34 0
35 0
36 0
37 0
38 0
39 0
40 0
41 370.63
42 370.478
43 370.343
44 370.497
45 375.841
46 377.867
47 372.077
48 373.731
49 371.459
50 372.551
51 372.439
52 374.19
53 372.601
54 370.84
55 370.53
56 379.126
57 370.248
58 370.311
59 370.351
60 372.55
61 347.18
62 251.087
63 251.09
64 251.204
65 251.03
66 250.902
67 251.073
68 250.933
69 250.988
70 250.975
71 0
72 0
73 0
74 0
75 0
76 0
77 0
78 0
79 0
80 0
81 0
82 0
83 0
84 0
85 0
86 0
87 0
88 0
89 0
90 0
91 0
92 0
93 0
94 0
95 0
96 0
97 0
98 0
99 0
100 0
};
\addplot [semithick, lightpink245169188, dashed, mark=pentagon*, mark size=1, mark options={solid}]
table {%
1 1.454
2 1.365
3 1.36
4 1.135
5 1.306
6 1.362
7 1.373
8 1.372
9 1.218
10 1.389
11 0.944
12 1.38
13 1.406
14 1.322
15 1.34
16 1.341
17 1.264
18 1.341
19 1.345
20 1.39
21 194.104
22 194.065
23 194.235
24 194.17
25 194.306
26 194.238
27 194.246
28 194.346
29 194.3
30 194.207
31 194.113
32 194.185
33 194.34
34 194.382
35 194.257
36 194.412
37 194.316
38 194.142
39 194.399
40 194.154
41 371.697
42 370.398
43 370.435
44 370.751
45 375.969
46 377.308
47 371.993
48 373.343
49 371.902
50 372.945
51 372.788
52 374.195
53 372.077
54 370.492
55 370.746
56 378.86
57 370.145
58 370.615
59 370.443
60 379.333
61 374.782
62 251.915
63 251.261
64 250.987
65 251.227
66 251.167
67 251.507
68 251.111
69 251.033
70 251.013
71 250.952
72 133.968
73 133.512
74 131.951
75 132.276
76 131.725
77 131.703
78 132.053
79 132.101
80 131.846
81 131.64
82 127.342
83 127.298
84 124.2
85 123.691
86 123.187
87 122.604
88 122.114
89 122.012
90 121.33
91 120.791
92 120.514
93 120.388
94 120.101
95 119.532
96 118.977
97 118.856
98 118.372
99 117.974
100 117.224
};
\addplot [semithick, blue064255, mark=*, mark size=1, mark options={solid}]
table {%
1 0
2 0
3 0
4 0
5 0
6 0
7 0
8 0
9 0
10 0
11 0
12 0
13 0
14 0
15 0
16 0
17 0
18 0
19 0
20 0
21 0
22 0
23 0
24 0
25 0
26 0
27 0
28 0
29 0
30 0
31 0
32 0
33 0
34 0
35 0
36 0
37 0
38 0
39 0
40 0
41 370.119
42 370.222
43 370.22
44 370.168
45 370.168
46 370.323
47 370.428
48 370.4
49 370.315
50 370.266
51 370.294
52 370.294
53 370.292
54 370.467
55 370.379
56 370.302
57 370.093
58 370.214
59 370.202
60 370.403
61 369.994
62 250.831
63 251.002
64 250.886
65 250.826
66 250.758
67 250.979
68 250.889
69 250.756
70 250.979
71 250.753
72 133.459
73 133.3
74 131.814
75 131.662
76 131.795
77 131.91
78 131.711
79 131.815
80 131.688
81 0
82 0
83 0
84 0
85 0
86 0
87 0
88 0
89 0
90 0
91 0
92 0
93 0
94 0
95 0
96 0
97 0
98 0
99 0
100 0
};
\end{axis}

\end{tikzpicture}
        \hspace*{3.35em}
        \centering \small (d) IETF: Latency Evolution.
    \end{minipage}
    \vspace*{0.25em}
    \hfill
    \begin{minipage}[t]{0.3\textwidth}
        \centering
\begin{tikzpicture}

\definecolor{blue064255}{RGB}{0,64,255}
\definecolor{darkgrey176}{RGB}{176,176,176}
\definecolor{lightpink245169188}{RGB}{245,169,188}
\definecolor{maroon971111}{RGB}{97,11,11}
\definecolor{orange}{RGB}{255,165,0}
\definecolor{palegreen190247129}{RGB}{190,247,129}

\begin{axis}[
height=0.8\textwidth,
tick align=outside,
tick pos=left,
width=\textwidth,
x grid style={darkgrey176},
xlabel={t (s)},
xmajorgrids,
xmin=0, xmax=100,
xtick style={color=black},
y grid style={darkgrey176},
ylabel={Latency (ms)},
ymajorgrids,
ymin=-61.83325, ymax=1298.49825,
ytick style={color=black}
]
\addplot [semithick, maroon971111, dash pattern=on 1pt off 3pt on 3pt off 3pt, mark=asterisk, mark size=1, mark options={solid}]
table {%
1 0
2 0
3 0
4 0
5 0
6 0
7 0
8 0
9 0
10 0
11 0
12 0
13 0
14 0
15 0
16 0
17 0
18 0
19 0
20 0
21 194.464
22 194.214
23 194.35
24 194.264
25 194.251
26 194.253
27 194.175
28 194.366
29 194.051
30 194.303
31 194.355
32 194.057
33 194.112
34 193.918
35 194.026
36 194.107
37 194.05
38 194.115
39 194.204
40 216.27
41 1057.862
42 1218.795
43 1218.126
44 1224.748
45 1222.01
46 1226.141
47 1229.411
48 1224.545
49 1231.188
50 1226.972
51 1232.534
52 1231.475
53 1231.673
54 1231.609
55 1231.177
56 1222.99
57 1222.497
58 1231.761
59 1186.632
60 1179.723
61 0
62 0
63 0
64 0
65 0
66 0
67 0
68 0
69 0
70 0
71 0
72 0
73 0
74 0
75 0
76 0
77 0
78 0
79 0
80 0
81 0
82 0
83 0
84 0
85 0
86 0
87 0
88 0
89 0
90 0
91 0
92 0
93 0
94 0
95 0
96 0
97 0
98 0
99 0
100 0
};
\addplot [semithick, palegreen190247129, dash pattern=on 1pt off 3pt on 3pt off 3pt, mark=diamond*, mark size=1, mark options={solid}]
table {%
1 0
2 0
3 0
4 0
5 0
6 0
7 0
8 0
9 0
10 0
11 0
12 0
13 0
14 0
15 0
16 0
17 0
18 0
19 0
20 0
21 194.308
22 194.533
23 194.197
24 194.417
25 194.205
26 194.172
27 194.109
28 194.094
29 194.02
30 194.333
31 193.822
32 194.022
33 193.981
34 194.145
35 194.041
36 194.315
37 194.106
38 194.084
39 194.213
40 210.191
41 1060.449
42 1221.111
43 1219.714
44 1223.165
45 1222.485
46 1227.322
47 1228.497
48 1226.531
49 1233.244
50 1228.399
51 1232.391
52 1231.398
53 1231.462
54 1231.445
55 1231.319
56 1222.702
57 1220.835
58 1231.338
59 1187.788
60 1178.719
61 1176.766
62 294.864
63 294.989
64 294.978
65 294.957
66 294.92
67 294.994
68 295.089
69 294.79
70 294.879
71 256.633
72 131.595
73 131.727
74 131.723
75 131.707
76 131.725
77 131.724
78 131.773
79 131.658
80 131.726
81 0
82 0
83 0
84 0
85 0
86 0
87 0
88 0
89 0
90 0
91 0
92 0
93 0
94 0
95 0
96 0
97 0
98 0
99 0
100 0
};
\addplot [semithick, orange, dashed, mark=triangle*, mark size=1, mark options={solid}]
table {%
1 0
2 0
3 0
4 0
5 0
6 0
7 0
8 0
9 0
10 0
11 0
12 0
13 0
14 0
15 0
16 0
17 0
18 0
19 0
20 0
21 0
22 0
23 0
24 0
25 0
26 0
27 0
28 0
29 0
30 0
31 0
32 0
33 0
34 0
35 0
36 0
37 0
38 0
39 0
40 0
41 2.644
42 1218.795
43 1218.891
44 1224.526
45 1223.736
46 1226.614
47 1228.495
48 1226.531
49 1231.073
50 1226.413
51 1236.665
52 1231.617
53 1231.93
54 1232.077
55 1231.269
56 1222.926
57 1221.236
58 1231.916
59 1186.513
60 1179.14
61 1178.443
62 295.021
63 294.952
64 295.138
65 295.033
66 294.953
67 295.028
68 295.325
69 294.884
70 294.873
71 0
72 0
73 0
74 0
75 0
76 0
77 0
78 0
79 0
80 0
81 0
82 0
83 0
84 0
85 0
86 0
87 0
88 0
89 0
90 0
91 0
92 0
93 0
94 0
95 0
96 0
97 0
98 0
99 0
100 0
};
\addplot [semithick, lightpink245169188, dashed, mark=pentagon*, mark size=1, mark options={solid}]
table {%
1 1.481
2 1.337
3 1.755
4 1.365
5 1.398
6 1.255
7 1.335
8 1.401
9 1.269
10 1.283
11 1.43
12 1.307
13 1.159
14 1.354
15 1.26
16 2.324
17 2.076
18 1.836
19 1.288
20 1.357
21 194.228
22 194.496
23 194.016
24 194.128
25 194.293
26 194.271
27 194.223
28 194.137
29 194.233
30 194.113
31 194.14
32 193.98
33 194.078
34 193.918
35 193.995
36 194.008
37 194.112
38 194.134
39 194.224
40 194.161
41 1022.914
42 1220.248
43 1219.398
44 1223.39
45 1222.209
46 1228.294
47 1228.635
48 1226.14
49 1230.943
50 1227.219
51 1235.507
52 1231.769
53 1231.394
54 1231.652
55 1231.579
56 1225.757
57 1222.037
58 1233.772
59 1186.207
60 1179.047
61 1174.286
62 294.94
63 295.016
64 295.129
65 294.956
66 294.918
67 294.923
68 295.164
69 294.869
70 294.877
71 294.383
72 131.767
73 131.781
74 131.555
75 131.764
76 131.808
77 131.781
78 131.701
79 131.814
80 131.753
81 131.665
82 123.88
83 122.892
84 122.764
85 122.017
86 121.908
87 121.721
88 121.406
89 120.035
90 119.864
91 119.713
92 119.335
93 119.154
94 118.813
95 118.869
96 118.612
97 118.314
98 118.175
99 117.669
100 117.51
};
\addplot [semithick, blue064255, mark=*, mark size=1, mark options={solid}]
table {%
1 0
2 0
3 0
4 0
5 0
6 0
7 0
8 0
9 0
10 0
11 0
12 0
13 0
14 0
15 0
16 0
17 0
18 0
19 0
20 0
21 0
22 0
23 0
24 0
25 0
26 0
27 0
28 0
29 0
30 0
31 0
32 0
33 0
34 0
35 0
36 0
37 0
38 0
39 0
40 0
41 1.946
42 1218.644
43 1218.703
44 1224.876
45 1222.2
46 1225.923
47 1231.052
48 1225.895
49 1230.867
50 1226.17
51 1234.563
52 1231.916
53 1232.321
54 1231.539
55 1231.221
56 1224.596
57 1222.45
58 1231.577
59 1185.696
60 1178.933
61 1176.702
62 295.004
63 294.973
64 294.914
65 294.993
66 295.068
67 295.002
68 295.287
69 294.999
70 294.914
71 293.275
72 131.589
73 131.758
74 131.552
75 131.647
76 131.557
77 131.916
78 131.595
79 131.895
80 131.628
81 0
82 0
83 0
84 0
85 0
86 0
87 0
88 0
89 0
90 0
91 0
92 0
93 0
94 0
95 0
96 0
97 0
98 0
99 0
100 0
};
\end{axis}

\end{tikzpicture}
        \hspace*{5em}
        \centering \small (e) \cite{Lin2021}: Latency Evolution.
    \end{minipage}
    \vspace*{0.25em}
    \hfill
    \begin{minipage}[t]{0.3\textwidth} 
        \centering
\begin{tikzpicture}

\definecolor{blue064255}{RGB}{0,64,255}
\definecolor{darkgrey176}{RGB}{176,176,176}
\definecolor{lightpink245169188}{RGB}{245,169,188}
\definecolor{maroon971111}{RGB}{97,11,11}
\definecolor{orange}{RGB}{255,165,0}
\definecolor{palegreen190247129}{RGB}{190,247,129}

\begin{axis}[
height=0.8\textwidth,
tick align=outside,
tick pos=left,
width=\textwidth,
x grid style={darkgrey176},
xlabel={t (s)},
xmajorgrids,
xmin=0, xmax=100,
xtick style={color=black},
y grid style={darkgrey176},
ylabel={Latency (ms)},
ymajorgrids,
ymin=-0.54155, ymax=11.37255,
ytick style={color=black}
]
\addplot [semithick, maroon971111, dash pattern=on 1pt off 3pt on 3pt off 3pt, mark=asterisk, mark size=1, mark options={solid}]
table {%
1 0
2 0
3 0
4 0
5 0
6 0
7 0
8 0
9 0
10 0
11 0
12 0
13 0
14 0
15 0
16 0
17 0
18 0
19 0
20 0
21 1.615
22 1.496
23 1.593
24 1.616
25 1.459
26 1.63
27 1.497
28 1.577
29 1.523
30 1.658
31 1.533
32 1.608
33 1.633
34 1.701
35 1.393
36 1.451
37 1.759
38 1.41
39 1.433
40 1.624
41 1.752
42 1.542
43 1.68
44 1.465
45 1.644
46 1.422
47 1.907
48 1.667
49 1.577
50 1.539
51 1.611
52 1.517
53 1.488
54 1.643
55 1.623
56 1.568
57 1.578
58 1.473
59 1.637
60 1.433
61 0
62 0
63 0
64 0
65 0
66 0
67 0
68 0
69 0
70 0
71 0
72 0
73 0
74 0
75 0
76 0
77 0
78 0
79 0
80 0
81 0
82 0
83 0
84 0
85 0
86 0
87 0
88 0
89 0
90 0
91 0
92 0
93 0
94 0
95 0
96 0
97 0
98 0
99 0
100 0
};
\addplot [semithick, palegreen190247129, dash pattern=on 1pt off 3pt on 3pt off 3pt, mark=diamond*, mark size=1, mark options={solid}]
table {%
1 0
2 0
3 0
4 0
5 0
6 0
7 0
8 0
9 0
10 0
11 0
12 0
13 0
14 0
15 0
16 0
17 0
18 0
19 0
20 0
21 3.724
22 3.297
23 3.338
24 3.316
25 3.8
26 3.168
27 3.22
28 3.062
29 3.545
30 2.914
31 2.959
32 3.88
33 3.423
34 3.308
35 3.077
36 3.073
37 3.17
38 3.078
39 3.667
40 3.11
41 4.225
42 4.032
43 4.161
44 4.136
45 4.114
46 4.298
47 4.079
48 4.082
49 4.038
50 4.032
51 4.234
52 4.041
53 4.091
54 4.339
55 4.033
56 4.068
57 4.047
58 4.002
59 4.038
60 4.012
61 1.637
62 1.812
63 1.722
64 1.575
65 1.657
66 1.571
67 1.565
68 1.608
69 1.574
70 1.561
71 1.861
72 1.638
73 1.499
74 1.498
75 1.531
76 1.592
77 1.474
78 1.651
79 1.44
80 1.607
81 0
82 0
83 0
84 0
85 0
86 0
87 0
88 0
89 0
90 0
91 0
92 0
93 0
94 0
95 0
96 0
97 0
98 0
99 0
100 0
};
\addplot [semithick, orange, dashed, mark=triangle*, mark size=1, mark options={solid}]
table {%
1 0
2 0
3 0
4 0
5 0
6 0
7 0
8 0
9 0
10 0
11 0
12 0
13 0
14 0
15 0
16 0
17 0
18 0
19 0
20 0
21 0
22 0
23 0
24 0
25 0
26 0
27 0
28 0
29 0
30 0
31 0
32 0
33 0
34 0
35 0
36 0
37 0
38 0
39 0
40 0
41 1.624
42 1.358
43 1.429
44 1.331
45 1.398
46 1.424
47 1.682
48 1.586
49 1.357
50 1.403
51 1.366
52 1.473
53 1.468
54 1.411
55 1.479
56 1.56
57 1.443
58 1.512
59 1.365
60 1.333
61 1.497
62 1.357
63 1.42
64 1.54
65 1.43
66 1.413
67 1.338
68 1.417
69 1.424
70 1.352
71 0
72 0
73 0
74 0
75 0
76 0
77 0
78 0
79 0
80 0
81 0
82 0
83 0
84 0
85 0
86 0
87 0
88 0
89 0
90 0
91 0
92 0
93 0
94 0
95 0
96 0
97 0
98 0
99 0
100 0
};
\addplot [semithick, lightpink245169188, dashed, mark=pentagon*, mark size=1, mark options={solid}]
table {%
1 1.391
2 1.4
3 1.37
4 1.373
5 1.308
6 1.434
7 1.338
8 1.27
9 1.349
10 1.303
11 1.335
12 1.333
13 1.226
14 1.383
15 1.187
16 1.317
17 1.16
18 1.267
19 1.327
20 1.691
21 1.408
22 1.448
23 1.349
24 1.413
25 1.412
26 1.462
27 1.53
28 1.371
29 1.513
30 1.436
31 1.392
32 1.32
33 1.361
34 1.344
35 1.48
36 1.42
37 1.44
38 1.429
39 1.443
40 1.395
41 10.831
42 7.775
43 7.986
44 7.79
45 8.221
46 7.821
47 7.784
48 8.308
49 8.144
50 8.108
51 8.215
52 8.032
53 8.103
54 7.964
55 8.002
56 7.982
57 7.979
58 8.087
59 8.025
60 7.859
61 7.133
62 5.905
63 6.048
64 5.934
65 5.778
66 5.802
67 5.54
68 6.352
69 5.621
70 5.714
71 5.637
72 1.386
73 1.371
74 1.374
75 1.336
76 1.484
77 1.425
78 1.393
79 1.399
80 1.289
81 1.415
82 1.292
83 1.343
84 1.193
85 1.281
86 1.186
87 1.303
88 1.298
89 1.322
90 1.28
91 1.202
92 1.368
93 1.245
94 1.248
95 1.15
96 1.267
97 1.332
98 1.124
99 1.312
100 1.217
};
\addplot [semithick, blue064255, mark=*, mark size=1, mark options={solid}]
table {%
1 0
2 0
3 0
4 0
5 0
6 0
7 0
8 0
9 0
10 0
11 0
12 0
13 0
14 0
15 0
16 0
17 0
18 0
19 0
20 0
21 0
22 0
23 0
24 0
25 0
26 0
27 0
28 0
29 0
30 0
31 0
32 0
33 0
34 0
35 0
36 0
37 0
38 0
39 0
40 0
41 7.707
42 7.486
43 7.402
44 7.367
45 7.371
46 7.343
47 7.287
48 6.985
49 7.013
50 7.239
51 7.196
52 6.916
53 6.927
54 6.881
55 7.102
56 6.868
57 7.538
58 7.581
59 6.946
60 7.444
61 5.045
62 5.077
63 4.849
64 4.512
65 4.643
66 4.6
67 4.677
68 4.477
69 4.951
70 5.161
71 2.438
72 2.403
73 2.395
74 2.415
75 2.472
76 2.484
77 2.534
78 2.784
79 2.86
80 3.375
81 0
82 0
83 0
84 0
85 0
86 0
87 0
88 0
89 0
90 0
91 0
92 0
93 0
94 0
95 0
96 0
97 0
98 0
99 0
100 0
};
\end{axis}

\end{tikzpicture}
        \hspace*{2.75em}
        \centering \small (f) \gls{hctns}: Latency Evolution.
    \end{minipage}
    \vspace*{0.25em}

    \vspace{0.5em} 
    \includegraphics[width=0.8\textwidth]{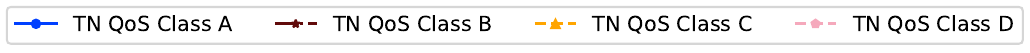}
    \vspace{0.3em}
    
    \begin{minipage}[t]{0.3\textwidth}
        \centering
\begin{tikzpicture}

\definecolor{blue064255}{RGB}{0,64,255}
\definecolor{darkgrey176}{RGB}{176,176,176}
\definecolor{lightpink245169188}{RGB}{245,169,188}
\definecolor{maroon971111}{RGB}{97,11,11}
\definecolor{orange}{RGB}{255,165,0}

\begin{axis}[
height=0.6\textwidth,
tick align=outside,
tick pos=left,
width=\textwidth,
x grid style={darkgrey176},
xlabel={t (s)},
xmajorgrids,
xmin=0, xmax=100,
xtick style={color=black},
y grid style={darkgrey176},
ylabel={Packet Loss},
ymajorgrids,
ymin=-20046.3, ymax=420972.3,
ytick style={color=black}
]
\addplot [semithick, blue064255, mark=*, mark size=1, mark options={solid}]
table {%
0 0
1 0
2 0
3 0
4 0
5 0
6 0
7 0
8 0
9 0
10 0
11 0
12 0
13 0
14 0
15 0
16 0
17 0
18 0
19 0
20 0
21 0
22 0
23 0
24 0
25 0
26 0
27 0
28 0
29 0
30 0
31 0
32 0
33 0
34 0
35 0
36 0
37 0
38 0
39 0
40 0
41 0
42 0
43 0
44 0
45 0
46 0
47 0
48 0
49 0
50 0
51 0
52 0
53 0
54 0
55 0
56 0
57 0
58 0
59 0
60 0
61 0
62 0
63 0
64 0
65 0
66 0
67 0
68 0
69 0
70 0
71 0
72 0
73 0
74 0
75 0
76 0
77 0
78 0
79 0
80 0
81 0
82 0
83 0
84 0
85 0
86 0
87 0
88 0
89 0
90 0
91 0
92 0
93 0
94 0
95 0
96 0
97 0
98 0
99 0
};
\addplot [semithick, maroon971111, dash pattern=on 1pt off 3pt on 3pt off 3pt, mark=asterisk, mark size=1, mark options={solid}]
table {%
0 0
1 0
2 0
3 0
4 0
5 0
6 0
7 0
8 0
9 0
10 0
11 0
12 0
13 0
14 0
15 0
16 0
17 0
18 0
19 0
20 0
21 0
22 0
23 0
24 0
25 0
26 0
27 0
28 0
29 0
30 0
31 0
32 0
33 0
34 0
35 0
36 0
37 0
38 0
39 0
40 0
41 0
42 0
43 0
44 0
45 0
46 0
47 0
48 0
49 0
50 0
51 0
52 0
53 0
54 0
55 0
56 0
57 0
58 0
59 0
60 0
61 0
62 0
63 0
64 0
65 0
66 0
67 0
68 0
69 0
70 0
71 0
72 0
73 0
74 0
75 0
76 0
77 0
78 0
79 0
80 0
81 0
82 0
83 0
84 0
85 0
86 0
87 0
88 0
89 0
90 0
91 0
92 0
93 0
94 0
95 0
96 0
97 0
98 0
99 0
};
\addplot [semithick, orange, dashed, mark=triangle*, mark size=1, mark options={solid}]
table {%
0 0
1 0
2 0
3 0
4 0
5 0
6 0
7 0
8 0
9 0
10 0
11 0
12 0
13 0
14 0
15 0
16 0
17 0
18 0
19 0
20 0
21 0
22 0
23 0
24 0
25 0
26 0
27 0
28 0
29 0
30 0
31 0
32 0
33 0
34 0
35 0
36 0
37 0
38 0
39 0
40 0
41 1038
42 2937
43 4822
44 6701
45 8578
46 10478
47 12373
48 14272
49 16172
50 18068
51 19962
52 21864
53 23765
54 24222
55 26120
56 28013
57 29881
58 31778
59 33634
60 34999
61 35577
62 36152
63 36747
64 37343
65 37931
66 38528
67 39109
68 39706
69 40296
70 40842
71 40842
72 40842
73 40842
74 40842
75 40842
76 40842
77 40842
78 40842
79 40842
80 40842
81 40842
82 40842
83 40842
84 40842
85 40842
86 40842
87 40842
88 40842
89 40842
90 40842
91 40842
92 40842
93 40842
94 40842
95 40842
96 40842
97 40842
98 40842
99 40842
};
\addplot [semithick, lightpink245169188, dashed, mark=pentagon*, mark size=1, mark options={solid}]
table {%
0 0
1 0
2 0
3 0
4 0
5 0
6 0
7 0
8 0
9 0
10 0
11 0
12 0
13 0
14 0
15 0
16 0
17 0
18 0
19 0
20 0
21 2050
22 4972
23 7895
24 10816
25 13736
26 16643
27 19553
28 22477
29 25400
30 28329
31 31244
32 34165
33 37080
34 39991
35 42909
36 45828
37 48755
38 51669
39 54582
40 57922
41 67079
42 76244
43 85406
44 94568
45 103694
46 112838
47 121983
48 131112
49 140255
50 149418
51 158564
52 167718
53 176859
54 182804
55 191950
56 201109
57 210280
58 219442
59 228535
60 237453
61 245318
62 253171
63 261022
64 268881
65 276745
66 284595
67 292463
68 300325
69 308185
70 316433
71 325484
72 333898
73 342347
74 350762
75 359201
76 367649
77 376059
78 384485
79 392919
80 400551
81 400745
82 400926
83 400926
84 400926
85 400926
86 400926
87 400926
88 400926
89 400926
90 400926
91 400926
92 400926
93 400926
94 400926
95 400926
96 400926
97 400926
98 400926
99 400926
};
\end{axis}

\end{tikzpicture}
        \hspace*{0.5em}   
        \centering \small (g) IETF: Packet Loss in output PE port.
    \end{minipage}
    \vspace*{0.25cm}
    \hfill
    \begin{minipage}[t]{0.3\textwidth}
        \centering
\begin{tikzpicture}

\definecolor{blue064255}{RGB}{0,64,255}
\definecolor{darkgrey176}{RGB}{176,176,176}
\definecolor{maroon971111}{RGB}{97,11,11}

\begin{axis}[
height=0.6\textwidth,
tick align=outside,
tick pos=left,
width=\textwidth,
x grid style={darkgrey176},
xlabel={t (s)},
xmajorgrids,
xmin=0, xmax=100,
xtick style={color=black},
y grid style={darkgrey176},
ylabel={Packet Loss},
ymajorgrids,
ymin=-68652.2, ymax=1441696.2,
ytick style={color=black}
]
\addplot [semithick, blue064255, mark=*, mark size=1, mark options={solid}]
table {%
0 0
1 0
2 0
3 0
4 0
5 0
6 0
7 0
8 0
9 0
10 0
11 0
12 0
13 0
14 0
15 0
16 0
17 0
18 0
19 0
20 0
21 0
22 0
23 0
24 0
25 0
26 0
27 0
28 0
29 0
30 0
31 0
32 0
33 0
34 0
35 0
36 0
37 0
38 0
39 0
40 0
41 0
42 0
43 0
44 0
45 0
46 0
47 0
48 0
49 0
50 0
51 0
52 0
53 0
54 0
55 0
56 0
57 0
58 0
59 0
60 0
61 0
62 0
63 0
64 0
65 0
66 0
67 0
68 0
69 0
70 0
71 0
72 0
73 0
74 0
75 0
76 0
77 0
78 0
79 0
80 0
81 0
82 0
83 0
84 0
85 0
86 0
87 0
88 0
89 0
90 0
91 0
92 0
93 0
94 0
95 0
96 0
97 0
98 0
99 0
};
\addplot [semithick, maroon971111, dash pattern=on 1pt off 3pt on 3pt off 3pt, mark=asterisk, mark size=1, mark options={solid}]
table {%
0 0
1 0
2 0
3 0
4 0
5 0
6 0
7 0
8 0
9 0
10 0
11 0
12 0
13 0
14 0
15 0
16 0
17 0
18 0
19 0
20 0
21 15137
22 31016
23 47111
24 63173
25 79182
26 95159
27 111177
28 127332
29 143342
30 159332
31 175350
32 191424
33 207498
34 223744
35 239981
36 256053
37 272294
38 288456
39 304614
40 321503
41 353955
42 386445
43 418780
44 451303
45 483731
46 516225
47 548690
48 581026
49 613490
50 645964
51 678445
52 710954
53 743437
54 775956
55 808313
56 840751
57 873269
58 905591
59 937842
60 969918
61 994275
62 1018586
63 1042858
64 1067219
65 1091574
66 1115942
67 1140269
68 1164638
69 1188976
70 1212696
71 1228881
72 1244913
73 1261144
74 1277373
75 1293595
76 1309729
77 1325940
78 1342022
79 1358107
80 1373044
81 1373044
82 1373044
83 1373044
84 1373044
85 1373044
86 1373044
87 1373044
88 1373044
89 1373044
90 1373044
91 1373044
92 1373044
93 1373044
94 1373044
95 1373044
96 1373044
97 1373044
98 1373044
99 1373044
};
\end{axis}

\end{tikzpicture}
        \hspace*{1.75em}
        \centering \small (h) \cite{Lin2021}: Packet Loss in output PE port.
    \end{minipage}
    \vspace*{0.25cm}
    \hfill
    \begin{minipage}[t]{0.3\textwidth} 
        \centering
\begin{tikzpicture}

\definecolor{blue064255}{RGB}{0,64,255}
\definecolor{darkgrey176}{RGB}{176,176,176}
\definecolor{lightpink245169188}{RGB}{245,169,188}
\definecolor{maroon971111}{RGB}{97,11,11}
\definecolor{orange}{RGB}{255,165,0}

\begin{axis}[
height=0.6\textwidth,
tick align=outside,
tick pos=left,
width=\textwidth,
x grid style={darkgrey176},
xlabel={t (s)},
xmajorgrids,
xmin=0, xmax=100,
xtick style={color=black},
y grid style={darkgrey176},
ylabel={Packet Loss},
ymajorgrids,
ymin=-0.1, ymax=1,
ytick style={color=black}
]
\addplot [semithick, blue064255, mark=*, mark size=1, mark options={solid}]
table {%
0 0
1 0
2 0
3 0
4 0
5 0
6 0
7 0
8 0
9 0
10 0
11 0
12 0
13 0
14 0
15 0
16 0
17 0
18 0
19 0
20 0
21 0
22 0
23 0
24 0
25 0
26 0
27 0
28 0
29 0
30 0
31 0
32 0
33 0
34 0
35 0
36 0
37 0
38 0
39 0
40 0
41 0
42 0
43 0
44 0
45 0
46 0
47 0
48 0
49 0
50 0
51 0
52 0
53 0
54 0
55 0
56 0
57 0
58 0
59 0
60 0
61 0
62 0
63 0
64 0
65 0
66 0
67 0
68 0
69 0
70 0
71 0
72 0
73 0
74 0
75 0
76 0
77 0
78 0
79 0
80 0
81 0
82 0
83 0
84 0
85 0
86 0
87 0
88 0
89 0
90 0
91 0
92 0
93 0
94 0
95 0
96 0
97 0
98 0
99 0
};
\addplot [semithick, maroon971111, dash pattern=on 1pt off 3pt on 3pt off 3pt, mark=asterisk, mark size=1, mark options={solid}]
table {%
0 0
1 0
2 0
3 0
4 0
5 0
6 0
7 0
8 0
9 0
10 0
11 0
12 0
13 0
14 0
15 0
16 0
17 0
18 0
19 0
20 0
21 0
22 0
23 0
24 0
25 0
26 0
27 0
28 0
29 0
30 0
31 0
32 0
33 0
34 0
35 0
36 0
37 0
38 0
39 0
40 0
41 0
42 0
43 0
44 0
45 0
46 0
47 0
48 0
49 0
50 0
51 0
52 0
53 0
54 0
55 0
56 0
57 0
58 0
59 0
60 0
61 0
62 0
63 0
64 0
65 0
66 0
67 0
68 0
69 0
70 0
71 0
72 0
73 0
74 0
75 0
76 0
77 0
78 0
79 0
80 0
81 0
82 0
83 0
84 0
85 0
86 0
87 0
88 0
89 0
90 0
91 0
92 0
93 0
94 0
95 0
96 0
97 0
98 0
99 0
};
\addplot [semithick, orange, dashed, mark=triangle*, mark size=1, mark options={solid}]
table {%
0 0
1 0
2 0
3 0
4 0
5 0
6 0
7 0
8 0
9 0
10 0
11 0
12 0
13 0
14 0
15 0
16 0
17 0
18 0
19 0
20 0
21 0
22 0
23 0
24 0
25 0
26 0
27 0
28 0
29 0
30 0
31 0
32 0
33 0
34 0
35 0
36 0
37 0
38 0
39 0
40 0
41 0
42 0
43 0
44 0
45 0
46 0
47 0
48 0
49 0
50 0
51 0
52 0
53 0
54 0
55 0
56 0
57 0
58 0
59 0
60 0
61 0
62 0
63 0
64 0
65 0
66 0
67 0
68 0
69 0
70 0
71 0
72 0
73 0
74 0
75 0
76 0
77 0
78 0
79 0
80 0
81 0
82 0
83 0
84 0
85 0
86 0
87 0
88 0
89 0
90 0
91 0
92 0
93 0
94 0
95 0
96 0
97 0
98 0
99 0
};
\addplot [semithick, lightpink245169188, dashed, mark=pentagon*, mark size=1, mark options={solid}]
table {%
0 0
1 0
2 0
3 0
4 0
5 0
6 0
7 0
8 0
9 0
10 0
11 0
12 0
13 0
14 0
15 0
16 0
17 0
18 0
19 0
20 0
21 0
22 0
23 0
24 0
25 0
26 0
27 0
28 0
29 0
30 0
31 0
32 0
33 0
34 0
35 0
36 0
37 0
38 0
39 0
40 0
41 0
42 0
43 0
44 0
45 0
46 0
47 0
48 0
49 0
50 0
51 0
52 0
53 0
54 0
55 0
56 0
57 0
58 0
59 0
60 0
61 0
62 0
63 0
64 0
65 0
66 0
67 0
68 0
69 0
70 0
71 0
72 0
73 0
74 0
75 0
76 0
77 0
78 0
79 0
80 0
81 0
82 0
83 0
84 0
85 0
86 0
87 0
88 0
89 0
90 0
91 0
92 0
93 0
94 0
95 0
96 0
97 0
98 0
99 0
};
\end{axis}

\end{tikzpicture}
        \hspace*{0.75em}
        \centering \small (i) \gls{hctns}: Packet Loss in output PE port.
    \end{minipage}
    \vspace*{0.25cm}
    \begin{minipage}[t]{0.3\textwidth}
        \centering
\begin{tikzpicture}

\definecolor{blue064255}{RGB}{0,64,255}
\definecolor{darkgrey176}{RGB}{176,176,176}
\definecolor{lightpink245169188}{RGB}{245,169,188}
\definecolor{maroon971111}{RGB}{97,11,11}
\definecolor{orange}{RGB}{255,165,0}

\begin{axis}[
height=0.6\textwidth,
tick align=outside,
tick pos=left,
width=\textwidth,
x grid style={darkgrey176},
xlabel={t (s)},
xmajorgrids,
xmin=0, xmax=100,
xtick style={color=black},
y grid style={darkgrey176},
ylabel={Packets Queued},
ymajorgrids,
ymin=-50, ymax=1050,
ytick style={color=black}
]
\addplot [semithick, blue064255, mark=*, mark size=1, mark options={solid}]
table {%
0 0
1 0
2 0
3 0
4 0
5 0
6 0
7 0
8 0
9 0
10 0
11 0
12 0
13 0
14 0
15 0
16 0
17 0
18 0
19 0
20 0
21 0
22 0
23 0
24 0
25 0
26 0
27 0
28 0
29 0
30 0
31 0
32 0
33 0
34 0
35 0
36 0
37 0
38 0
39 0
40 0
41 0
42 0
43 0
44 0
45 0
46 0
47 0
48 0
49 0
50 0
51 0
52 0
53 0
54 0
55 0
56 0
57 0
58 0
59 0
60 0
61 0
62 0
63 0
64 0
65 0
66 0
67 0
68 0
69 0
70 0
71 0
72 0
73 0
74 0
75 0
76 0
77 0
78 0
79 0
80 0
81 0
82 0
83 0
84 0
85 0
86 0
87 0
88 0
89 0
90 0
91 0
92 0
93 0
94 0
95 0
96 0
97 0
98 0
99 0
};
\addplot [semithick, maroon971111, dash pattern=on 1pt off 3pt on 3pt off 3pt, mark=asterisk, mark size=1, mark options={solid}]
table {%
0 0
1 0
2 0
3 0
4 0
5 0
6 0
7 0
8 0
9 0
10 0
11 0
12 0
13 0
14 0
15 0
16 0
17 0
18 0
19 0
20 0
21 0
22 0
23 0
24 0
25 0
26 0
27 0
28 0
29 0
30 0
31 0
32 0
33 1
34 1
35 1
36 0
37 1
38 0
39 0
40 0
41 0
42 0
43 1
44 1
45 0
46 0
47 1
48 1
49 1
50 0
51 0
52 1
53 0
54 0
55 1
56 1
57 2
58 1
59 1
60 0
61 0
62 0
63 0
64 0
65 0
66 0
67 0
68 0
69 0
70 0
71 0
72 0
73 0
74 0
75 0
76 0
77 0
78 0
79 0
80 0
81 0
82 0
83 0
84 0
85 0
86 0
87 0
88 0
89 0
90 0
91 0
92 0
93 0
94 0
95 0
96 0
97 0
98 0
99 0
};
\addplot [semithick, orange, dashed, mark=triangle*, mark size=1, mark options={solid}]
table {%
0 0
1 0
2 0
3 0
4 0
5 0
6 0
7 0
8 0
9 0
10 0
11 0
12 0
13 0
14 0
15 0
16 0
17 0
18 0
19 0
20 0
21 0
22 0
23 0
24 0
25 0
26 1
27 0
28 0
29 0
30 0
31 1
32 0
33 0
34 0
35 0
36 0
37 0
38 1
39 0
40 129
41 1000
42 1000
43 999
44 1000
45 1000
46 1000
47 1000
48 999
49 999
50 1000
51 1000
52 999
53 1000
54 1000
55 1000
56 1000
57 993
58 1000
59 962
60 999
61 1000
62 1000
63 1000
64 999
65 1000
66 999
67 1000
68 999
69 999
70 676
71 0
72 0
73 0
74 0
75 0
76 0
77 0
78 0
79 0
80 0
81 0
82 0
83 0
84 0
85 0
86 0
87 0
88 0
89 0
90 0
91 0
92 0
93 0
94 0
95 0
96 0
97 0
98 0
99 0
};
\addplot [semithick, lightpink245169188, dashed, mark=pentagon*, mark size=1, mark options={solid}]
table {%
0 0
1 0
2 0
3 0
4 0
5 0
6 0
7 0
8 0
9 0
10 0
11 0
12 0
13 0
14 0
15 0
16 0
17 0
18 0
19 0
20 127
21 999
22 1000
23 1000
24 1000
25 1000
26 1000
27 999
28 1000
29 1000
30 999
31 999
32 1000
33 1000
34 1000
35 1000
36 999
37 1000
38 1000
39 1000
40 1000
41 1000
42 1000
43 1000
44 1000
45 1000
46 1000
47 1000
48 1000
49 1000
50 1000
51 1000
52 1000
53 1000
54 999
55 1000
56 1000
57 1000
58 1000
59 1000
60 1000
61 999
62 999
63 1000
64 1000
65 1000
66 999
67 999
68 1000
69 999
70 1000
71 1000
72 999
73 999
74 999
75 1000
76 1000
77 1000
78 1000
79 1000
80 999
81 1000
82 999
83 992
84 988
85 984
86 979
87 976
88 974
89 971
90 967
91 965
92 963
93 959
94 954
95 951
96 949
97 944
98 941
99 936
};
\end{axis}

\end{tikzpicture}
        \hspace*{0.4em}   
        \centering \small (j) IETF: Packets Queued in output PE port.
    \end{minipage}
    \hfill
    \begin{minipage}[t]{0.3\textwidth}
        \centering
\begin{tikzpicture}

\definecolor{blue064255}{RGB}{0,64,255}
\definecolor{darkgrey176}{RGB}{176,176,176}
\definecolor{maroon971111}{RGB}{97,11,11}

\begin{axis}[
height=0.6\textwidth,
tick align=outside,
tick pos=left,
width=\textwidth,
x grid style={darkgrey176},
xlabel={t (s)},
xmajorgrids,
xmin=0, xmax=100,
xtick style={color=black},
y grid style={darkgrey176},
ylabel={Packets Queued},
ymajorgrids,
ymin=-50, ymax=1050,
ytick style={color=black}
]
\addplot [semithick, blue064255, mark=*, mark size=1, mark options={solid}]
table {%
0 0
1 0
2 0
3 0
4 0
5 0
6 0
7 0
8 0
9 0
10 0
11 0
12 0
13 0
14 0
15 0
16 0
17 0
18 0
19 0
20 0
21 0
22 0
23 0
24 0
25 0
26 0
27 0
28 0
29 0
30 0
31 0
32 0
33 0
34 0
35 0
36 1
37 0
38 0
39 0
40 0
41 2
42 1
43 2
44 1
45 3
46 0
47 1
48 1
49 1
50 2
51 0
52 1
53 1
54 2
55 1
56 0
57 2
58 1
59 1
60 0
61 0
62 0
63 1
64 0
65 0
66 0
67 1
68 0
69 0
70 0
71 0
72 0
73 0
74 0
75 0
76 0
77 0
78 0
79 0
80 0
81 0
82 0
83 0
84 0
85 0
86 0
87 0
88 0
89 0
90 0
91 0
92 0
93 0
94 0
95 0
96 0
97 0
98 0
99 0
};
\addplot [semithick, maroon971111, dash pattern=on 1pt off 3pt on 3pt off 3pt, mark=asterisk, mark size=1, mark options={solid}]
table {%
0 0
1 0
2 0
3 0
4 0
5 0
6 0
7 0
8 0
9 0
10 0
11 0
12 0
13 0
14 0
15 0
16 0
17 0
18 0
19 0
20 0
21 1000
22 1000
23 1000
24 1000
25 999
26 999
27 1000
28 999
29 999
30 999
31 1000
32 999
33 999
34 1000
35 1000
36 1000
37 999
38 1000
39 1000
40 1000
41 1000
42 1000
43 1000
44 1000
45 1000
46 1000
47 1000
48 1000
49 1000
50 1000
51 1000
52 1000
53 1000
54 1000
55 1000
56 1000
57 1000
58 1000
59 1000
60 1000
61 1000
62 1000
63 1000
64 1000
65 1000
66 1000
67 1000
68 1000
69 1000
70 1000
71 1000
72 999
73 1000
74 999
75 999
76 999
77 1000
78 999
79 999
80 995
81 989
82 982
83 976
84 975
85 974
86 973
87 970
88 959
89 957
90 955
91 954
92 953
93 952
94 949
95 948
96 947
97 943
98 941
99 937
};
\end{axis}

\end{tikzpicture}
        \hspace*{0.5em}
        \centering \small (k) \cite{Lin2021}: Packets Queued in output PE port.
    \end{minipage}
    \hfill
    \begin{minipage}[t]{0.3\textwidth} 
        \centering
\begin{tikzpicture}

\definecolor{blue064255}{RGB}{0,64,255}
\definecolor{darkgrey176}{RGB}{176,176,176}
\definecolor{lightpink245169188}{RGB}{245,169,188}
\definecolor{maroon971111}{RGB}{97,11,11}
\definecolor{orange}{RGB}{255,165,0}

\begin{axis}[
height=0.6\textwidth,
tick align=outside,
tick pos=left,
width=\textwidth,
x grid style={darkgrey176},
xlabel={t (s)},
xmajorgrids,
xmin=0, xmax=100,
xtick style={color=black},
y grid style={darkgrey176},
ylabel={Packets Queued},
ymajorgrids,
ymin=-0.1, ymax=1,
ytick style={color=black}
]
\addplot [semithick, blue064255, mark=*, mark size=1, mark options={solid}]
table {%
0 0
1 0
2 0
3 0
4 0
5 0
6 0
7 0
8 0
9 0
10 0
11 0
12 0
13 0
14 0
15 0
16 0
17 0
18 0
19 0
20 0
21 0
22 0
23 0
24 0
25 0
26 0
27 0
28 0
29 0
30 0
31 0
32 0
33 0
34 0
35 0
36 0
37 0
38 0
39 0
40 0
41 0
42 0
43 0
44 0
45 0
46 0
47 0
48 0
49 0
50 0
51 0
52 0
53 0
54 0
55 0
56 0
57 0
58 0
59 0
60 0
61 0
62 0
63 0
64 0
65 0
66 0
67 0
68 0
69 0
70 0
71 0
72 0
73 0
74 0
75 0
76 0
77 0
78 0
79 0
80 0
81 0
82 0
83 0
84 0
85 0
86 0
87 0
88 0
89 0
90 0
91 0
92 0
93 0
94 0
95 0
96 0
97 0
98 0
99 0
};
\addplot [semithick, maroon971111, dash pattern=on 1pt off 3pt on 3pt off 3pt, mark=asterisk, mark size=1, mark options={solid}]
table {%
0 0
1 0
2 0
3 0
4 0
5 0
6 0
7 0
8 0
9 0
10 0
11 0
12 0
13 0
14 0
15 0
16 0
17 0
18 0
19 0
20 0
21 0
22 0
23 0
24 0
25 0
26 0
27 0
28 0
29 0
30 0
31 0
32 0
33 0
34 0
35 0
36 0
37 0
38 0
39 0
40 0
41 0
42 0
43 0
44 0
45 0
46 0
47 0
48 0
49 0
50 0
51 0
52 0
53 0
54 0
55 0
56 0
57 0
58 0
59 0
60 0
61 0
62 0
63 0
64 0
65 0
66 0
67 0
68 0
69 0
70 0
71 0
72 0
73 0
74 0
75 0
76 0
77 0
78 0
79 0
80 0
81 0
82 0
83 0
84 0
85 0
86 0
87 0
88 0
89 0
90 0
91 0
92 0
93 0
94 0
95 0
96 0
97 0
98 0
99 0
};
\addplot [semithick, orange, dashed, mark=triangle*, mark size=1, mark options={solid}]
table {%
0 0
1 0
2 0
3 0
4 0
5 0
6 0
7 0
8 0
9 0
10 0
11 0
12 0
13 0
14 0
15 0
16 0
17 0
18 0
19 0
20 0
21 0
22 0
23 0
24 0
25 0
26 0
27 0
28 0
29 0
30 0
31 0
32 0
33 0
34 0
35 0
36 0
37 0
38 0
39 0
40 0
41 0
42 0
43 0
44 0
45 0
46 0
47 0
48 0
49 0
50 0
51 0
52 0
53 0
54 0
55 0
56 0
57 0
58 0
59 0
60 0
61 0
62 0
63 0
64 0
65 0
66 0
67 0
68 0
69 0
70 0
71 0
72 0
73 0
74 0
75 0
76 0
77 0
78 0
79 0
80 0
81 0
82 0
83 0
84 0
85 0
86 0
87 0
88 0
89 0
90 0
91 0
92 0
93 0
94 0
95 0
96 0
97 0
98 0
99 0
};
\addplot [semithick, lightpink245169188, dashed, mark=pentagon*, mark size=1, mark options={solid}]
table {%
0 0
1 0
2 0
3 0
4 0
5 0
6 0
7 0
8 0
9 0
10 0
11 0
12 0
13 0
14 0
15 0
16 0
17 0
18 0
19 0
20 0
21 0
22 0
23 0
24 0
25 0
26 0
27 0
28 0
29 0
30 0
31 0
32 0
33 0
34 0
35 0
36 0
37 0
38 0
39 0
40 0
41 0
42 0
43 0
44 0
45 0
46 0
47 0
48 0
49 0
50 0
51 0
52 0
53 0
54 0
55 0
56 0
57 0
58 0
59 0
60 0
61 0
62 0
63 0
64 0
65 0
66 0
67 0
68 0
69 0
70 0
71 0
72 0
73 0
74 0
75 0
76 0
77 0
78 0
79 0
80 0
81 0
82 0
83 0
84 0
85 0
86 0
87 0
88 0
89 0
90 0
91 0
92 0
93 0
94 0
95 0
96 0
97 0
98 0
99 0
};
\end{axis}

\end{tikzpicture}
        \hspace*{-0.2cm}
        \centering \small (l) \gls{hctns}: Packets Queued in output PE port.
    \end{minipage}
    \caption{Experiment A: \gls{ietf} and \cite{Lin2021} network slicing models versus \gls{hctns}.}
    \label{fig:expa-ietf-proposal}
\end{figure*}
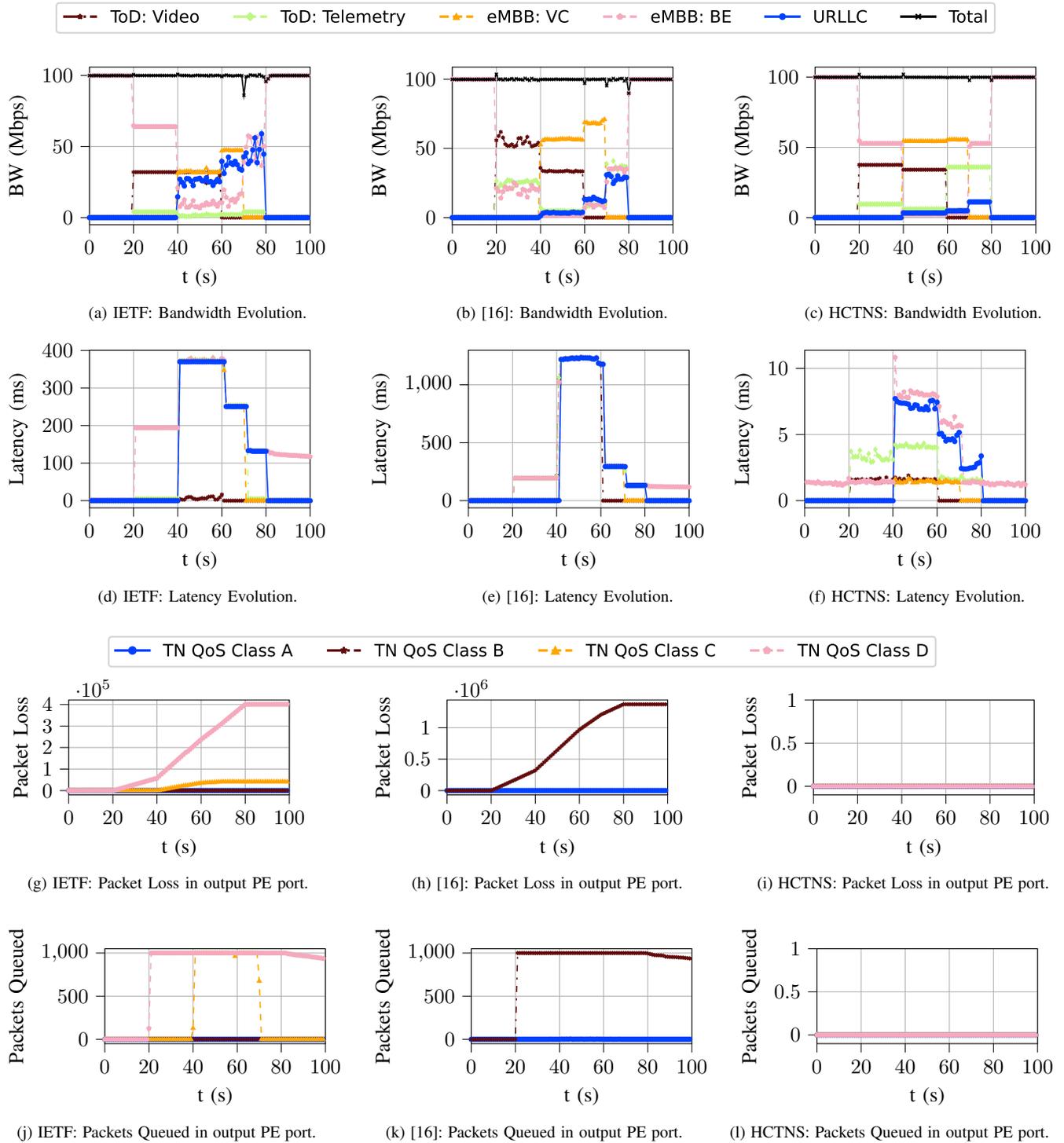
}

Fig.~\ref{fig:expa-ietf-proposal} presents the results obtained with the \gls{ietf} network slicing model alongside the results achieved with~\rev{the \cite{Lin2021} model and} \gls{hctns}. As discussed below, significant differences in the \gls{qos} provided to the traffic flows are observed between the three approaches.  

In \textit{Interval 1}, only BE traffic is active, fully utilizing the available bandwidth (100\,Mbps of the link between the PE1 and the P routers). All traffic is admitted by the ingress policers in all the models (Figures~\ref{fig:expa-ietf-proposal}a, \ref{fig:expa-ietf-proposal}b and \ref{fig:expa-ietf-proposal}c). Since the accepted BE traffic does not exceed the link capacity, packet delay remains low (Figures~\ref{fig:expa-ietf-proposal}d, \ref{fig:expa-ietf-proposal}e and \ref{fig:expa-ietf-proposal}f), and no packets are lost (Figures~\ref{fig:expa-ietf-proposal}g, \ref{fig:expa-ietf-proposal}h  and \ref{fig:expa-ietf-proposal}i) or enqueued (Figures~\ref{fig:expa-ietf-proposal}j, \ref{fig:expa-ietf-proposal}k and \ref{fig:expa-ietf-proposal}l) in the output port of \gls{pe}1. For this interval, all models show the same behavior.

In \textit{Interval 2}, BE traffic from the \gls{embb} slice, along with video and telemetry traffic from the \gls{tod} slice simultaneously enter the transport network.
\rev{The \gls{ietf} and \gls{hctns} models, by utilizing hierarchical policers, not only ensure the configured \gls{cir} for each traffic flow but also guarantee it at the slice level. Consequently, even though BE traffic does not have explicit bandwidth guarantees, it belongs to the \gls{embb} slice, which has a \gls{cir} of 52.8\,Mbps. If no other traffic in the slice is using that bandwidth, BE traffic can take advantage of the slice's guaranteed bandwidth.
In the \gls{ietf} model, traffic classes with bandwidth guarantees cannot exceed their assigned \gls{cir}. As a result, video traffic is limited to 32\,Mbps, aligning with the \gls{cir} values defined in Table~\ref{tab:expa-params}. Similarly, telemetry traffic is restricted to 4 Mbps. However, the 100\,Mbps of BE traffic are accepted by the policer, therefore, the remaining bandwidth is fully utilized by BE traffic (achieving 64 Mbps of the bandwidth in the link between the PE1 and the P routers). However, in the \gls{hctns} model, all the active flows can consume part of the excess bandwidth exceeding their respective \glspl{cir}. \rev{The \gls{htb} Linux tool implementation, based on a \gls{drr} scheduler that we configured with equal quantum values for all leaf nodes, assigns the excess bandwidth, in this case, as follows: the telemetry and video classes receive an additional 5.6 Mbps, while the BE traffic, which is set without a \gls{cir}, consume the 52.8 Mbps allocated to the \gls{embb} slice.} On the other hand, in the~\cite{Lin2021} model, BE traffic is treated as an independent slice without any bandwidth guarantees. Consequently, in the \cite{Lin2021} model, traffic flows can secure their \gls{cir} bandwidth (the policer marks this traffic as ``green") and other traffic (BE and ``yellow" traffic, i.e., the traffic above the \gls{cir} and under the \gls{pir} of other flows) compete for the remaining available bandwidth, achieving a total of 54\,Mbps for video, 26\,Mbps for telemetry, and 20\,Mbps for BE traffic.}

The coarse-grained resource control plays a crucial role in achieving these results. Packet loss in the \gls{ietf} model occurs when traffic admitted by the ingress policer attempts to access a \gls{drr} queue that is already full, as shown in Fig.~\ref{fig:expa-ietf-proposal}j, in the egress port of \gls{pe}1. \rev{The same occurs with the \cite{Lin2021} model, where BE and ``yellow" packets wait in the lower priority queue, as shown in Fig.~\ref{fig:expa-ietf-proposal}k}. In contrast, the operation of the global policer in \gls{hctns} prevents packets from waiting in the output port queues of the \gls{pe}1 (Fig.~\ref{fig:expa-ietf-proposal}l). This is achieved because our fine-grained resource control mechanism only accepts traffic that can be processed within the transport network with \gls{qos} guarantees.

 As shown in Table~\ref{tab:expa-params}, the same quantums values were configured for the \gls{drr} queues of the \gls{ietf} and \gls{hctns} models, corresponding to \gls{tn} classes B, C, and D. This configuration does not apply to \gls{tn} class A, as this traffic class utilizes the priority queue within the coarse-grained resource control queuing system, \rev{neither to the queues implemented in the~\cite{Lin2021} model, since~\cite{Lin2021} uses two priority queues.} For the quantum values used in this experiment, and without traffic of \gls{tn} \gls{qos} class A, the packet scheduler can serve packets from the \gls{drr} queues at an average transmission rate of 33.3\,Mbps for each of the \gls{tn} \gls{qos} classes B, C and D, when there are packets of all these \gls{tn} classes waiting in the queues to be transmitted.  

During \textit{Interval 2} of the Experiment A, the \gls{ietf} \gls{drr} queue corresponding to the \gls{tn} \gls{qos} class B exclusively receives video traffic from the \gls{tod} slice at a rate of 32\,Mbps, which explains why packets from this flow are neither dropped nor enqueued (Figs.~\ref{fig:expa-ietf-proposal}g and~\ref{fig:expa-ietf-proposal}j). Similarly, the \gls{drr} queue assigned to \gls{tn} \gls{qos} class C receives telemetry traffic at a rate of 4\,Mbps, a rate that is also below the threshold of 33.3\,Mbps. The problem arises with BE traffic, which is enqueued in the \gls{drr} for the \gls{tn} class D. In this case, BE traffic is received at a rate of 100\,Mbps. As previously mentioned, when all \gls{drr} queues contain packets waiting for transmission, the packet scheduler can only serve packets at an average rate of 33.3\,Mbps per queue. However, since the video and telemetry rates remain below their thresholds, the BE traffic can utilize the remaining available bandwidth, achieving a transmission rate of 64\,Mbps. As BE traffic is admitted by the \gls{ietf} ingress policer at a rate of 100\,Mbps, which exceeds the available 64\,Mbps, packet losses occur, and some packets experience queuing delays. \rev{On the other hand, the \cite{Lin2021} model forwards all the ``green" traffic (i.e., traffic under \gls{cir}) to the highest priority queue. Since the sum of the \gls{cir} values of the current traffic, 34\,Mbps in this interval, is lower than the link capacity, packets are not enqueued nor dropped. However, this model forwards all ``yellow'' traffic (i.e., between \gls{cir} and \gls{pir} rate) and the BE traffic to the lowest priority queue. As a consequence, more traffic arrives at the lowest priority queue than it can handle, leading to queue filling and packet losses, as shown in Figs.~\ref{fig:expa-ietf-proposal}k and ~\ref{fig:expa-ietf-proposal}h}. In contrast to both the IETF and the~\cite{Lin2021} models, \gls{hctns} includes a global policer that limits the total traffic entering the \gls{tn}. This mechanism ensures that traffic does not exceed the global available bandwidth at the output of the \gls{pe}1 router. As a result, packets are not enqueued nor dropped, as illustrated in Figs.~\ref{fig:expa-ietf-proposal}l and ~\ref{fig:expa-ietf-proposal}i.

Queuing delays determine the packet delay measurements shown in Figs.~\ref{fig:expa-ietf-proposal}d, \ref{fig:expa-ietf-proposal}e and \ref{fig:expa-ietf-proposal}f. During \textit{Interval 2}, in the \gls{ietf} model, packets from the \gls{tod} slice are not significantly affected by queuing delays, because traffic is generated at rates below the threshold of 33.3\,Mbps \rev{and there is not other traffic in their associated \gls{tn} \gls{qos} classes}. However, BE traffic is accumulated (also discarded) in the \gls{drr} queue of the \gls{tn} class D. \rev{In the \cite{Lin2021} model, since all flows have packets marked as ``yellow'' by the ingress policer, the maximum latency experienced by a flow is determined by the waiting time of the lowest priority queue, leading to similar latencies for all flows as BE traffic, around 195\,ms.} In contrast, in our proposed model, packet delay is influenced by the waiting time in the queue associated with the input port. This waiting time depends on the token arrival rate defined for each traffic class, i.e., the speed at which tokens are received in the \textit{\gls{cir}} and \textit{\gls{pir}} buckets defined in Section~\ref{sec:proposal}.\ref{subsec:proposal-finegraine}. As soon as there are tokens available in these buckets, the packet waiting in the queue of the \gls{htb} can be transmitted to the switching fabric of the \gls{pe} router. 

\textit{Interval 3} is particularly insightful for comparing our proposal with the other models. During this time interval, all traffic classes from the defined slices are transmitting traffic to the transport network. Unlike \textit{Interval 2}, for the \gls{ietf} and \gls{hctns} models, the Video Conferencing (VC) and the BE traffic classes are now sharing the network resources allocated to the \gls{embb} slice. However, VC traffic requires \gls{qos} guarantees, unlike the BE traffic class. The \gls{ietf} policer allows VC traffic to be transmitted at its \gls{cir} (52.8\,Mbps as shown in Table.~\ref{tab:expa-params}), leaving  47.2\,Mbps of remaining bandwidth of the slice available for BE traffic. 

During this interval, traffic from the \gls{urllc} slice is also generated at 100\,Mbps. Although the \gls{cir} of the \gls{urllc} slice policer is set to 1.2\,Mbps (Table~\ref{tab:expa-params}), Fig.~\ref{fig:expa-ietf-proposal}a reveals that, in the \gls{ietf} model, more bandwidth than this limit is being processed in the transport network (around 25\,Mbps). This occurs because traffic exceeding the \gls{cir} but under the \gls{pir} undergoes a process of ''de-prioritization'', according to the \gls{ietf} terminology. In practical terms, these packets will be treated as if they belonged to the BE traffic class, entering the \gls{drr} queue associated with the \gls{tn} \gls{qos} class D. This severely impacts the \gls{urllc} traffic, as shown in Fig.~\ref{fig:expa-ietf-proposal}d. Since a portion of the \gls{urllc} traffic is being enqueued in the \gls{drr} queue of the \gls{tn} \gls{qos} class D, the maximum packet delay for this traffic class rises to levels (more than 350\,ms) that make the \gls{urllc} slice unusable. Fig.~\ref{fig:expa-ietf-proposal}d also illustrates that BE traffic experiences similar delays, as it is enqueued in the same queue. Another problem arising from this situation is that, as more packets access the same \gls{drr} queue, the number  of discarded packets increases. From $t=40$ to $t=60$ there is a significant increase in the number of packets lost.

The queue associated with \gls{tn} \gls{qos} class C, which is assigned to the VC and telemetry traffic classes, also overflows during this interval in the \gls{ietf} model. This is explained because, when all the \gls{drr} queues contain packets waiting for transmission and there is also \gls{urllc} traffic (\gls{tn} class A), the packet scheduler delivers $\frac{100-1.2}{3}$\,Mbps$=32.9$\,Mbps per \gls{tn} \gls{qos} class using the \gls{drr} queues. Fig.~\ref{fig:expa-ietf-proposal}g shows that packets from the \gls{tn} \gls{qos} class C are dropped in the corresponding \gls{drr} queue, which is further confirmed by Fig.~\ref{fig:expa-ietf-proposal}j, where the queue is shown to be full. The same happens for \gls{tn} \gls{qos} class D. As a consequence, the guaranteed bandwidth for the \gls{embb} and \gls{tod} slices, which is 52.8\,Mbps and 36\,Mbps respectively (see Table \ref{tab:expa-params}), is not respected and the agreed bandwidth specified in the \glspl{sla} is violated.

Regarding the packet delay experienced during this time interval in the \gls{ietf} model, Fig.~\ref{fig:expa-ietf-proposal}d shows worse results compared to \textit{Interval 2}. This is because more traffic is sharing the \gls{drr} queues associated with the \gls{tn} \gls{qos} classes than in \textit{Interval 2}. The \gls{tn} \gls{qos} class A only receives \gls{urllc} traffic, but as explained earlier, a portion of this traffic is treated as best effort, leading to unacceptable delay levels. \gls{tn} \gls{qos} class B receives only video generated by the \gls{tod} slice. Since this video is generated at a rate of 32\,Mbps, below the 32.9\,Mbps threshold, the packet delay for this type of traffic remains very low during this interval. However, both \gls{tn} \gls{qos} classes C and D receive traffic at an aggregated rate that exceeds the 32.9\,Mbps limit, resulting in increased delays for the corresponding traffic classes. 

\rev{During this interval, the~\cite{Lin2021} model improves the bandwidth performance of the \gls{ietf} model but not its latency performance. As shown in Fig.~\ref{fig:expa-ietf-proposal}b, the model processes more bandwidth than the configured \gls{cir} values (Table~\ref{tab:expa-params}) for all flows. However, similar to the \gls{ietf} model, traffic exceeding the \gls{cir} but below the \gls{pir} undergoes a process of "de-prioritization," where packets are marked as ``yellow" and directed to the lowest-priority queue, negatively impacting their latencies. This occurs because these packets from all flows are enqueued in the same low priority queue.}

\rev{Since ``green" packets of all the flows are forwarded to the highest priority queue, this queue utilizes 90\,Mbps (i.e., the sum of the \gls{cir} values for each traffic class), leaving only 10\,Mbps for the lowest-priority queue. Additionally, due to the ingress policing mechanism, all traffic between the \gls{cir} and \gls{pir} is accepted and marked as ``yellow," to be discarded later if necessary. Consequently, the lowest priority queue receives traffic at the sum of the \gls{pir}-\gls{cir} values for \gls{urllc}, video, telemetry, and VC traffic, as well as best-effort (BE) traffic, reaching an arrival rate of 410\,Mbps and saturating the queue. In \textit{Interval 2}, the lowest priority queue has 66\,Mbps available for transmission with packets arriving at 264\,Mbps, whereas in \textit{Interval 3}, the lower-priority queue has only 10\,Mbps for transmission while receiving traffic at 410\,Mbps. This results in a substantial increase in packet losses and latency, making the ``yellow" traffic of all flows unusable and reaching levels exceeding 1200\,ms, as shown in Figs.~\ref{fig:expa-ietf-proposal}h and \ref{fig:expa-ietf-proposal}e.} 

In contrast, \gls{hctns} incorporates a global policer that limits the traffic admitted at the transport network's ingress, ensuring that the accepted traffic rate does not exceed the capacity of the link connecting the PE1 and P routers (100\,Mbps). The proposed three-level \gls{htb} efficiently shares bandwidth among slices, and further distributes it within slices according to their \gls{cir} and \gls{pir} values. As shown in Fig.~\ref{fig:expa-ietf-proposal}c, each traffic class is provided with more than its guaranteed bandwidth. For instance, VC traffic is transmitted at 54.3\,Mbps, exceeding the \gls{cir} of 52.8\,Mbps. The results obtained by our model in terms of latency (Fig.~\ref{fig:expa-ietf-proposal}f) are remarkable, with values approximately two orders of magnitude lower than those observed with the \gls{ietf} \rev{and the~\cite{Lin2021} models}. For example, the packet delay perceived by the \gls{urllc} packets is approximately 7\,ms when using our proposal, compared to over 350\,ms with the \gls{ietf} model \rev{and 1230\,ms with the~\cite{Lin2021} model}. This improvement is due to absence of queuing delays in the \gls{drr} queues associated with the output port of the \gls{pe}. In \gls{hctns}, packet delay under saturation scenarios is primarily influenced by the token arrival rate for each traffic class. 

During \textit{Interval 4}, all traffic classes remain active except the \gls{tod} video. In the \gls{ietf} model, this results in the \gls{drr} queue associated with the \gls{tn} \gls{qos} class B becoming empty. Consequently, the bandwidth previously allocated to this traffic flow is now available to be shared between \gls{tn} \gls{qos} classes C and D. As a result, the packet scheduler can serve packets from the two remaining \gls{drr} queues at a rate higher than 32.9\,Mbps, in particular, at 49.4\,Mbps. This allows packets waiting in the \gls{drr} queues associated with \gls{tn} \gls{qos} classes C and D to be dequeued more quickly. Fig.~\ref{fig:expa-ietf-proposal}d illustrates the resulting reduction in latency in the \gls{ietf} model for all active traffic classes, which decreases to approximately to 200\,ms.

\rev{In the~\cite{Lin2021} model, unlike the \gls{ietf} and \gls{hctns} models, the bandwidth released by the video traffic is not directly consumed by the telemetry traffic, since the~\cite{Lin2021} model treats telemetry and video traffic as separated slices. The bandwidth below the \gls{cir} released by the video is freed in the higher priority queue and added to the lower priority queue. Moreover, the bandwidth between the \gls{cir} and the \gls{pir} released by the video also becomes available in the lower priority queue and can be leveraged by ``yellow'' traffic from other traffic flows. However, traffic continues to arrive at a higher rate than the lower priority queue can transmit now, i.e., 42\,Mbps. As a result, the queue remains full, although packet loss rates and latencies decrease.}

\rev{In \gls{hctns}, since the \gls{tod} video flow is deactivated, the remaining traffic classes gain a slightly larger share of the available bandwidth. The total \gls{cir} across all defined traffic classes in this interval amounts to 58\,Mbps, leaving 42\,Mbps for redistribution by the global policer among all active flows. Therefore, in this interval, telemetry traffic, which belongs to the \gls{tod} slice, consumes the \gls{cir} in the slice released by the video, reaching 36\,Mbps, while \gls{urllc}, VC, and BE traffic consume the rest of the bandwidth, reaching 4.8\,Mbps, 3.6\,Mbps, and 55.5\,Mbps, respectively. }

During \textit{Interval 5}, the VC traffic flow is deactivated, causing the \gls{drr} queue associated with the \gls{tn} \gls{qos} class C in the \gls{ietf} model to become empty (Fig.~\ref{fig:expa-ietf-proposal}j), as it only receives packets from the telemetry traffic class. Consequently, packet delay for telemetry packets decreases notably during this interval, as shown in Fig.~\ref{fig:expa-ietf-proposal}d. Additionally, the number of packets lost for \gls{tn} \gls{qos} class C stops increasing, as Fig.~\ref{fig:expa-ietf-proposal}g shows. In contrast, traffic continues to arrive at \gls{tn} \gls{qos} class D faster than it can be transmitted, causing the corresponding \gls{drr} queue to remain full. This explains the ongoing enqueuing and packet loss observed for \gls{tn} \gls{qos} class D. \rev{In the \cite{Lin2021} model, the lowest priority queue gains more bandwidth that is shared between the BE and the ``yellow" traffic of the \gls{urllc} and telemetry flows, increasing their bandwidths as shown in Fig.~\ref{fig:expa-ietf-proposal}b. However, traffic continues to arrive at a higher rate than the lower priority queue can transmit now, 294.8\,Mbps versus 94.8\,Mbps. As a result, the queue remains full, but packet loss rates and latencies decrease, as shown in Figs.~\ref{fig:expa-ietf-proposal}h and \ref{fig:expa-ietf-proposal}e.} In \gls{hctns}, the active classes achieve a larger share of bandwidth without packet losses or added delay.  

Finally, \textit{Interval 6} returns to the initial state, where only BE traffic is active in the experiment. In the \gls{ietf} model, this traffic encounters the \gls{drr} queue associated with \gls{tn} \gls{qos} class D already full of packets. Fig.~\ref{fig:expa-ietf-proposal}j shows that packets from this class remain enqueued during this interval. This results in a larger latency than in \textit{Interval 1}. However, Fig.~\ref{fig:expa-ietf-proposal}e indicates that no additional packets are lost from this traffic class. This is because BE traffic is being received at the transmission rate. The results achieved in terms of bandwidth are consistent with those observed during \textit{Interval 1}. \rev{Similarly to the \gls{ietf} model, in the \cite{Lin2021} model, BE traffic encounters the lower priority queue already full of packets, as Fig.~\ref{fig:expa-ietf-proposal}k shows, resulting in a higher experienced latency than in \textit{Interval 1}. Nevertheless, Fig.~\ref{fig:expa-ietf-proposal}h indicates that no additional packets are lost. This is because BE traffic is received at the transmission rate available in the lower priority queue. The results achieved in terms of bandwidth are also consistent with those observed during \textit{Interval 1}}. In \gls{hctns}, the results in terms of both bandwidth and latency are the same as in \textit{Interval 1}.

An important conclusion from this experiment is that, \rev{unlike the \gls{ietf} model, the~\cite{Lin2021} model and \gls{hctns} comply with the \glspl{cir} defined (Table~\ref{tab:expa-params}) in all the intervals.  However, the results obtained with the~\cite{Lin2021} model show that the traffic above the \gls{cir} experiments large latencies, while our model ensures that all the traffic from the slices is treated with the expected \gls{qos} level. Moreover, our model allows that the bandwidth unused by any traffic class from a slice to be shared among the other classes. In contrast, the \gls{ietf} model restricts the sharing of unused bandwidth in the slice to the best-effort classes only, and the~\cite{Lin2021} model does not distinguish between slices and classes, treating each flow as a separated slice.} Moreover, our proposed model provides flexibility for network operators to configure the bandwidth sharing mechanism according to their specific needs, enabling adaptation to the slices defined in their network scenarios.

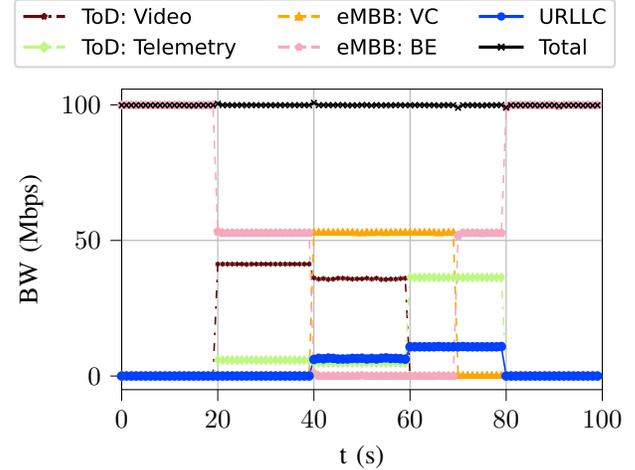
\begin{figure}[t]
    \vspace*{-0.75em}
    \centering
    \includegraphics[width=0.47\textwidth, trim=210 10 210 10, clip]{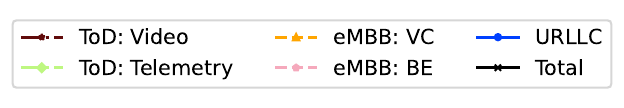}
    \centering
    \begin{minipage}[t]{0.45\textwidth} 
        \centering
\begin{tikzpicture}

\definecolor{blue064255}{RGB}{0,64,255}
\definecolor{darkgrey176}{RGB}{176,176,176}
\definecolor{lightpink245169188}{RGB}{245,169,188}
\definecolor{maroon971111}{RGB}{97,11,11}
\definecolor{orange}{RGB}{255,165,0}
\definecolor{palegreen190247129}{RGB}{190,247,129}

\begin{axis}[
height=0.7\textwidth,
tick align=outside,
tick pos=left,
width=\textwidth,
x grid style={darkgrey176},
xlabel={t (s)},
xmajorgrids,
xmin=0, xmax=100,
xtick style={color=black},
y grid style={darkgrey176},
ylabel={BW (Mbps)},
ymajorgrids,
ymin=-5.03872622282609, ymax=105.813250679348,
ytick style={color=black}
]
\addplot [semithick, maroon971111, dash pattern=on 1pt off 3pt on 3pt off 3pt, mark=asterisk, mark size=1, mark options={solid}]
table {%
0 0
1 0
2 0
3 0
4 0
5 0
6 0
7 0
8 0
9 0
10 0
11 0
12 0
13 0
14 0
15 0
16 0
17 0
18 0
19 0
20 41.2710597826087
21 41.2710597826087
22 41.2710597826087
23 41.2710597826087
24 41.2710597826087
25 41.2710597826087
26 41.1665760869565
27 41.2710597826087
28 41.2710597826087
29 41.2710597826087
30 41.2710597826087
31 41.2710597826087
32 41.2710597826087
33 41.2710597826087
34 41.2710597826087
35 41.2710597826087
36 41.2710597826087
37 41.2710597826087
38 41.2710597826087
39 41.2710597826087
40 36.1513586956522
41 35.7334239130435
42 35.9423913043478
43 35.6289402173913
44 35.7334239130435
45 36.046875
46 36.046875
47 35.9423913043478
48 35.7334239130435
49 35.8379076086956
50 35.8379076086956
51 36.046875
52 35.7334239130435
53 36.046875
54 35.6289402173913
55 35.6289402173913
56 35.7334239130435
57 35.7334239130435
58 35.9423913043478
59 36.046875
60 0
61 0
62 0
63 0
64 0
65 0
66 0
67 0
68 0
69 0
70 0
71 0
72 0
73 0
74 0
75 0
76 0
77 0
78 0
79 0
80 0
81 0
82 0
83 0
84 0
85 0
86 0
87 0
88 0
89 0
90 0
91 0
92 0
93 0
94 0
95 0
96 0
97 0
98 0
99 0
};
\addplot [semithick, palegreen190247129, dash pattern=on 1pt off 3pt on 3pt off 3pt, mark=diamond*, mark size=1.5, mark options={solid}]
table {%
0 0
1 0
2 0
3 0
4 0
5 0
6 0
7 0
8 0
9 0
10 0
11 0
12 0
13 0
14 0
15 0
16 0
17 0
18 0
19 0
20 5.88243206521739
21 5.87198369565217
22 5.88243206521739
23 5.86153532608696
24 5.86153532608696
25 5.89288043478261
26 5.91377717391304
27 5.88243206521739
28 5.86153532608696
29 5.87198369565217
30 5.86153532608696
31 5.86153532608696
32 5.84063858695652
33 5.86153532608696
34 5.86153532608696
35 5.86153532608696
36 5.86153532608696
37 5.86153532608696
38 5.86153532608696
39 5.85108695652174
40 4.85849184782609
41 4.87938858695652
42 4.85849184782609
43 4.90028532608696
44 4.85849184782609
45 4.82714673913043
46 4.85849184782609
47 4.82714673913043
48 4.87938858695652
49 4.87938858695652
50 4.81669836956522
51 4.85849184782609
52 4.8689402173913
53 4.83759510869565
54 4.87938858695652
55 4.87938858695652
56 4.83759510869565
57 4.87938858695652
58 4.85849184782609
59 4.83759510869565
60 36.3603260869565
61 36.3603260869565
62 36.3603260869565
63 36.3603260869565
64 36.3603260869565
65 36.3603260869565
66 36.2558423913043
67 36.3603260869565
68 36.3603260869565
69 36.3603260869565
70 36.3603260869565
71 36.2558423913043
72 36.3603260869565
73 36.3603260869565
74 36.3603260869565
75 36.3603260869565
76 36.3603260869565
77 36.3603260869565
78 36.3603260869565
79 36.3603260869565
80 0
81 0
82 0
83 0
84 0
85 0
86 0
87 0
88 0
89 0
90 0
91 0
92 0
93 0
94 0
95 0
96 0
97 0
98 0
99 0
};
\addplot [semithick, orange, dashed, mark=triangle*, mark size=1.5, mark options={solid}]
table {%
0 0
1 0
2 0
3 0
4 0
5 0
6 0
7 0
8 0
9 0
10 0
11 0
12 0
13 0
14 0
15 0
16 0
17 0
18 0
19 0
20 0
21 0
22 0
23 0
24 0
25 0
26 0
27 0
28 0
29 0
30 0
31 0
32 0
33 0
34 0
35 0
36 0
37 0
38 0
39 0
40 52.7642663043478
41 52.7642663043478
42 52.7642663043478
43 52.7642663043478
44 52.7642663043478
45 52.7642663043478
46 52.7642663043478
47 52.7642663043478
48 52.7642663043478
49 52.6597826086956
50 52.5552989130435
51 52.7642663043478
52 52.6597826086956
53 52.7642663043478
54 52.7642663043478
55 52.6597826086956
56 52.7642663043478
57 52.7642663043478
58 52.7642663043478
59 52.7642663043478
60 52.7642663043478
61 52.7642663043478
62 52.7642663043478
63 52.7642663043478
64 52.7642663043478
65 52.7642663043478
66 52.5552989130435
67 52.7642663043478
68 52.7642663043478
69 52.7642663043478
70 0
71 0
72 0
73 0
74 0
75 0
76 0
77 0
78 0
79 0
80 0
81 0
82 0
83 0
84 0
85 0
86 0
87 0
88 0
89 0
90 0
91 0
92 0
93 0
94 0
95 0
96 0
97 0
98 0
99 0
};
\addplot [semithick, lightpink245169188, dashed, mark=pentagon*, mark size=1.5, mark options={solid}]
table {%
0 99.8864130434782
1 99.8864130434782
2 99.8864130434782
3 99.8864130434782
4 99.8864130434782
5 99.8864130434782
6 99.8864130434782
7 99.8864130434782
8 99.8864130434782
9 99.8864130434782
10 99.8864130434782
11 99.8864130434782
12 99.8864130434782
13 99.8864130434782
14 99.7819293478261
15 99.7819293478261
16 99.8864130434782
17 99.8864130434782
18 99.8864130434782
19 99.8864130434782
20 53.2866847826087
21 52.7642663043478
22 52.7642663043478
23 52.7642663043478
24 52.7642663043478
25 52.7642663043478
26 52.7642663043478
27 52.7642663043478
28 52.7642663043478
29 52.7642663043478
30 52.7642663043478
31 52.7642663043478
32 52.7642663043478
33 52.7642663043478
34 52.7642663043478
35 52.7642663043478
36 52.7642663043478
37 52.7642663043478
38 52.7642663043478
39 52.7642663043478
40 0.909008152173913
41 0.0114932065217391
42 0
43 0.0114932065217391
44 0.0114932065217391
45 0.0114932065217391
46 0.0114932065217391
47 0.0114932065217391
48 0
49 0.0114932065217391
50 0.183891304347826
51 0.0114932065217391
52 0.0365692934782609
53 0
54 0.0114932065217391
55 0.0365692934782609
56 0.02403125
57 0.0114932065217391
58 0
59 0.0114932065217391
60 0.0114932065217391
61 0.0114932065217391
62 0.0114932065217391
63 0.0114932065217391
64 0
65 0.0114932065217391
66 0.110752717391304
67 0.0114932065217391
68 0.0114932065217391
69 0
70 51.8239130434783
71 52.7642663043478
72 52.7642663043478
73 52.7642663043478
74 52.7642663043478
75 52.6597826086956
76 52.7642663043478
77 52.7642663043478
78 52.7642663043478
79 52.7642663043478
80 99.0505434782609
81 99.8864130434782
82 99.8864130434782
83 99.8864130434782
84 99.8864130434782
85 99.8864130434782
86 99.8864130434782
87 99.7819293478261
88 99.8864130434782
89 99.8864130434782
90 99.8864130434782
91 99.5729619565217
92 99.8864130434782
93 99.8864130434782
94 99.8864130434782
95 99.8864130434782
96 99.7819293478261
97 99.8864130434782
98 99.7819293478261
99 99.8864130434782
};
\addplot [semithick, blue064255, mark=*, mark size=1.5, mark options={solid}]
table {%
0 0
1 0
2 0
3 0
4 0
5 0
6 0
7 0
8 0
9 0
10 0
11 0
12 0
13 0
14 0
15 0
16 0
17 0
18 0
19 0
20 0
21 0
22 0
23 0
24 0
25 0
26 0
27 0
28 0
29 0
30 0
31 0
32 0
33 0
34 0
35 0
36 0
37 0
38 0
39 0
40 6.09139945652174
41 6.50933423913043
42 6.35260869565217
43 6.61381793478261
44 6.50933423913043
45 6.26902173913043
46 6.27947010869565
47 6.27947010869565
48 6.49888586956522
49 6.43619565217391
50 6.56157608695652
51 6.26902173913043
52 6.4884375
53 6.30036684782609
54 6.5511277173913
55 6.67650815217391
56 6.4884375
57 6.49888586956522
58 6.36305706521739
59 6.21677989130435
60 10.7618206521739
61 10.7618206521739
62 10.7618206521739
63 10.7618206521739
64 10.7618206521739
65 10.7618206521739
66 10.8663043478261
67 10.7618206521739
68 10.7618206521739
69 10.7618206521739
70 10.7618206521739
71 10.8663043478261
72 10.7618206521739
73 10.7618206521739
74 10.7618206521739
75 10.7618206521739
76 10.7618206521739
77 10.7618206521739
78 10.7618206521739
79 10.8663043478261
80 0
81 0
82 0
83 0
84 0
85 0
86 0
87 0
88 0
89 0
90 0
91 0
92 0
93 0
94 0
95 0
96 0
97 0
98 0
99 0
};
\addplot [semithick, black, mark=x, mark size=1.5, mark options={solid}]
table {%
0 99.8864130434782
1 99.8864130434782
2 99.8864130434782
3 99.8864130434782
4 99.8864130434782
5 99.8864130434782
6 99.8864130434782
7 99.8864130434782
8 99.8864130434782
9 99.8864130434782
10 99.8864130434782
11 99.8864130434782
12 99.8864130434782
13 99.8864130434782
14 99.7819293478261
15 99.7819293478261
16 99.8864130434782
17 99.8864130434782
18 99.8864130434782
19 99.8864130434782
20 100.440176630435
21 99.9073097826087
22 99.9177581521739
23 99.8968614130435
24 99.8968614130435
25 99.9282065217391
26 99.8446195652174
27 99.9177581521739
28 99.8968614130435
29 99.9073097826087
30 99.8968614130435
31 99.8968614130435
32 99.875964673913
33 99.8968614130435
34 99.8968614130435
35 99.8968614130435
36 99.8968614130435
37 99.8968614130435
38 99.8968614130435
39 99.8864130434783
40 100.774524456522
41 99.89790625
42 99.9177581521739
43 99.9188029891304
44 99.8770095108696
45 99.9188029891304
46 99.9605964673913
47 99.8247676630435
48 99.875964673913
49 99.8247676630434
50 99.9553722826087
51 99.9501480978261
52 99.7871535326087
53 99.9491032608695
54 99.8352160326087
55 99.8811888586957
56 99.847754076087
57 99.8874578804348
58 99.9282065217391
59 99.8770095108695
60 99.89790625
61 99.89790625
62 99.89790625
63 99.89790625
64 99.8864130434782
65 99.89790625
66 99.7881983695652
67 99.89790625
68 99.89790625
69 99.8864130434782
70 98.9460597826087
71 99.8864130434783
72 99.8864130434783
73 99.8864130434783
74 99.8864130434783
75 99.7819293478261
76 99.8864130434783
77 99.8864130434783
78 99.8864130434783
79 99.9908967391304
80 99.0505434782609
81 99.8864130434782
82 99.8864130434782
83 99.8864130434782
84 99.8864130434782
85 99.8864130434782
86 99.8864130434782
87 99.7819293478261
88 99.8864130434782
89 99.8864130434782
90 99.8864130434782
91 99.5729619565217
92 99.8864130434782
93 99.8864130434782
94 99.8864130434782
95 99.8864130434782
96 99.7819293478261
97 99.8864130434782
98 99.7819293478261
99 99.8864130434782
};
\end{axis}

\end{tikzpicture}
    \end{minipage}
    \vspace{-0.35cm}
    \caption{Bandwidth Behaviour with Table~\ref{tab:expa-params-sharing} configuration.}
    \label{other-bw-config}
\end{figure}

The results described above were obtained using the Linux \gls{htb} bandwidth sharing mechanism with all leaf nodes quantum set to the same value (the packet size). However, \gls{hctns}, unlike the other models, introduces two alternative methods for bandwidth sharing within the ingress policer: Weighted/Quantum and Priority based. As previously mentioned, the Linux \gls{tc} tool currently allows configuring these parameters only at the child nodes of the \gls{htb} tree. Our proposal extends this capability to any node of the tree. To evaluate the potential utility of these alternative bandwidth sharing methods, we conducted an additional experiment leveraging the capabilities of the Linux \gls{tc} tool. Table~\ref{tab:expa-params-sharing} shows the priority and quantums values set for the defined traffic classes in our network scenario. The results, depicted in Fig.~\ref{other-bw-config}, show the bandwidth evolution over the same timeline as in the previous experiment. 

\begin{table}[t]
\vspace{0.5em}
\hspace*{6em}
\resizebox{0.33\textwidth}{!}{%
\begin{tabular}{cccll}
\cline{1-3}
\multicolumn{1}{|c|}{\cellcolor[HTML]{FFFFFF}\textbf{Traffic Class}} & \multicolumn{1}{c|}{\cellcolor[HTML]{FFFFFF}\textbf{PRIO}} & \multicolumn{1}{c|}{\cellcolor[HTML]{FFFFFF}\textbf{Quantum}} &  &  \\ \cline{1-3}
\multicolumn{1}{|c|}{\cellcolor[HTML]{FFFFFF}URLLC}                  & \multicolumn{1}{c|}{\cellcolor[HTML]{FFFFFF}0}             & \multicolumn{1}{c|}{\cellcolor[HTML]{FFFFFF}18456 B}           &  &  \\ \cline{1-3}
\multicolumn{1}{|c|}{\cellcolor[HTML]{FFFFFF}Video}                  & \multicolumn{1}{c|}{\cellcolor[HTML]{FFFFFF}0}             & \multicolumn{1}{c|}{\cellcolor[HTML]{FFFFFF}15380 B}           &  &  \\ \cline{1-3}
\multicolumn{1}{|c|}{\cellcolor[HTML]{FFFFFF}Telemetry}              & \multicolumn{1}{c|}{\cellcolor[HTML]{FFFFFF}0}             & \multicolumn{1}{c|}{\cellcolor[HTML]{FFFFFF}3076 B}            &  &  \\ \cline{1-3}
\multicolumn{1}{|c|}{\cellcolor[HTML]{FFFFFF}VC}                     & \multicolumn{1}{c|}{\cellcolor[HTML]{FFFFFF}7}             & \multicolumn{1}{c|}{\cellcolor[HTML]{FFFFFF}12304 B}           &  &  \\ \cline{1-3}
\multicolumn{1}{|c|}{\cellcolor[HTML]{FFFFFF}BE}                     & \multicolumn{1}{c|}{\cellcolor[HTML]{FFFFFF}7}             & \multicolumn{1}{c|}{\cellcolor[HTML]{FFFFFF}6152 B}            &  &  \\ \cline{1-3}
\multicolumn{1}{l}{}                                                 & \multicolumn{1}{l}{}                                       & \multicolumn{1}{l}{}                                          &  &  
\end{tabular}%
}
\vspace*{0.5em}
\caption{New bandwidth sharing configuration in the ingress policer.}
\label{tab:expa-params-sharing}
\vspace*{-0.75em}
\end{table}

During \textit{Interval 1}, when only BE traffic is active, the bandwidth evolution matches that shown in Fig.~\ref{fig:expa-ietf-proposal}b. In \textit{Interval 2}, however, differences appear. Traffic classes from the \gls{tod} slice are given higher priority than those from the \gls{embb} slice. As a result, when tokens are available in the global policer, \gls{tod} slice traffic classes are prioritized for consuming the remaining bandwidth. Additionally, the \gls{tod} video traffic class is assigned a quantum value five times higher than that of the telemetry traffic class, enabling video traffic to borrow up to five times more bandwidth than telemetry traffic. Nonetheless, the \gls{cir} set for all of the traffic classes and slices is guaranteed. For instance, while BE traffic cannot borrow bandwidth due to its low priority, the \gls{cir} of the \gls{embb} slice is still respected.

During \textit{Interval 3}, all traffic classes have their \gls{cir} guaranteed. However, as shown in Table~\ref{tab:expa-params-sharing}, the \gls{urllc} and the \gls{tod} traffic classes are assigned higher priority. This ensures that any unused bandwidth is exclusively shared between these two classes according to the configured quantum values. In this scenario, since BE traffic does not have a guaranteed rate and is assigned the lowest priority, it does not consume any bandwidth, with packets from this class being denied access by the ingress policer.

In \textit{Interval 4}, the telemetry class takes over the bandwidth initially allocated to the \gls{tod} slice, while the \gls{urllc} traffic takes over the remaining available bandwidth. During the \textit{Interval 5}, when the \gls{tod} video and \gls{embb} VC traffic classes are deactivated, the active traffic flows obtain the \gls{cir} allocated to their slices, with the \gls{urllc} traffic consuming the remaining bandwidth. Finally, in \textit{Interval 6}, where only BE traffic is active in the \gls{tn}, the behavior mirrors that of \textit{Interval 1}.

The results demonstrate that, regardless of the bandwidth sharing method used, the global policer ensures that traffic is transmitted to the \gls{tn} without exceeding the global available bandwidth. This prevents packets from waiting in the \gls{drr} and priority queues of the coarse-grained resource control mechanism, while also avoiding packet losses, the same as in Figs.~\ref{fig:expa-ietf-proposal}f and \ref{fig:expa-ietf-proposal}h.

\begin{table*}
\centering
\resizebox{\textwidth}{!}{%
\begin{tabular}{cclccclccccccccll}
\cline{2-15}
\multicolumn{1}{l|}{}                                                                  & \multicolumn{10}{c|}{\cellcolor[HTML]{F5F5F5}\textbf{Fine-grained Ingress Policer Resource Control}}                                                                                                                                                                                                                                                                                                                                                                                                                                                                            & \multicolumn{4}{c|}{\cellcolor[HTML]{DAE8FC}\textbf{Coarse-grained Resource Control}}                                                                                                                                                                                                       &  &  \\ \cline{1-15}
\multicolumn{1}{|c|}{\cellcolor[HTML]{FFFFFF}}                                         & \multicolumn{4}{c|}{\cellcolor[HTML]{FFFFFF}\textbf{Class Policer}}                                                                                                                               & \multicolumn{4}{c|}{\cellcolor[HTML]{FFFFFF}\textbf{Slice Policer}}                                                                                                                                                          & \multicolumn{2}{c|}{\cellcolor[HTML]{FFFFFF}}                                                                                                & \multicolumn{4}{c|}{\cellcolor[HTML]{FFFFFF}}                                                                                                                                                                                                                                               &  &  \\ \cline{2-9}
\multicolumn{1}{|c|}{\cellcolor[HTML]{FFFFFF}}                                         & \multicolumn{2}{c|}{\cellcolor[HTML]{FFFFFF}\textbf{CIR}}           & \multicolumn{2}{c|}{\cellcolor[HTML]{FFFFFF}\textbf{PIR}}                                                                   & \multicolumn{2}{c|}{\cellcolor[HTML]{FFFFFF}\textbf{CIR}}                & \multicolumn{2}{c|}{\cellcolor[HTML]{FFFFFF}\textbf{PIR}}                                                                                         & \multicolumn{2}{c|}{\multirow{-2}{*}{\cellcolor[HTML]{FFFFFF}\textbf{Global Policer}}}                                                       & \multicolumn{4}{c|}{\multirow{-2}{*}{\cellcolor[HTML]{FFFFFF}\textbf{Quantum Parameters}}}                                                                                                                                                                                                  &  &  \\ \cline{2-15}
\multicolumn{1}{|c|}{\multirow{-3}{*}{\cellcolor[HTML]{FFFFFF}\textbf{Traffic Class}}} & \multicolumn{2}{c|}{\cellcolor[HTML]{FFFFFF}\textbf{IETF/\gls{hctns}}} & \multicolumn{1}{c|}{\cellcolor[HTML]{FFFFFF}\textbf{IETF}} & \multicolumn{1}{c|}{\cellcolor[HTML]{FFFFFF}\textbf{\gls{hctns}}} & \multicolumn{2}{c|}{\cellcolor[HTML]{FFFFFF}\textbf{IETF/\gls{hctns}}}      & \multicolumn{1}{c|}{\cellcolor[HTML]{FFFFFF}\textbf{IETF}}              & \multicolumn{1}{c|}{\cellcolor[HTML]{FFFFFF}\textbf{\gls{hctns}}}          & \multicolumn{1}{c|}{\cellcolor[HTML]{FFFFFF}\textbf{IETF}}         & \multicolumn{1}{c|}{\cellcolor[HTML]{FFFFFF}\textbf{\gls{hctns}}}          & \multicolumn{1}{c|}{\cellcolor[HTML]{FFFFFF}\textbf{Conf1}}          & \multicolumn{1}{c|}{\cellcolor[HTML]{FFFFFF}\textbf{Conf2}}           & \multicolumn{1}{c|}{\cellcolor[HTML]{FFFFFF}\textbf{Conf3}}          & \multicolumn{1}{c|}{\cellcolor[HTML]{FFFFFF}\textbf{Conf4}}           &  &  \\ \cline{1-15}
\multicolumn{1}{|c|}{\cellcolor[HTML]{FFFFFF}URLLC}                                    & \multicolumn{2}{c|}{\cellcolor[HTML]{FFFFFF}N/A}                    & \multicolumn{1}{c|}{\cellcolor[HTML]{FFFFFF}N/A}           & \multicolumn{1}{c|}{\cellcolor[HTML]{FFFFFF}N/A}               & \multicolumn{2}{c|}{\cellcolor[HTML]{FFFFFF}1.2 Mbps}                    & \multicolumn{1}{c|}{\cellcolor[HTML]{FFFFFF}100 Mbps}                   & \multicolumn{1}{c|}{\cellcolor[HTML]{FFFFFF}100 Mbps}                   & \multicolumn{1}{c|}{\cellcolor[HTML]{FFFFFF}}                      & \multicolumn{1}{c|}{\cellcolor[HTML]{FFFFFF}}                           & \multicolumn{1}{c|}{\cellcolor[HTML]{FFFFFF}PQ}                     & \multicolumn{1}{c|}{\cellcolor[HTML]{FFFFFF}PQ}                      & \multicolumn{1}{c|}{\cellcolor[HTML]{FFFFFF}PQ}                     & \multicolumn{1}{c|}{\cellcolor[HTML]{FFFFFF}PQ}                      &  &  \\ \cline{1-9} \cline{12-15}
\multicolumn{1}{|c|}{\cellcolor[HTML]{FFFFFF}Video}                                    & \multicolumn{2}{c|}{\cellcolor[HTML]{FFFFFF}32 Mbps}                & \multicolumn{1}{c|}{\cellcolor[HTML]{FFFFFF}N/A}           & \multicolumn{1}{c|}{\cellcolor[HTML]{FFFFFF}100 Mbps}          & \multicolumn{2}{c|}{\cellcolor[HTML]{FFFFFF}}                            & \multicolumn{1}{c|}{\cellcolor[HTML]{FFFFFF}}                           & \multicolumn{1}{c|}{\cellcolor[HTML]{FFFFFF}}                           & \multicolumn{1}{c|}{\cellcolor[HTML]{FFFFFF}}                      & \multicolumn{1}{c|}{\cellcolor[HTML]{FFFFFF}}                           & \multicolumn{1}{c|}{\cellcolor[HTML]{FFFFFF}1538 B}                   & \multicolumn{1}{c|}{\cellcolor[HTML]{FFFFFF}1538 B}                    & \multicolumn{1}{c|}{\cellcolor[HTML]{FFFFFF}15380 B}                  & \multicolumn{1}{c|}{\cellcolor[HTML]{FFFFFF}15380 B}                   &  &  \\ \cline{1-5} \cline{12-15}
\multicolumn{1}{|c|}{\cellcolor[HTML]{FFFFFF}Telemetry}                                & \multicolumn{2}{c|}{\cellcolor[HTML]{FFFFFF}4 Mbps}                 & \multicolumn{1}{c|}{\cellcolor[HTML]{FFFFFF}N/A}           & \multicolumn{1}{c|}{\cellcolor[HTML]{FFFFFF}100 Mbps}          & \multicolumn{2}{c|}{\multirow{-2}{*}{\cellcolor[HTML]{FFFFFF}36 Mbps}}   & \multicolumn{1}{c|}{\multirow{-2}{*}{\cellcolor[HTML]{FFFFFF}N/A}}      & \multicolumn{1}{c|}{\multirow{-2}{*}{\cellcolor[HTML]{FFFFFF}100 Mbps}} & \multicolumn{1}{c|}{\cellcolor[HTML]{FFFFFF}}                      & \multicolumn{1}{c|}{\cellcolor[HTML]{FFFFFF}}                           & \multicolumn{1}{c|}{\cellcolor[HTML]{FFFFFF}}                        & \multicolumn{1}{c|}{\cellcolor[HTML]{FFFFFF}}                         & \multicolumn{1}{c|}{\cellcolor[HTML]{FFFFFF}}                        & \multicolumn{1}{c|}{\cellcolor[HTML]{FFFFFF}}                         &  &  \\ \cline{1-9}
\multicolumn{1}{|c|}{\cellcolor[HTML]{FFFFFF}VC}                                       & \multicolumn{2}{c|}{\cellcolor[HTML]{FFFFFF}52.8 Mbps}              & \multicolumn{1}{c|}{\cellcolor[HTML]{FFFFFF}N/A}           & \multicolumn{1}{c|}{\cellcolor[HTML]{FFFFFF}100 Mbps}          & \multicolumn{2}{c|}{\cellcolor[HTML]{FFFFFF}}                            & \multicolumn{1}{c|}{\cellcolor[HTML]{FFFFFF}}                           & \multicolumn{1}{c|}{\cellcolor[HTML]{FFFFFF}}                           & \multicolumn{1}{c|}{\cellcolor[HTML]{FFFFFF}}                      & \multicolumn{1}{c|}{\cellcolor[HTML]{FFFFFF}}                           & \multicolumn{1}{c|}{\multirow{-2}{*}{\cellcolor[HTML]{FFFFFF}1538 B}} & \multicolumn{1}{c|}{\multirow{-2}{*}{\cellcolor[HTML]{FFFFFF}15380 B}} & \multicolumn{1}{c|}{\multirow{-2}{*}{\cellcolor[HTML]{FFFFFF}1538 B}} & \multicolumn{1}{c|}{\multirow{-2}{*}{\cellcolor[HTML]{FFFFFF}10766 B}} &  &  \\ \cline{1-5} \cline{12-15}
\multicolumn{1}{|c|}{\cellcolor[HTML]{FFFFFF}BE}                                       & \multicolumn{2}{c|}{\cellcolor[HTML]{FFFFFF}N/A}                    & \multicolumn{1}{c|}{\cellcolor[HTML]{FFFFFF}N/A}           & \multicolumn{1}{c|}{\cellcolor[HTML]{FFFFFF}N/A}               & \multicolumn{2}{c|}{\multirow{-2}{*}{\cellcolor[HTML]{FFFFFF}52.8 Mbps}} & \multicolumn{1}{c|}{\multirow{-2}{*}{\cellcolor[HTML]{FFFFFF}100 Mbps}} & \multicolumn{1}{c|}{\multirow{-2}{*}{\cellcolor[HTML]{FFFFFF}100 Mbps}} & \multicolumn{1}{c|}{\multirow{-5}{*}{\cellcolor[HTML]{FFFFFF}N/A}} & \multicolumn{1}{c|}{\multirow{-5}{*}{\cellcolor[HTML]{FFFFFF}100 Mbps}} & \multicolumn{1}{c|}{\cellcolor[HTML]{FFFFFF}15380 B}                  & \multicolumn{1}{c|}{\cellcolor[HTML]{FFFFFF}1538 B}                    & \multicolumn{1}{c|}{\cellcolor[HTML]{FFFFFF}1538 B}                   & \multicolumn{1}{c|}{\cellcolor[HTML]{FFFFFF}1538 B}                    &  &  \\ \cline{1-15}
\multicolumn{1}{l}{}                                                                   & \multicolumn{1}{l}{}                       &                        & \multicolumn{1}{l}{}                                       & \multicolumn{1}{l}{}                                           & \multicolumn{1}{l}{}                          &                          & \multicolumn{1}{l}{}                                                    & \multicolumn{1}{l}{}                                                    & \multicolumn{1}{l}{}                                               & \multicolumn{1}{l}{}                                                    & \multicolumn{1}{l}{}                                                 & \multicolumn{1}{l}{}                                                  & \multicolumn{1}{l}{}                                                 & \multicolumn{1}{l}{}                                                  &  &  \\
\multicolumn{1}{l}{}                                                                   & \multicolumn{1}{l}{}                       &                        & \multicolumn{1}{l}{}                                       & \multicolumn{1}{l}{}                                           & \multicolumn{1}{l}{}                          &                          & \multicolumn{1}{l}{}                                                    & \multicolumn{1}{l}{}                                                    & \multicolumn{1}{l}{}                                               & \multicolumn{1}{l}{}                                                    & \multicolumn{1}{l}{}                                                 & \multicolumn{1}{l}{}                                                  & \multicolumn{1}{l}{}                                                 & \multicolumn{1}{l}{}                                                  &  & 
\end{tabular}%
}
\vspace*{-0.25cm}
\caption{Experiment B System Parameter Configuration.}
\label{tab:expb-params}
\end{table*}

\subsubsection{Experiment B: Teleoperated Driving Slice Performance}
\label{expb}

In this experiment, we analyze the impact of coarse-grained resource control configuration on the bandwidth consumption and packet losses of the telemetry and video traffic classes within the ToD slice. The \gls{pe}1 router of our network scenario has been configured using the parameters and values listed in Table~\ref{tab:expb-params}. While the configuration of the fine-grained resource control mechanism remains unchanged, four different configurations of the coarse-grained resource control mechanism are defined in this experiment. The \gls{ietf} network slicing model and \gls{hctns} are compared. 

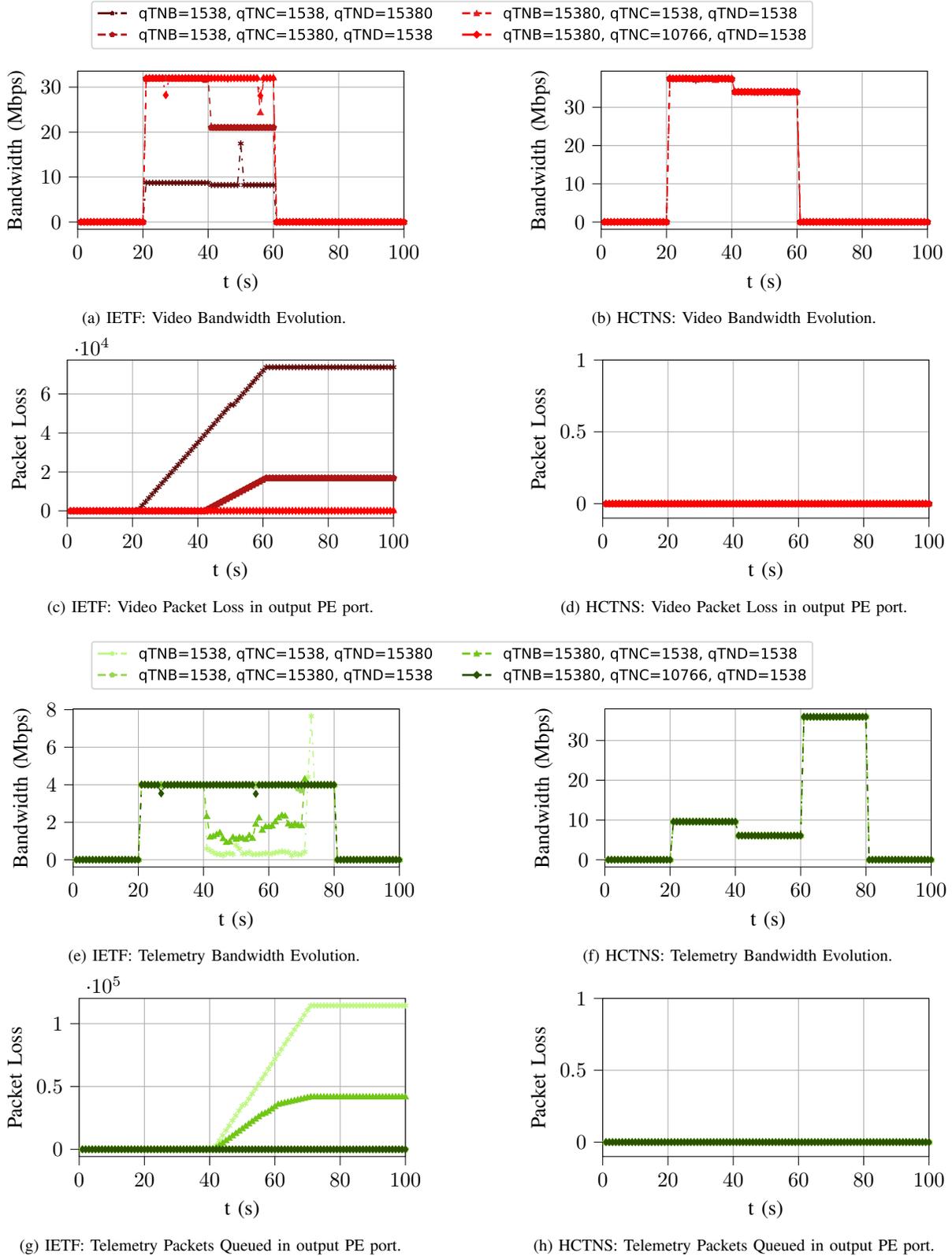
\begin{figure*}[!h]
    \centering
        \vspace*{-2.5em}
        \hspace*{-0.8cm}\includegraphics[width=1\textwidth, trim=0 10 0 10, clip]{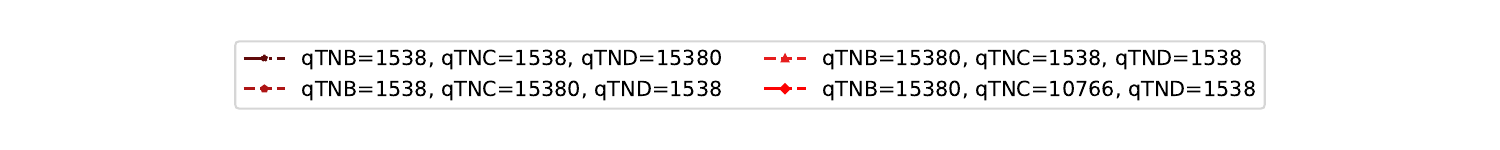}
    \begin{minipage}[t]{0.4\textwidth} 
        \centering
\begin{tikzpicture}

\definecolor{crimson2302929}{RGB}{230,29,29}
\definecolor{darkgrey176}{RGB}{176,176,176}
\definecolor{firebrick1742121}{RGB}{174,21,21}
\definecolor{maroon971111}{RGB}{97,11,11}
\definecolor{red25233}{RGB}{252,3,3}

\begin{axis}[
height=0.6\textwidth,
tick align=outside,
tick pos=left,
width=\textwidth,
x grid style={darkgrey176},
xlabel={t (s)},
xmajorgrids,
xmin=0, xmax=100,
xtick style={color=black},
y grid style={darkgrey176},
ylabel={Bandwidth (Mbps)},
ymajorgrids,
ymin=-1.59860054347826, ymax=33.5706114130435,
ytick style={color=black}
]
\addplot [semithick, maroon971111, dash pattern=on 1pt off 3pt on 3pt off 3pt, mark=asterisk, mark size=1.5, mark options={solid}]
table {%
1 0
2 0
3 0
4 0
5 0
6 0
7 0
8 0
9 0
10 0
11 0
12 0
13 0
14 0
15 0
16 0
17 0
18 0
19 0
20 0
21 8.74528532608696
22 8.7139402173913
23 8.72438858695652
24 8.7139402173913
25 8.7139402173913
26 8.72438858695652
27 8.7139402173913
28 8.7139402173913
29 8.72438858695652
30 8.7139402173913
31 8.7139402173913
32 8.72438858695652
33 8.7139402173913
34 8.72438858695652
35 8.72438858695652
36 8.7139402173913
37 8.7139402173913
38 8.72438858695652
39 8.7139402173913
40 8.7139402173913
41 8.22286684782609
42 8.2333152173913
43 8.22286684782609
44 8.2333152173913
45 8.22286684782609
46 8.2333152173913
47 8.22286684782609
48 8.22286684782609
49 8.2333152173913
50 17.448777173913
51 8.22286684782609
52 8.2333152173913
53 8.22286684782609
54 8.2333152173913
55 8.22286684782609
56 8.22286684782609
57 8.2333152173913
58 8.22286684782609
59 8.2333152173913
60 8.22286684782609
61 0
62 0
63 0
64 0
65 0
66 0
67 0
68 0
69 0
70 0
71 0
72 0
73 0
74 0
75 0
76 0
77 0
78 0
79 0
80 0
81 0
82 0
83 0
84 0
85 0
86 0
87 0
88 0
89 0
90 0
91 0
92 0
93 0
94 0
95 0
96 0
97 0
98 0
99 0
100 0
};
\addplot [semithick, firebrick1742121, dashed, mark=pentagon*, mark size=1.5, mark options={solid}]
table {%
1 0
2 0
3 0
4 0
5 0
6 0
7 0
8 0
9 0
10 0
11 0
12 0
13 0
14 0
15 0
16 0
17 0
18 0
19 0
20 0
21 31.867527173913
22 31.9720108695652
23 31.9720108695652
24 31.9720108695652
25 31.9720108695652
26 31.9720108695652
27 31.9720108695652
28 31.9720108695652
29 31.9720108695652
30 31.9720108695652
31 31.9720108695652
32 31.9720108695652
33 31.9720108695652
34 31.9720108695652
35 31.9720108695652
36 31.9720108695652
37 31.9720108695652
38 31.9720108695652
39 31.6585597826087
40 31.9720108695652
41 21.001222826087
42 21.001222826087
43 21.001222826087
44 21.001222826087
45 21.001222826087
46 21.001222826087
47 21.001222826087
48 21.001222826087
49 21.001222826087
50 21.001222826087
51 21.001222826087
52 21.001222826087
53 21.001222826087
54 21.001222826087
55 21.001222826087
56 21.001222826087
57 21.001222826087
58 21.001222826087
59 21.001222826087
60 21.001222826087
61 0
62 0
63 0
64 0
65 0
66 0
67 0
68 0
69 0
70 0
71 0
72 0
73 0
74 0
75 0
76 0
77 0
78 0
79 0
80 0
81 0
82 0
83 0
84 0
85 0
86 0
87 0
88 0
89 0
90 0
91 0
92 0
93 0
94 0
95 0
96 0
97 0
98 0
99 0
100 0
};
\addplot [semithick, crimson2302929, dashed, mark=triangle*, mark size=1.5, mark options={solid}]
table {%
1 0
2 0
3 0
4 0
5 0
6 0
7 0
8 0
9 0
10 0
11 0
12 0
13 0
14 0
15 0
16 0
17 0
18 0
19 0
20 0
21 31.9720108695652
22 31.9720108695652
23 31.9720108695652
24 31.9720108695652
25 31.9720108695652
26 31.9720108695652
27 31.9720108695652
28 31.9720108695652
29 31.9720108695652
30 31.9720108695652
31 31.9720108695652
32 31.9720108695652
33 31.9720108695652
34 31.9720108695652
35 31.9720108695652
36 31.9720108695652
37 31.9720108695652
38 31.9720108695652
39 31.9720108695652
40 31.867527173913
41 31.9720108695652
42 31.9720108695652
43 31.9720108695652
44 31.9720108695652
45 31.9720108695652
46 31.9720108695652
47 31.9720108695652
48 31.9720108695652
49 31.9720108695652
50 31.9720108695652
51 31.9720108695652
52 31.9720108695652
53 31.9720108695652
54 31.9720108695652
55 31.9720108695652
56 24.2402173913043
57 31.9720108695652
58 31.9720108695652
59 31.9720108695652
60 31.9720108695652
61 0
62 0
63 0
64 0
65 0
66 0
67 0
68 0
69 0
70 0
71 0
72 0
73 0
74 0
75 0
76 0
77 0
78 0
79 0
80 0
81 0
82 0
83 0
84 0
85 0
86 0
87 0
88 0
89 0
90 0
91 0
92 0
93 0
94 0
95 0
96 0
97 0
98 0
99 0
100 0
};
\addplot [semithick, red25233, dash pattern=on 1pt off 3pt on 3pt off 3pt, mark=diamond*, mark size=1.5, mark options={solid}]
table {%
1 0
2 0
3 0
4 0
5 0
6 0
7 0
8 0
9 0
10 0
11 0
12 0
13 0
14 0
15 0
16 0
17 0
18 0
19 0
20 0
21 31.9720108695652
22 31.9720108695652
23 31.9720108695652
24 31.9720108695652
25 31.9720108695652
26 31.9720108695652
27 28.210597826087
28 31.9720108695652
29 31.9720108695652
30 31.9720108695652
31 31.9720108695652
32 31.9720108695652
33 31.9720108695652
34 31.9720108695652
35 31.9720108695652
36 31.9720108695652
37 31.9720108695652
38 31.9720108695652
39 31.9720108695652
40 31.867527173913
41 31.9720108695652
42 31.9720108695652
43 31.9720108695652
44 31.9720108695652
45 31.9720108695652
46 31.867527173913
47 31.9720108695652
48 31.9720108695652
49 31.9720108695652
50 31.9720108695652
51 31.9720108695652
52 31.9720108695652
53 31.9720108695652
54 31.9720108695652
55 31.9720108695652
56 28.1061141304348
57 31.9720108695652
58 31.9720108695652
59 31.9720108695652
60 31.9720108695652
61 0
62 0
63 0
64 0
65 0
66 0
67 0
68 0
69 0
70 0
71 0
72 0
73 0
74 0
75 0
76 0
77 0
78 0
79 0
80 0
81 0
82 0
83 0
84 0
85 0
86 0
87 0
88 0
89 0
90 0
91 0
92 0
93 0
94 0
95 0
96 0
97 0
98 0
99 0
100 0
};
\end{axis}

\end{tikzpicture}\vspace*{-0.1cm}
        \vspace*{-0.2cm} \centering \small (a) IETF: Video Bandwidth Evolution.
    \end{minipage}
    \vspace*{0.2cm}
    \hspace{1.5cm}
    \begin{minipage}[t]{0.4\textwidth} 
        \centering
\begin{tikzpicture}

\definecolor{crimson2302929}{RGB}{230,29,29}
\definecolor{darkgrey176}{RGB}{176,176,176}
\definecolor{firebrick1742121}{RGB}{174,21,21}
\definecolor{maroon971111}{RGB}{97,11,11}
\definecolor{red25233}{RGB}{252,3,3}

\begin{axis}[
height=0.6\textwidth,
tick align=outside,
tick pos=left,
width=\textwidth,
x grid style={darkgrey176},
xlabel={t (s)},
xmajorgrids,
xmin=0, xmax=100,
xtick style={color=black},
y grid style={darkgrey176},
ylabel={Bandwidth (Mbps)},
ymajorgrids,
ymin=-1.88070652173913, ymax=39.4948369565217,
ytick style={color=black}
]
\addplot [semithick, maroon971111, dash pattern=on 1pt off 3pt on 3pt off 3pt, mark=asterisk, mark size=1.5, mark options={solid}]
table {%
1 0
2 0
3 0
4 0
5 0
6 0
7 0
8 0
9 0
10 0
11 0
12 0
13 0
14 0
15 0
16 0
17 0
18 0
19 0
20 0
21 37.6141304347826
22 37.5096467391304
23 37.5096467391304
24 37.5096467391304
25 37.5096467391304
26 37.5096467391304
27 37.5096467391304
28 37.5096467391304
29 36.9872282608696
30 37.5096467391304
31 37.5096467391304
32 37.6141304347826
33 37.5096467391304
34 37.5096467391304
35 37.5096467391304
36 37.5096467391304
37 37.4051630434783
38 37.3006793478261
39 37.4051630434783
40 37.4051630434783
41 33.9572010869565
42 34.0616847826087
43 33.9572010869565
44 34.0616847826087
45 34.0616847826087
46 34.0616847826087
47 34.0616847826087
48 33.9572010869565
49 33.9572010869565
50 34.0616847826087
51 33.8527173913043
52 34.0616847826087
53 34.0616847826087
54 33.9572010869565
55 33.9572010869565
56 34.0616847826087
57 34.0616847826087
58 33.9572010869565
59 34.0616847826087
60 33.9572010869565
61 0
62 0
63 0
64 0
65 0
66 0
67 0
68 0
69 0
70 0
71 0
72 0
73 0
74 0
75 0
76 0
77 0
78 0
79 0
80 0
81 0
82 0
83 0
84 0
85 0
86 0
87 0
88 0
89 0
90 0
91 0
92 0
93 0
94 0
95 0
96 0
97 0
98 0
99 0
100 0
};
\addplot [semithick, firebrick1742121, dashed, mark=pentagon*, mark size=1.5, mark options={solid}]
table {%
1 0
2 0
3 0
4 0
5 0
6 0
7 0
8 0
9 0
10 0
11 0
12 0
13 0
14 0
15 0
16 0
17 0
18 0
19 0
20 0
21 37.5096467391304
22 37.5096467391304
23 37.5096467391304
24 37.5096467391304
25 37.5096467391304
26 37.6141304347826
27 37.5096467391304
28 37.5096467391304
29 37.5096467391304
30 37.5096467391304
31 37.5096467391304
32 37.6141304347826
33 37.5096467391304
34 37.5096467391304
35 37.5096467391304
36 37.6141304347826
37 37.5096467391304
38 37.5096467391304
39 37.5096467391304
40 37.4051630434783
41 34.0616847826087
42 34.0616847826087
43 34.0616847826087
44 34.0616847826087
45 34.0616847826087
46 34.0616847826087
47 33.9572010869565
48 34.0616847826087
49 34.0616847826087
50 34.0616847826087
51 34.0616847826087
52 34.0616847826087
53 34.0616847826087
54 34.0616847826087
55 34.0616847826087
56 34.0616847826087
57 33.9572010869565
58 33.9572010869565
59 34.0616847826087
60 33.9572010869565
61 0
62 0
63 0
64 0
65 0
66 0
67 0
68 0
69 0
70 0
71 0
72 0
73 0
74 0
75 0
76 0
77 0
78 0
79 0
80 0
81 0
82 0
83 0
84 0
85 0
86 0
87 0
88 0
89 0
90 0
91 0
92 0
93 0
94 0
95 0
96 0
97 0
98 0
99 0
100 0
};
\addplot [semithick, crimson2302929, dashed, mark=triangle*, mark size=1.5, mark options={solid}]
table {%
1 0
2 0
3 0
4 0
5 0
6 0
7 0
8 0
9 0
10 0
11 0
12 0
13 0
14 0
15 0
16 0
17 0
18 0
19 0
20 0
21 37.6141304347826
22 37.5096467391304
23 37.5096467391304
24 37.5096467391304
25 37.6141304347826
26 37.5096467391304
27 37.4051630434783
28 37.5096467391304
29 37.5096467391304
30 37.1961956521739
31 37.5096467391304
32 37.4051630434783
33 37.5096467391304
34 37.5096467391304
35 36.9872282608696
36 37.5096467391304
37 37.4051630434783
38 37.6141304347826
39 37.5096467391304
40 37.5096467391304
41 33.9572010869565
42 33.9572010869565
43 33.9572010869565
44 33.9572010869565
45 33.9572010869565
46 33.9572010869565
47 33.9572010869565
48 33.9572010869565
49 33.9572010869565
50 34.0616847826087
51 33.9572010869565
52 34.0616847826087
53 33.9572010869565
54 33.7482336956522
55 33.9572010869565
56 34.0616847826087
57 33.9572010869565
58 34.0616847826087
59 33.9572010869565
60 33.9572010869565
61 0
62 0
63 0
64 0
65 0
66 0
67 0
68 0
69 0
70 0
71 0
72 0
73 0
74 0
75 0
76 0
77 0
78 0
79 0
80 0
81 0
82 0
83 0
84 0
85 0
86 0
87 0
88 0
89 0
90 0
91 0
92 0
93 0
94 0
95 0
96 0
97 0
98 0
99 0
100 0
};
\addplot [semithick, red25233, dash pattern=on 1pt off 3pt on 3pt off 3pt, mark=diamond*, mark size=1.5, mark options={solid}]
table {%
1 0
2 0
3 0
4 0
5 0
6 0
7 0
8 0
9 0
10 0
11 0
12 0
13 0
14 0
15 0
16 0
17 0
18 0
19 0
20 0
21 37.6141304347826
22 37.5096467391304
23 37.5096467391304
24 37.6141304347826
25 37.5096467391304
26 37.5096467391304
27 37.5096467391304
28 37.5096467391304
29 37.4051630434783
30 37.5096467391304
31 37.5096467391304
32 37.5096467391304
33 37.5096467391304
34 37.6141304347826
35 37.5096467391304
36 37.5096467391304
37 37.5096467391304
38 37.5096467391304
39 37.5096467391304
40 37.5096467391304
41 34.0616847826087
42 34.0616847826087
43 34.0616847826087
44 34.0616847826087
45 34.0616847826087
46 34.0616847826087
47 34.0616847826087
48 33.9572010869565
49 34.0616847826087
50 33.9572010869565
51 34.0616847826087
52 34.0616847826087
53 33.9572010869565
54 34.0616847826087
55 34.0616847826087
56 33.9572010869565
57 34.0616847826087
58 34.0616847826087
59 34.0616847826087
60 34.0616847826087
61 0
62 0
63 0
64 0
65 0
66 0
67 0
68 0
69 0
70 0
71 0
72 0
73 0
74 0
75 0
76 0
77 0
78 0
79 0
80 0
81 0
82 0
83 0
84 0
85 0
86 0
87 0
88 0
89 0
90 0
91 0
92 0
93 0
94 0
95 0
96 0
97 0
98 0
99 0
100 0
};
\end{axis}

\end{tikzpicture}
        \vspace*{-0.2cm} \centering \small (b) \gls{hctns}: Video Bandwidth Evolution.
    \end{minipage}
    \vspace*{0.2cm}

    \begin{minipage}[t]{0.4\textwidth}
        \centering
\begin{tikzpicture}

\definecolor{crimson2302929}{RGB}{230,29,29}
\definecolor{darkgrey176}{RGB}{176,176,176}
\definecolor{firebrick1742121}{RGB}{174,21,21}
\definecolor{maroon971111}{RGB}{97,11,11}
\definecolor{red25233}{RGB}{252,3,3}

\begin{axis}[
height=0.6\textwidth,
tick align=outside,
tick pos=left,
width=\textwidth,
x grid style={darkgrey176},
xlabel={t (s)},
xmajorgrids,
xmin=0, xmax=100,
xtick style={color=black},
y grid style={darkgrey176},
ylabel={Packet Loss},
ymajorgrids,
ymin=-3686.95, ymax=77425.95,
ytick style={color=black}
]
\addplot [semithick, maroon971111, dash pattern=on 1pt off 3pt on 3pt off 3pt, mark=asterisk, mark size=1.5, mark options={solid}]
table {%
1 0
2 0
3 0
4 0
5 0
6 0
7 0
8 0
9 0
10 0
11 0
12 0
13 0
14 0
15 0
16 0
17 0
18 0
19 0
20 0
21 0
22 953
23 2845
24 4740
25 6633
26 8526
27 10419
28 12311
29 14203
30 16094
31 17985
32 19877
33 21770
34 23664
35 25557
36 27450
37 29340
38 31232
39 33127
40 35020
41 36910
42 38839
43 40771
44 42706
45 44637
46 46570
47 48504
48 50438
49 52373
50 54260
51 54536
52 56470
53 58403
54 60335
55 62269
56 64203
57 66135
58 68069
59 70003
60 71938
61 73739
62 73739
63 73739
64 73739
65 73739
66 73739
67 73739
68 73739
69 73739
70 73739
71 73739
72 73739
73 73739
74 73739
75 73739
76 73739
77 73739
78 73739
79 73739
80 73739
81 73739
82 73739
83 73739
84 73739
85 73739
86 73739
87 73739
88 73739
89 73739
90 73739
91 73739
92 73739
93 73739
94 73739
95 73739
96 73739
97 73739
98 73739
99 73739
100 73739
};
\addplot [semithick, firebrick1742121, dashed, mark=pentagon*, mark size=1.5, mark options={solid}]
table {%
1 0
2 0
3 0
4 0
5 0
6 0
7 0
8 0
9 0
10 0
11 0
12 0
13 0
14 0
15 0
16 0
17 0
18 0
19 0
20 0
21 0
22 0
23 0
24 0
25 0
26 0
27 0
28 0
29 0
30 0
31 0
32 0
33 0
34 0
35 0
36 0
37 0
38 0
39 0
40 0
41 0
42 0
43 822
44 1717
45 2610
46 3499
47 4387
48 5279
49 6172
50 7064
51 7957
52 8849
53 9740
54 10632
55 11523
56 12415
57 13309
58 14195
59 15086
60 15980
61 16829
62 16829
63 16829
64 16829
65 16829
66 16829
67 16829
68 16829
69 16829
70 16829
71 16829
72 16829
73 16829
74 16829
75 16829
76 16829
77 16829
78 16829
79 16829
80 16829
81 16829
82 16829
83 16829
84 16829
85 16829
86 16829
87 16829
88 16829
89 16829
90 16829
91 16829
92 16829
93 16829
94 16829
95 16829
96 16829
97 16829
98 16829
99 16829
100 16829
};
\addplot [semithick, crimson2302929, dashed, mark=triangle*, mark size=1.5, mark options={solid}]
table {%
1 0
2 0
3 0
4 0
5 0
6 0
7 0
8 0
9 0
10 0
11 0
12 0
13 0
14 0
15 0
16 0
17 0
18 0
19 0
20 0
21 0
22 0
23 0
24 0
25 0
26 0
27 0
28 0
29 0
30 0
31 0
32 0
33 0
34 0
35 0
36 0
37 0
38 0
39 0
40 0
41 0
42 0
43 0
44 0
45 0
46 0
47 0
48 0
49 0
50 0
51 0
52 0
53 0
54 0
55 0
56 0
57 0
58 0
59 0
60 0
61 0
62 0
63 0
64 0
65 0
66 0
67 0
68 0
69 0
70 0
71 0
72 0
73 0
74 0
75 0
76 0
77 0
78 0
79 0
80 0
81 0
82 0
83 0
84 0
85 0
86 0
87 0
88 0
89 0
90 0
91 0
92 0
93 0
94 0
95 0
96 0
97 0
98 0
99 0
100 0
};
\addplot [semithick, red25233, dash pattern=on 1pt off 3pt on 3pt off 3pt, mark=diamond*, mark size=1.5, mark options={solid}]
table {%
1 0
2 0
3 0
4 0
5 0
6 0
7 0
8 0
9 0
10 0
11 0
12 0
13 0
14 0
15 0
16 0
17 0
18 0
19 0
20 0
21 0
22 0
23 0
24 0
25 0
26 0
27 0
28 0
29 0
30 0
31 0
32 0
33 0
34 0
35 0
36 0
37 0
38 0
39 0
40 0
41 0
42 0
43 0
44 0
45 0
46 0
47 0
48 0
49 0
50 0
51 0
52 0
53 0
54 0
55 0
56 0
57 0
58 0
59 0
60 0
61 0
62 0
63 0
64 0
65 0
66 0
67 0
68 0
69 0
70 0
71 0
72 0
73 0
74 0
75 0
76 0
77 0
78 0
79 0
80 0
81 0
82 0
83 0
84 0
85 0
86 0
87 0
88 0
89 0
90 0
91 0
92 0
93 0
94 0
95 0
96 0
97 0
98 0
99 0
100 0
};
\end{axis}

\end{tikzpicture}
        \vspace*{-0.2cm} \centering \small (c) IETF: Video Packet Loss in output PE port.
    \end{minipage}
    \vspace*{0.2cm}
    \hspace{1.5cm}
    \begin{minipage}[t]{0.4\textwidth} 
        \centering
\begin{tikzpicture}

\definecolor{crimson2302929}{RGB}{230,29,29}
\definecolor{darkgrey176}{RGB}{176,176,176}
\definecolor{firebrick1742121}{RGB}{174,21,21}
\definecolor{maroon971111}{RGB}{97,11,11}
\definecolor{red25233}{RGB}{252,3,3}

\begin{axis}[
height=0.6\textwidth,
tick align=outside,
tick pos=left,
width=\textwidth,
x grid style={darkgrey176},
xlabel={t (s)},
xmajorgrids,
xmin=0, xmax=100,
xtick style={color=black},
y grid style={darkgrey176},
ylabel={Packet Loss},
ymajorgrids,
ymin=-0.1, ymax=1,
ytick style={color=black}
]
\addplot [semithick, maroon971111, dash pattern=on 1pt off 3pt on 3pt off 3pt, mark=asterisk, mark size=1.5, mark options={solid}]
table {%
1 0
2 0
3 0
4 0
5 0
6 0
7 0
8 0
9 0
10 0
11 0
12 0
13 0
14 0
15 0
16 0
17 0
18 0
19 0
20 0
21 0
22 0
23 0
24 0
25 0
26 0
27 0
28 0
29 0
30 0
31 0
32 0
33 0
34 0
35 0
36 0
37 0
38 0
39 0
40 0
41 0
42 0
43 0
44 0
45 0
46 0
47 0
48 0
49 0
50 0
51 0
52 0
53 0
54 0
55 0
56 0
57 0
58 0
59 0
60 0
61 0
62 0
63 0
64 0
65 0
66 0
67 0
68 0
69 0
70 0
71 0
72 0
73 0
74 0
75 0
76 0
77 0
78 0
79 0
80 0
81 0
82 0
83 0
84 0
85 0
86 0
87 0
88 0
89 0
90 0
91 0
92 0
93 0
94 0
95 0
96 0
97 0
98 0
99 0
100 0
};
\addplot [semithick, firebrick1742121, dashed, mark=pentagon*, mark size=1.5, mark options={solid}]
table {%
1 0
2 0
3 0
4 0
5 0
6 0
7 0
8 0
9 0
10 0
11 0
12 0
13 0
14 0
15 0
16 0
17 0
18 0
19 0
20 0
21 0
22 0
23 0
24 0
25 0
26 0
27 0
28 0
29 0
30 0
31 0
32 0
33 0
34 0
35 0
36 0
37 0
38 0
39 0
40 0
41 0
42 0
43 0
44 0
45 0
46 0
47 0
48 0
49 0
50 0
51 0
52 0
53 0
54 0
55 0
56 0
57 0
58 0
59 0
60 0
61 0
62 0
63 0
64 0
65 0
66 0
67 0
68 0
69 0
70 0
71 0
72 0
73 0
74 0
75 0
76 0
77 0
78 0
79 0
80 0
81 0
82 0
83 0
84 0
85 0
86 0
87 0
88 0
89 0
90 0
91 0
92 0
93 0
94 0
95 0
96 0
97 0
98 0
99 0
100 0
};
\addplot [semithick, crimson2302929, dashed, mark=triangle*, mark size=1.5, mark options={solid}]
table {%
1 0
2 0
3 0
4 0
5 0
6 0
7 0
8 0
9 0
10 0
11 0
12 0
13 0
14 0
15 0
16 0
17 0
18 0
19 0
20 0
21 0
22 0
23 0
24 0
25 0
26 0
27 0
28 0
29 0
30 0
31 0
32 0
33 0
34 0
35 0
36 0
37 0
38 0
39 0
40 0
41 0
42 0
43 0
44 0
45 0
46 0
47 0
48 0
49 0
50 0
51 0
52 0
53 0
54 0
55 0
56 0
57 0
58 0
59 0
60 0
61 0
62 0
63 0
64 0
65 0
66 0
67 0
68 0
69 0
70 0
71 0
72 0
73 0
74 0
75 0
76 0
77 0
78 0
79 0
80 0
81 0
82 0
83 0
84 0
85 0
86 0
87 0
88 0
89 0
90 0
91 0
92 0
93 0
94 0
95 0
96 0
97 0
98 0
99 0
100 0
};
\addplot [semithick, red25233, dash pattern=on 1pt off 3pt on 3pt off 3pt, mark=diamond*, mark size=1.5, mark options={solid}]
table {%
1 0
2 0
3 0
4 0
5 0
6 0
7 0
8 0
9 0
10 0
11 0
12 0
13 0
14 0
15 0
16 0
17 0
18 0
19 0
20 0
21 0
22 0
23 0
24 0
25 0
26 0
27 0
28 0
29 0
30 0
31 0
32 0
33 0
34 0
35 0
36 0
37 0
38 0
39 0
40 0
41 0
42 0
43 0
44 0
45 0
46 0
47 0
48 0
49 0
50 0
51 0
52 0
53 0
54 0
55 0
56 0
57 0
58 0
59 0
60 0
61 0
62 0
63 0
64 0
65 0
66 0
67 0
68 0
69 0
70 0
71 0
72 0
73 0
74 0
75 0
76 0
77 0
78 0
79 0
80 0
81 0
82 0
83 0
84 0
85 0
86 0
87 0
88 0
89 0
90 0
91 0
92 0
93 0
94 0
95 0
96 0
97 0
98 0
99 0
100 0
};
\end{axis}

\end{tikzpicture}
        \vspace*{-0.2cm} \centering \small (d) \gls{hctns}: Video Packet Loss in output PE port.
    \end{minipage}
    \vspace*{0.2cm}

    \hspace*{-0.8cm}\includegraphics[width=1\textwidth, trim=0 15 0 10, clip]{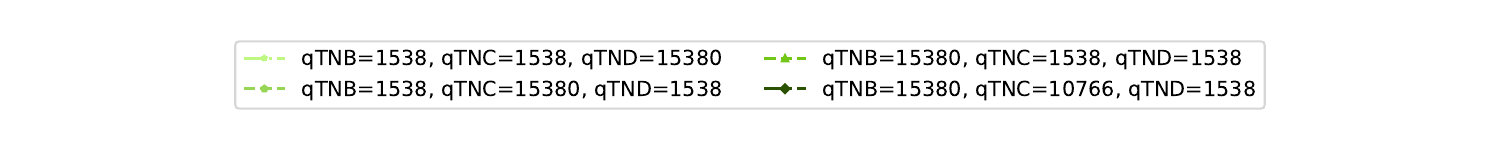}
    \begin{minipage}[t]{0.4\textwidth}
        \centering
\begin{tikzpicture}

\definecolor{darkgreen42820}{RGB}{42,82,0}
\definecolor{darkgrey176}{RGB}{176,176,176}
\definecolor{palegreen190247129}{RGB}{190,247,129}
\definecolor{yellowgreen11319822}{RGB}{113,198,22}
\definecolor{yellowgreen15121484}{RGB}{151,214,84}

\begin{axis}[
height=0.6\textwidth,
tick align=outside,
tick pos=left,
width=\textwidth,
x grid style={darkgrey176},
xlabel={t (s)},
xmajorgrids,
xmin=0, xmax=100,
xtick style={color=black},
y grid style={darkgrey176},
ylabel={Bandwidth (Mbps)},
ymajorgrids,
ymin=-0.382932744565217, ymax=8.04158763586956,
ytick style={color=black}
]
\addplot [semithick, palegreen190247129, dash pattern=on 1pt off 3pt on 3pt off 3pt, mark=asterisk, mark size=1.5, mark options={solid}]
table {%
1 0
2 0
3 0
4 0
5 0
6 0
7 0
8 0
9 0
10 0
11 0
12 0
13 0
14 0
15 0
16 0
17 0
18 0
19 0
20 0
21 4.00172554347826
22 4.00172554347826
23 3.98082880434783
24 3.99127717391304
25 4.00172554347826
26 3.99127717391304
27 3.99127717391304
28 3.99127717391304
29 3.99127717391304
30 3.99127717391304
31 3.99127717391304
32 3.99127717391304
33 3.99127717391304
34 4.00172554347826
35 3.99127717391304
36 3.99127717391304
37 3.99127717391304
38 3.99127717391304
39 3.99127717391304
40 3.99127717391304
41 0.589288043478261
42 0.491073369565217
43 0.392858695652174
44 0.307182065217391
45 0.29464402173913
46 0.232998641304348
47 0.332258152173913
48 0.343751358695652
49 0.29464402173913
50 0.774224184782609
51 0.577794836956522
52 0.319720108695652
53 0.332258152173913
54 0.405396739130435
55 0.28210597826087
56 0.29464402173913
57 0.29464402173913
58 0.319720108695652
59 0.356289402173913
60 0.319720108695652
61 0.319720108695652
62 0.381365489130435
63 0.405396739130435
64 0.454504076086956
65 0.405396739130435
66 0.405396739130435
67 0.221505434782609
68 0.356289402173913
69 0.27061277173913
70 0.307182065217391
71 0.417934782608696
72 4.40921195652174
73 7.65865489130435
74 3.99127717391304
75 4.00172554347826
76 3.99127717391304
77 3.99127717391304
78 3.99127717391304
79 3.99127717391304
80 3.99127717391304
81 0
82 0
83 0
84 0
85 0
86 0
87 0
88 0
89 0
90 0
91 0
92 0
93 0
94 0
95 0
96 0
97 0
98 0
99 0
100 0
};
\addplot [semithick, yellowgreen15121484, dashed, mark=pentagon*, mark size=1.5, mark options={solid}]
table {%
1 0
2 0
3 0
4 0
5 0
6 0
7 0
8 0
9 0
10 0
11 0
12 0
13 0
14 0
15 0
16 0
17 0
18 0
19 0
20 0
21 4.00172554347826
22 4.00172554347826
23 3.99127717391304
24 3.99127717391304
25 3.99127717391304
26 3.99127717391304
27 3.99127717391304
28 3.99127717391304
29 3.99127717391304
30 3.99127717391304
31 3.99127717391304
32 4.00172554347826
33 3.99127717391304
34 3.99127717391304
35 3.99127717391304
36 3.99127717391304
37 3.99127717391304
38 3.99127717391304
39 3.99127717391304
40 3.99127717391304
41 3.99127717391304
42 3.99127717391304
43 4.00172554347826
44 3.99127717391304
45 3.99127717391304
46 3.99127717391304
47 3.99127717391304
48 3.99127717391304
49 3.99127717391304
50 3.99127717391304
51 3.99127717391304
52 3.99127717391304
53 4.00172554347826
54 3.99127717391304
55 3.99127717391304
56 3.99127717391304
57 3.99127717391304
58 3.99127717391304
59 3.99127717391304
60 3.99127717391304
61 3.99127717391304
62 3.99127717391304
63 4.00172554347826
64 3.99127717391304
65 3.99127717391304
66 3.99127717391304
67 3.99127717391304
68 3.99127717391304
69 3.81365489130435
70 3.70917119565217
71 3.99127717391304
72 4.00172554347826
73 3.99127717391304
74 3.99127717391304
75 3.99127717391304
76 3.99127717391304
77 3.99127717391304
78 3.99127717391304
79 3.99127717391304
80 3.99127717391304
81 0
82 0
83 0
84 0
85 0
86 0
87 0
88 0
89 0
90 0
91 0
92 0
93 0
94 0
95 0
96 0
97 0
98 0
99 0
100 0
};
\addplot [semithick, yellowgreen11319822, dashed, mark=triangle*, mark size=1.5, mark options={solid}]
table {%
1 0
2 0
3 0
4 0
5 0
6 0
7 0
8 0
9 0
10 0
11 0
12 0
13 0
14 0
15 0
16 0
17 0
18 0
19 0
20 0
21 4.0226222826087
22 3.99127717391304
23 3.99127717391304
24 3.99127717391304
25 3.99127717391304
26 3.99127717391304
27 3.99127717391304
28 3.99127717391304
29 3.99127717391304
30 3.99127717391304
31 4.00172554347826
32 3.99127717391304
33 3.99127717391304
34 3.99127717391304
35 3.99127717391304
36 3.99127717391304
37 3.99127717391304
38 3.99127717391304
39 3.99127717391304
40 3.99127717391304
41 2.28819293478261
42 1.19111413043478
43 1.23290760869565
44 1.32694293478261
45 1.42097826086957
46 1.10752717391304
47 0.909008152173913
48 0.934084239130435
49 1.18066576086956
50 1.07618206521739
51 1.17021739130435
52 1.12842391304348
53 1.06573369565217
54 1.27470108695652
55 1.12842391304348
56 1.88070652173913
57 2.21505434782609
58 1.55680706521739
59 1.7448777173913
60 1.7448777173913
61 1.78667119565217
62 2.00608695652174
63 2.21505434782609
64 2.35088315217391
65 2.30908967391304
66 1.89115489130435
67 1.79711956521739
68 1.89115489130435
69 1.81801630434783
70 1.79711956521739
71 4.28383152173913
72 3.99127717391304
73 3.99127717391304
74 4.00172554347826
75 3.99127717391304
76 3.99127717391304
77 3.99127717391304
78 3.99127717391304
79 3.99127717391304
80 3.99127717391304
81 0
82 0
83 0
84 0
85 0
86 0
87 0
88 0
89 0
90 0
91 0
92 0
93 0
94 0
95 0
96 0
97 0
98 0
99 0
100 0
};
\addplot [semithick, darkgreen42820, dash pattern=on 1pt off 3pt on 3pt off 3pt, mark=diamond*, mark size=1.5, mark options={solid}]
table {%
1 0
2 0
3 0
4 0
5 0
6 0
7 0
8 0
9 0
10 0
11 0
12 0
13 0
14 0
15 0
16 0
17 0
18 0
19 0
20 0
21 4.00172554347826
22 4.00172554347826
23 3.99127717391304
24 3.99127717391304
25 3.99127717391304
26 3.99127717391304
27 3.53154891304348
28 3.99127717391304
29 3.99127717391304
30 3.99127717391304
31 3.99127717391304
32 4.00172554347826
33 3.99127717391304
34 3.99127717391304
35 3.99127717391304
36 3.99127717391304
37 3.99127717391304
38 3.99127717391304
39 3.99127717391304
40 3.99127717391304
41 3.99127717391304
42 3.99127717391304
43 3.99127717391304
44 4.00172554347826
45 3.99127717391304
46 3.99127717391304
47 3.99127717391304
48 3.99127717391304
49 3.99127717391304
50 3.99127717391304
51 3.99127717391304
52 3.99127717391304
53 4.00172554347826
54 3.99127717391304
55 3.99127717391304
56 3.51065217391304
57 3.99127717391304
58 3.99127717391304
59 3.99127717391304
60 3.99127717391304
61 3.99127717391304
62 3.99127717391304
63 4.00172554347826
64 3.99127717391304
65 3.99127717391304
66 3.99127717391304
67 3.99127717391304
68 3.99127717391304
69 3.99127717391304
70 3.99127717391304
71 3.99127717391304
72 4.00172554347826
73 3.99127717391304
74 3.99127717391304
75 3.99127717391304
76 3.99127717391304
77 3.99127717391304
78 3.99127717391304
79 3.99127717391304
80 3.99127717391304
81 0
82 0
83 0
84 0
85 0
86 0
87 0
88 0
89 0
90 0
91 0
92 0
93 0
94 0
95 0
96 0
97 0
98 0
99 0
100 0
};
\end{axis}

\end{tikzpicture}
        \centering \small (e) IETF: Telemetry Bandwidth Evolution.
    \end{minipage}
    \vspace*{0.2cm}
    \hspace{1.5cm}
    \begin{minipage}[t]{0.4\textwidth} 
        \centering
\begin{tikzpicture}

\definecolor{darkgreen42820}{RGB}{42,82,0}
\definecolor{darkgrey176}{RGB}{176,176,176}
\definecolor{palegreen190247129}{RGB}{190,247,129}
\definecolor{yellowgreen11319822}{RGB}{113,198,22}
\definecolor{yellowgreen15121484}{RGB}{151,214,84}

\begin{axis}[
height=0.6\textwidth,
tick align=outside,
tick pos=left,
width=\textwidth,
x grid style={darkgrey176},
xlabel={t (s)},
xmajorgrids,
xmin=0, xmax=100,
xtick style={color=black},
y grid style={darkgrey176},
ylabel={Bandwidth (Mbps)},
ymajorgrids,
ymin=-1.80756793478261, ymax=37.9589266304348,
ytick style={color=black}
]
\addplot [semithick, palegreen190247129, dash pattern=on 1pt off 3pt on 3pt off 3pt, mark=asterisk, mark size=1.5, mark options={solid}]
table {%
1 0
2 0
3 0
4 0
5 0
6 0
7 0
8 0
9 0
10 0
11 0
12 0
13 0
14 0
15 0
16 0
17 0
18 0
19 0
20 0
21 9.58115489130435
22 9.58115489130435
23 9.59160326086956
24 9.58115489130435
25 9.58115489130435
26 9.58115489130435
27 9.58115489130435
28 9.58115489130435
29 9.51846467391304
30 9.58115489130435
31 9.58115489130435
32 9.59160326086956
33 9.58115489130435
34 9.59160326086956
35 9.58115489130435
36 9.5498097826087
37 9.56025815217391
38 9.49756793478261
39 9.50801630434782
40 9.50801630434782
41 6.06005434782609
42 6.09139945652174
43 6.04960597826087
44 6.06005434782609
45 6.06005434782609
46 6.0705027173913
47 6.06005434782609
48 6.0705027173913
49 6.09139945652174
50 6.0705027173913
51 6.08095108695652
52 6.0705027173913
53 6.0705027173913
54 6.0705027173913
55 6.06005434782609
56 6.0705027173913
57 6.12274456521739
58 6.04960597826087
59 6.06005434782609
60 6.10184782608696
61 35.8379076086956
62 35.9423913043478
63 35.7334239130435
64 35.9423913043478
65 35.8379076086956
66 35.8379076086956
67 35.9423913043478
68 35.9423913043478
69 35.8379076086956
70 35.8379076086956
71 35.9423913043478
72 35.9423913043478
73 35.8379076086956
74 35.9423913043478
75 35.9423913043478
76 35.9423913043478
77 35.9423913043478
78 35.9423913043478
79 35.9423913043478
80 36.1513586956522
81 0
82 0
83 0
84 0
85 0
86 0
87 0
88 0
89 0
90 0
91 0
92 0
93 0
94 0
95 0
96 0
97 0
98 0
99 0
100 0
};
\addplot [semithick, yellowgreen15121484, dashed, mark=pentagon*, mark size=1.5, mark options={solid}]
table {%
1 0
2 0
3 0
4 0
5 0
6 0
7 0
8 0
9 0
10 0
11 0
12 0
13 0
14 0
15 0
16 0
17 0
18 0
19 0
20 0
21 9.58115489130435
22 9.58115489130435
23 9.58115489130435
24 9.59160326086956
25 9.58115489130435
26 9.58115489130435
27 9.58115489130435
28 9.58115489130435
29 9.59160326086956
30 9.58115489130435
31 9.59160326086956
32 9.58115489130435
33 9.57070652173913
34 9.59160326086956
35 9.58115489130435
36 9.58115489130435
37 9.58115489130435
38 9.59160326086956
39 9.5498097826087
40 9.53936141304348
41 6.0705027173913
42 6.08095108695652
43 6.0705027173913
44 6.0705027173913
45 6.08095108695652
46 6.06005434782609
47 6.04960597826087
48 6.04960597826087
49 6.08095108695652
50 6.13319293478261
51 6.08095108695652
52 6.10184782608696
53 6.06005434782609
54 6.06005434782609
55 6.09139945652174
56 6.06005434782609
57 6.02870923913043
58 6.08095108695652
59 6.06005434782609
60 6.04960597826087
61 35.9423913043478
62 35.9423913043478
63 35.9423913043478
64 35.9423913043478
65 35.9423913043478
66 35.9423913043478
67 35.9423913043478
68 35.9423913043478
69 35.9423913043478
70 35.9423913043478
71 35.9423913043478
72 35.9423913043478
73 35.9423913043478
74 35.9423913043478
75 35.9423913043478
76 35.9423913043478
77 35.9423913043478
78 35.9423913043478
79 35.9423913043478
80 36.046875
81 0
82 0
83 0
84 0
85 0
86 0
87 0
88 0
89 0
90 0
91 0
92 0
93 0
94 0
95 0
96 0
97 0
98 0
99 0
100 0
};
\addplot [semithick, yellowgreen11319822, dashed, mark=triangle*, mark size=1.5, mark options={solid}]
table {%
1 0
2 0
3 0
4 0
5 0
6 0
7 0
8 0
9 0
10 0
11 0
12 0
13 0
14 0
15 0
16 0
17 0
18 0
19 0
20 0
21 9.58115489130435
22 9.58115489130435
23 9.59160326086956
24 9.58115489130435
25 9.58115489130435
26 9.59160326086956
27 9.57070652173913
28 9.58115489130435
29 9.58115489130435
30 9.51846467391304
31 9.58115489130435
32 9.5498097826087
33 9.58115489130435
34 9.57070652173913
35 9.44532608695652
36 9.58115489130435
37 9.5498097826087
38 9.58115489130435
39 9.59160326086956
40 9.58115489130435
41 6.04960597826087
42 6.04960597826087
43 6.04960597826087
44 6.02870923913043
45 6.02870923913043
46 6.02870923913043
47 6.02870923913043
48 6.04960597826087
49 6.06005434782609
50 6.06005434782609
51 6.04960597826087
52 6.0705027173913
53 6.06005434782609
54 6.01826086956522
55 6.04960597826087
56 6.06005434782609
57 6.02870923913043
58 6.04960597826087
59 6.04960597826087
60 6.02870923913043
61 35.9423913043478
62 35.9423913043478
63 35.9423913043478
64 35.9423913043478
65 35.9423913043478
66 35.9423913043478
67 35.9423913043478
68 35.9423913043478
69 35.9423913043478
70 35.9423913043478
71 35.9423913043478
72 35.9423913043478
73 35.9423913043478
74 35.9423913043478
75 35.9423913043478
76 35.9423913043478
77 35.9423913043478
78 35.9423913043478
79 35.9423913043478
80 35.9423913043478
81 0
82 0
83 0
84 0
85 0
86 0
87 0
88 0
89 0
90 0
91 0
92 0
93 0
94 0
95 0
96 0
97 0
98 0
99 0
100 0
};
\addplot [semithick, darkgreen42820, dash pattern=on 1pt off 3pt on 3pt off 3pt, mark=diamond*, mark size=1.5, mark options={solid}]
table {%
1 0
2 0
3 0
4 0
5 0
6 0
7 0
8 0
9 0
10 0
11 0
12 0
13 0
14 0
15 0
16 0
17 0
18 0
19 0
20 0
21 9.58115489130435
22 9.58115489130435
23 9.58115489130435
24 9.59160326086956
25 9.58115489130435
26 9.58115489130435
27 9.58115489130435
28 9.58115489130435
29 9.58115489130435
30 9.59160326086956
31 9.58115489130435
32 9.58115489130435
33 9.57070652173913
34 9.58115489130435
35 9.57070652173913
36 9.58115489130435
37 9.59160326086956
38 9.58115489130435
39 9.58115489130435
40 9.59160326086956
41 6.08095108695652
42 6.06005434782609
43 6.0705027173913
44 6.0705027173913
45 6.06005434782609
46 6.0705027173913
47 6.06005434782609
48 6.02870923913043
49 6.04960597826087
50 6.02870923913043
51 6.08095108695652
52 6.0705027173913
53 6.06005434782609
54 6.04960597826087
55 6.06005434782609
56 6.01826086956522
57 6.04960597826087
58 6.08095108695652
59 6.06005434782609
60 6.08095108695652
61 35.9423913043478
62 35.9423913043478
63 35.9423913043478
64 35.9423913043478
65 35.9423913043478
66 35.9423913043478
67 35.9423913043478
68 35.9423913043478
69 35.9423913043478
70 35.9423913043478
71 35.9423913043478
72 35.9423913043478
73 35.9423913043478
74 35.9423913043478
75 35.9423913043478
76 35.9423913043478
77 35.9423913043478
78 35.9423913043478
79 35.9423913043478
80 35.9423913043478
81 0
82 0
83 0
84 0
85 0
86 0
87 0
88 0
89 0
90 0
91 0
92 0
93 0
94 0
95 0
96 0
97 0
98 0
99 0
100 0
};
\end{axis}

\end{tikzpicture}
        \centering \small (f) \gls{hctns}: Telemetry Bandwidth Evolution.
    \end{minipage}
    \vspace*{0.2cm}
    \begin{minipage}[t]{0.4\textwidth}
        \centering
\begin{tikzpicture}

\definecolor{darkgreen42820}{RGB}{42,82,0}
\definecolor{darkgrey176}{RGB}{176,176,176}
\definecolor{palegreen190247129}{RGB}{190,247,129}
\definecolor{yellowgreen11319822}{RGB}{113,198,22}
\definecolor{yellowgreen15121484}{RGB}{151,214,84}

\begin{axis}[
height=0.6\textwidth,
tick align=outside,
tick pos=left,
width=\textwidth,
x grid style={darkgrey176},
xlabel={t (s)},
xmajorgrids,
xmin=0, xmax=100,
xtick style={color=black},
y grid style={darkgrey176},
ylabel={Packet Loss},
ymajorgrids,
ymin=-5716.25, ymax=120041.25,
ytick style={color=black}
]
\addplot [semithick, palegreen190247129, dash pattern=on 1pt off 3pt on 3pt off 3pt, mark=asterisk, mark size=1.5, mark options={solid}]
table {%
1 0
2 0
3 0
4 0
5 0
6 0
7 0
8 0
9 0
10 0
11 0
12 0
13 0
14 0
15 0
16 0
17 0
18 0
19 0
20 0
21 0
22 0
23 0
24 0
25 0
26 0
27 0
28 0
29 0
30 0
31 0
32 0
33 0
34 0
35 0
36 0
37 0
38 0
39 0
40 0
41 0
42 3130
43 7077
44 11032
45 14977
46 18929
47 22880
48 26832
49 30786
50 34641
51 36244
52 40198
53 44147
54 48095
55 52047
56 56001
57 59948
58 63900
59 67853
60 71807
61 75753
62 79704
63 83631
64 87512
65 91391
66 95215
67 99098
68 102987
69 106870
70 110724
71 114325
72 114325
73 114325
74 114325
75 114325
76 114325
77 114325
78 114325
79 114325
80 114325
81 114325
82 114325
83 114325
84 114325
85 114325
86 114325
87 114325
88 114325
89 114325
90 114325
91 114325
92 114325
93 114325
94 114325
95 114325
96 114325
97 114325
98 114325
99 114325
100 114325
};
\addplot [semithick, yellowgreen15121484, dashed, mark=pentagon*, mark size=1.5, mark options={solid}]
table {%
1 0
2 0
3 0
4 0
5 0
6 0
7 0
8 0
9 0
10 0
11 0
12 0
13 0
14 0
15 0
16 0
17 0
18 0
19 0
20 0
21 0
22 0
23 0
24 0
25 0
26 0
27 0
28 0
29 0
30 0
31 0
32 0
33 0
34 0
35 0
36 0
37 0
38 0
39 0
40 0
41 0
42 0
43 0
44 0
45 0
46 0
47 0
48 0
49 0
50 0
51 0
52 0
53 0
54 0
55 0
56 0
57 0
58 0
59 0
60 0
61 0
62 0
63 0
64 0
65 0
66 0
67 0
68 0
69 0
70 0
71 0
72 0
73 0
74 0
75 0
76 0
77 0
78 0
79 0
80 0
81 0
82 0
83 0
84 0
85 0
86 0
87 0
88 0
89 0
90 0
91 0
92 0
93 0
94 0
95 0
96 0
97 0
98 0
99 0
100 0
};
\addplot [semithick, yellowgreen11319822, dashed, mark=triangle*, mark size=1.5, mark options={solid}]
table {%
1 0
2 0
3 0
4 0
5 0
6 0
7 0
8 0
9 0
10 0
11 0
12 0
13 0
14 0
15 0
16 0
17 0
18 0
19 0
20 0
21 0
22 0
23 0
24 0
25 0
26 0
27 0
28 0
29 0
30 0
31 0
32 0
33 0
34 0
35 0
36 0
37 0
38 0
39 0
40 0
41 0
42 972
43 2869
44 4770
45 6671
46 8571
47 10472
48 12375
49 14277
50 16178
51 18079
52 19981
53 21881
54 23786
55 25684
56 27582
57 28574
58 29942
59 31834
60 33728
61 35565
62 36166
63 36763
64 37363
65 37964
66 38563
67 39163
68 39765
69 40365
70 40961
71 41527
72 41527
73 41527
74 41527
75 41527
76 41527
77 41527
78 41527
79 41527
80 41527
81 41527
82 41527
83 41527
84 41527
85 41527
86 41527
87 41527
88 41527
89 41527
90 41527
91 41527
92 41527
93 41527
94 41527
95 41527
96 41527
97 41527
98 41527
99 41527
100 41527
};
\addplot [semithick, darkgreen42820, dash pattern=on 1pt off 3pt on 3pt off 3pt, mark=diamond*, mark size=1.5, mark options={solid}]
table {%
1 0
2 0
3 0
4 0
5 0
6 0
7 0
8 0
9 0
10 0
11 0
12 0
13 0
14 0
15 0
16 0
17 0
18 0
19 0
20 0
21 0
22 0
23 0
24 0
25 0
26 0
27 0
28 0
29 0
30 0
31 0
32 0
33 0
34 0
35 0
36 0
37 0
38 0
39 0
40 0
41 0
42 0
43 0
44 0
45 0
46 0
47 0
48 0
49 0
50 0
51 0
52 0
53 0
54 0
55 0
56 0
57 0
58 0
59 0
60 0
61 0
62 0
63 0
64 0
65 0
66 0
67 0
68 0
69 0
70 0
71 0
72 0
73 0
74 0
75 0
76 0
77 0
78 0
79 0
80 0
81 0
82 0
83 0
84 0
85 0
86 0
87 0
88 0
89 0
90 0
91 0
92 0
93 0
94 0
95 0
96 0
97 0
98 0
99 0
100 0
};
\end{axis}

\end{tikzpicture}
        \centering \small (g) IETF: Telemetry Packets Queued in output PE port.
    \end{minipage}
    \hspace{1.5cm}
    \begin{minipage}[t]{0.4\textwidth} 
        \centering
\begin{tikzpicture}

\definecolor{darkgreen42820}{RGB}{42,82,0}
\definecolor{darkgrey176}{RGB}{176,176,176}
\definecolor{palegreen190247129}{RGB}{190,247,129}
\definecolor{yellowgreen11319822}{RGB}{113,198,22}
\definecolor{yellowgreen15121484}{RGB}{151,214,84}

\begin{axis}[
height=0.6\textwidth,
tick align=outside,
tick pos=left,
width=\textwidth,
x grid style={darkgrey176},
xlabel={t (s)},
xmajorgrids,
xmin=0, xmax=100,
xtick style={color=black},
y grid style={darkgrey176},
ylabel={Packet Loss},
ymajorgrids,
ymin=-0.1, ymax=1,
ytick style={color=black}
]
\addplot [semithick, palegreen190247129, dash pattern=on 1pt off 3pt on 3pt off 3pt, mark=asterisk, mark size=1.5, mark options={solid}]
table {%
1 0
2 0
3 0
4 0
5 0
6 0
7 0
8 0
9 0
10 0
11 0
12 0
13 0
14 0
15 0
16 0
17 0
18 0
19 0
20 0
21 0
22 0
23 0
24 0
25 0
26 0
27 0
28 0
29 0
30 0
31 0
32 0
33 0
34 0
35 0
36 0
37 0
38 0
39 0
40 0
41 0
42 0
43 0
44 0
45 0
46 0
47 0
48 0
49 0
50 0
51 0
52 0
53 0
54 0
55 0
56 0
57 0
58 0
59 0
60 0
61 0
62 0
63 0
64 0
65 0
66 0
67 0
68 0
69 0
70 0
71 0
72 0
73 0
74 0
75 0
76 0
77 0
78 0
79 0
80 0
81 0
82 0
83 0
84 0
85 0
86 0
87 0
88 0
89 0
90 0
91 0
92 0
93 0
94 0
95 0
96 0
97 0
98 0
99 0
100 0
};
\addplot [semithick, yellowgreen15121484, dashed, mark=pentagon*, mark size=1.5, mark options={solid}]
table {%
1 0
2 0
3 0
4 0
5 0
6 0
7 0
8 0
9 0
10 0
11 0
12 0
13 0
14 0
15 0
16 0
17 0
18 0
19 0
20 0
21 0
22 0
23 0
24 0
25 0
26 0
27 0
28 0
29 0
30 0
31 0
32 0
33 0
34 0
35 0
36 0
37 0
38 0
39 0
40 0
41 0
42 0
43 0
44 0
45 0
46 0
47 0
48 0
49 0
50 0
51 0
52 0
53 0
54 0
55 0
56 0
57 0
58 0
59 0
60 0
61 0
62 0
63 0
64 0
65 0
66 0
67 0
68 0
69 0
70 0
71 0
72 0
73 0
74 0
75 0
76 0
77 0
78 0
79 0
80 0
81 0
82 0
83 0
84 0
85 0
86 0
87 0
88 0
89 0
90 0
91 0
92 0
93 0
94 0
95 0
96 0
97 0
98 0
99 0
100 0
};
\addplot [semithick, yellowgreen11319822, dashed, mark=triangle*, mark size=1.5, mark options={solid}]
table {%
1 0
2 0
3 0
4 0
5 0
6 0
7 0
8 0
9 0
10 0
11 0
12 0
13 0
14 0
15 0
16 0
17 0
18 0
19 0
20 0
21 0
22 0
23 0
24 0
25 0
26 0
27 0
28 0
29 0
30 0
31 0
32 0
33 0
34 0
35 0
36 0
37 0
38 0
39 0
40 0
41 0
42 0
43 0
44 0
45 0
46 0
47 0
48 0
49 0
50 0
51 0
52 0
53 0
54 0
55 0
56 0
57 0
58 0
59 0
60 0
61 0
62 0
63 0
64 0
65 0
66 0
67 0
68 0
69 0
70 0
71 0
72 0
73 0
74 0
75 0
76 0
77 0
78 0
79 0
80 0
81 0
82 0
83 0
84 0
85 0
86 0
87 0
88 0
89 0
90 0
91 0
92 0
93 0
94 0
95 0
96 0
97 0
98 0
99 0
100 0
};
\addplot [semithick, darkgreen42820, dash pattern=on 1pt off 3pt on 3pt off 3pt, mark=diamond*, mark size=1.5, mark options={solid}]
table {%
1 0
2 0
3 0
4 0
5 0
6 0
7 0
8 0
9 0
10 0
11 0
12 0
13 0
14 0
15 0
16 0
17 0
18 0
19 0
20 0
21 0
22 0
23 0
24 0
25 0
26 0
27 0
28 0
29 0
30 0
31 0
32 0
33 0
34 0
35 0
36 0
37 0
38 0
39 0
40 0
41 0
42 0
43 0
44 0
45 0
46 0
47 0
48 0
49 0
50 0
51 0
52 0
53 0
54 0
55 0
56 0
57 0
58 0
59 0
60 0
61 0
62 0
63 0
64 0
65 0
66 0
67 0
68 0
69 0
70 0
71 0
72 0
73 0
74 0
75 0
76 0
77 0
78 0
79 0
80 0
81 0
82 0
83 0
84 0
85 0
86 0
87 0
88 0
89 0
90 0
91 0
92 0
93 0
94 0
95 0
96 0
97 0
98 0
99 0
100 0
};
\end{axis}

\end{tikzpicture}
        \centering \small (h) \gls{hctns}: Telemetry Packets Queued in output PE port.
    \end{minipage}
    \vspace*{0.1cm}
    \caption{Experiment B: \gls{ietf} versus our proposed network slicing model.}
    \label{expb-ietf-proposal}
\end{figure*}

This experiment also spans 100\,s and it is divided into the same six time intervals defined for Experiment A. First, we analyze the performance of the \gls{ietf} model. As illustrated in Fig.~\ref{expb-ietf-proposal}a, with the third and fourth coarse-grained configurations, the \gls{tod} video traffic flow achieves its guaranteed rate throughout the entire experiment. In contrast, with the first two configurations, this guarantee is not met due to packet losses occurring at the output port of \gls{pe}1, as shown in Fig.~\ref{expb-ietf-proposal}c. This is caused by the low quantum value assigned to the queue of the \gls{tn} \gls{qos} class B (i.e., the \gls{tn} \gls{qos} class assigned to video) in the first two configurations. This quantum is significantly lower than the values assigned to the other \gls{drr} queues, resulting in a reduced transmission rate at the output port for the \gls{tn} \gls{qos} class B queue. Consequently, when there is traffic in the other \gls{drr} queues, more packets arrive in the \gls{drr} queue associated with \gls{tn} \gls{qos} class B than can be transmitted, leading to an accumulation of packets, overflow of the queue, and packet losses.

For telemetry traffic, with the \gls{ietf} model, as shown in Fig.~\ref{expb-ietf-proposal}e, with the second and fourth coarse-grained configurations, the telemetry traffic achieves its guaranteed rate throughout the entire experiment. However, with the first and third configurations, this guarantee is not met due to packet losses occurring at the output port of \gls{pe}1, as shown in Fig.~\ref{expb-ietf-proposal}g. The same reasons given in the previous point apply to this case. The quantum values set in the second and fourth configurations favor the transmission of the \gls{tn} \gls{qos} class C (i.e., the \gls{tn} \gls{qos} class associated to telemetry). 

These results highlight the importance of properly configuring the coarse-grained resource control parameters in transport network routers when employing the \gls{ietf} network slicing model. In contrast, the slice model proposed in this paper guarantees bandwidth levels for both video and telemetry traffic (Figs.~\ref{expb-ietf-proposal}b and~\ref{expb-ietf-proposal}f), while preventing packet losses (Figs.~\ref{expb-ietf-proposal}d and~\ref{expb-ietf-proposal}h). This is particularly crucial for a service like \gls{tod}, which demands high-reliable communications with guaranteed bandwidth to ensure road safety. \gls{hctns} demonstrates robustness and independence from the coarse-grained resource control configuration under normal traffic conditions (without traffic bursts), ensuring that traffic classes receive the \gls{qos} level they require. Experiment C illustrates how the proposed traffic burst control mechanism performs under different network scenarios.

\subsubsection{Experiment C: Traffic Burst Control}

This experiment analyzes the impact of traffic bursts on the latency experienced by packets from different traffic classes and slices. The \gls{ietf} model does not incorporate this function, so results only show the performance of \rev{the \cite{Lin2021} model and} our proposed traffic control burst mechanism for \gls{hctns}. For this purpose, the \gls{pe}1 router of our network scenario has been configured using the parameters and values listed in Table~\ref{tab:expc-params}. Unlike Experiment A, the fine-grained resource control mechanism in this experiment incorporates traffic burst control, with the \gls{cbs} and the \gls{pbs} parameters defined for both models. \rev{Although the work in~\cite{Lin2021} does not address traffic burst control, it mentions that the \gls{trtcm} mechanism they use has two parameters, \gls{cbs} and \gls{pbs}, that can be used to control traffic bursts. For Experiment C, we have configured these parameters in our implementation of the~\cite{Lin2021} model.  Table~\ref{tab:expc-params} shows some differences between the \gls{cir} and \gls{pir} parameters configured for \gls{hctns} and the~\cite{Lin2021} model. To make the closest comparison between the two models, since \gls{hctns} treats all the traffic accepted with the same \gls{qos}, we have configured the burst in \cite{Lin2021} using only the \gls{cbs} parameter, in order to treat all burst traffic as ``green". The \gls{pbs} values remain at their default setting, i.e, the maximum packet size. As   Table~\ref{tab:expc-params} shows, the \gls{cbs} and \gls{pbs} parameters have not been specified for BE traffic, which means that the bucket sizes can only store enough tokens to allow the transmission of a single packet. This is because BE traffic does not have a guaranteed rate.} Additionally, Table~\ref{tab:expc-params} indicates that two coarse-grained resource control configurations have been applied in our model to analyze how traffic burst control behaves under these two configurations. \rev{In \cite{Lin2021}, only two priority queues are used, so no other parameter configurations are possible.}

\begin{table*}
\centering
\resizebox{\textwidth}{!}{%
\begin{tabular}{ccccccccccccccccc}
\cline{2-17}
\multicolumn{1}{c|}{}                                                                  & \multicolumn{13}{c|}{\cellcolor[HTML]{F5F5F5}\textbf{Fine-grained Ingress Policer Resource Control}}                                                                                                                                                                                                                                                                                                                                                                                                                                                                                                                                                                                                                                                                                                                                                                                  & \multicolumn{3}{c|}{\cellcolor[HTML]{DAE8FC}\textbf{\begin{tabular}[c]{@{}c@{}}Coarse-grained \\ Resource Control\end{tabular}}}                                                                                                                                                                                                   \\ \cline{2-17} 
\multicolumn{1}{c|}{}                                                                  & \multicolumn{9}{c|}{\cellcolor[HTML]{FFFFFF}\textbf{HCTNS}}                                                                                                                                                                                                                                                                                                                                                                                                                                                                                                                                                                           & \multicolumn{4}{c|}{\cellcolor[HTML]{FFFFFF}}                                                                                                                                                                                                 & \multicolumn{3}{c|}{\cellcolor[HTML]{FFFFFF}\textbf{\begin{tabular}[c]{@{}c@{}}TN QoS Class\\ Parameters\end{tabular}}}                                                                                                                                                                \\ \cline{1-10} \cline{15-17} 
\rowcolor[HTML]{FFFFFF} 
\multicolumn{1}{|c|}{\cellcolor[HTML]{FFFFFF}}                                         & \multicolumn{4}{c|}{\cellcolor[HTML]{FFFFFF}\textbf{Class Policer}}                                                                                                                                                                           & \multicolumn{4}{c|}{\cellcolor[HTML]{FFFFFF}\textbf{Slice Policer}}                                                                                                                                                                                                                          & \multicolumn{1}{c|}{\cellcolor[HTML]{FFFFFF}}                                          & \multicolumn{4}{c|}{\multirow{-2}{*}{\cellcolor[HTML]{FFFFFF}\textbf{trTCM} \cite{Lin2021}}}                                                                                                                                 & \multicolumn{2}{c|}{\cellcolor[HTML]{FFFFFF}\textbf{HCTNS}}                                                                                  & \multicolumn{1}{c|}{\cellcolor[HTML]{FFFFFF} \cite{Lin2021}}                                                            \\ \cline{2-9} \cline{11-17} 
\rowcolor[HTML]{FFFFFF} 
\multicolumn{1}{|c|}{\multirow{-2}{*}{\cellcolor[HTML]{FFFFFF}\textbf{\begin{tabular}[c]{@{}c@{}}Traffic\\ Class\end{tabular}}}} & \multicolumn{1}{c|}{\cellcolor[HTML]{FFFFFF}\textbf{CIR}} & \multicolumn{1}{c|}{\cellcolor[HTML]{FFFFFF}\textbf{PIR}} & \multicolumn{1}{c|}{\cellcolor[HTML]{FFFFFF}\textbf{CBS}} & \multicolumn{1}{c|}{\cellcolor[HTML]{FFFFFF}\textbf{PBS}} & \multicolumn{1}{c|}{\cellcolor[HTML]{FFFFFF}\textbf{CIR}}                & \multicolumn{1}{c|}{\cellcolor[HTML]{FFFFFF}\textbf{PIR}}               & \multicolumn{1}{c|}{\cellcolor[HTML]{FFFFFF}\textbf{CBS}}          & \multicolumn{1}{c|}{\cellcolor[HTML]{FFFFFF}\textbf{PBS}}          & \multicolumn{1}{c|}{\multirow{-2}{*}{\cellcolor[HTML]{FFFFFF}\textbf{\begin{tabular}[c]{@{}c@{}}Global\\ Policer\end{tabular}}}} & \multicolumn{1}{c|}{\cellcolor[HTML]{FFFFFF}\textbf{CIR}} & \multicolumn{1}{c|}{\cellcolor[HTML]{FFFFFF}\textbf{PIR}} & \multicolumn{1}{c|}{\cellcolor[HTML]{FFFFFF}\textbf{CBS}} & \multicolumn{1}{c|}{\cellcolor[HTML]{FFFFFF}\textbf{PBS}} & \multicolumn{1}{c|}{\cellcolor[HTML]{FFFFFF}\textbf{Conf1}}          & \multicolumn{1}{c|}{\cellcolor[HTML]{FFFFFF}\textbf{Conf2}}           & \multicolumn{1}{c|}{\cellcolor[HTML]{FFFFFF}}                                                                                           \\ \cline{1-16}
\rowcolor[HTML]{FFFFFF} 
\multicolumn{1}{|c|}{\cellcolor[HTML]{FFFFFF}URLLC}                                    & \multicolumn{1}{c|}{\cellcolor[HTML]{FFFFFF}N/A}          & \multicolumn{1}{c|}{\cellcolor[HTML]{FFFFFF}N/A}          & \multicolumn{1}{c|}{\cellcolor[HTML]{FFFFFF}N/A}          & \multicolumn{1}{c|}{\cellcolor[HTML]{FFFFFF}N/A}          & \multicolumn{1}{c|}{\cellcolor[HTML]{FFFFFF}1.2 Mbps}                    & \multicolumn{1}{c|}{\cellcolor[HTML]{FFFFFF}100 Mbps}                   & \multicolumn{1}{c|}{\cellcolor[HTML]{FFFFFF}50 KB}                 & \multicolumn{1}{c|}{\cellcolor[HTML]{FFFFFF}50 KB}                 & \multicolumn{1}{c|}{\cellcolor[HTML]{FFFFFF}}                                          & \multicolumn{1}{c|}{\cellcolor[HTML]{FFFFFF}1.2 Mbps}     & \multicolumn{1}{c|}{\cellcolor[HTML]{FFFFFF}100 Mbps}     & \multicolumn{1}{c|}{\cellcolor[HTML]{FFFFFF}50 KB}        & \multicolumn{1}{c|}{\cellcolor[HTML]{FFFFFF}1538 B}        & \multicolumn{1}{c|}{\cellcolor[HTML]{FFFFFF}PQ}                     & \multicolumn{1}{c|}{\cellcolor[HTML]{FFFFFF}PQ}                      & \multicolumn{1}{c|}{\cellcolor[HTML]{FFFFFF}}                                                                                           \\ \cline{1-9} \cline{11-16}
\rowcolor[HTML]{FFFFFF} 
\multicolumn{1}{|c|}{\cellcolor[HTML]{FFFFFF}Video}                                    & \multicolumn{1}{c|}{\cellcolor[HTML]{FFFFFF}32 Mbps}      & \multicolumn{1}{c|}{\cellcolor[HTML]{FFFFFF}100 Mbps}     & \multicolumn{1}{c|}{\cellcolor[HTML]{FFFFFF}50 KB}        & \multicolumn{1}{c|}{\cellcolor[HTML]{FFFFFF}50 KB}        & \multicolumn{1}{c|}{\cellcolor[HTML]{FFFFFF}}                            & \multicolumn{1}{c|}{\cellcolor[HTML]{FFFFFF}}                           & \multicolumn{1}{c|}{\cellcolor[HTML]{FFFFFF}}                      & \multicolumn{1}{c|}{\cellcolor[HTML]{FFFFFF}}                      & \multicolumn{1}{c|}{\cellcolor[HTML]{FFFFFF}}                                          & \multicolumn{1}{c|}{\cellcolor[HTML]{FFFFFF}32 Mbps}      & \multicolumn{1}{c|}{\cellcolor[HTML]{FFFFFF}100 Mbps}     & \multicolumn{1}{c|}{\cellcolor[HTML]{FFFFFF}50 KB}        & \multicolumn{1}{c|}{\cellcolor[HTML]{FFFFFF}1538 B}        & \multicolumn{1}{c|}{\cellcolor[HTML]{FFFFFF}1538 B}                   & \multicolumn{1}{c|}{\cellcolor[HTML]{FFFFFF}15380 B}                   & \multicolumn{1}{c|}{\cellcolor[HTML]{FFFFFF}}                                                                                           \\ \cline{1-5} \cline{11-16}
\rowcolor[HTML]{FFFFFF} 
\multicolumn{1}{|c|}{\cellcolor[HTML]{FFFFFF}Telemetry}                                & \multicolumn{1}{c|}{\cellcolor[HTML]{FFFFFF}4 Mbps}       & \multicolumn{1}{c|}{\cellcolor[HTML]{FFFFFF}100 Mbps}     & \multicolumn{1}{c|}{\cellcolor[HTML]{FFFFFF}50 KB}        & \multicolumn{1}{c|}{\cellcolor[HTML]{FFFFFF}50 KB}        & \multicolumn{1}{c|}{\multirow{-2}{*}{\cellcolor[HTML]{FFFFFF}36 Mbps}}   & \multicolumn{1}{c|}{\multirow{-2}{*}{\cellcolor[HTML]{FFFFFF}100 Mbps}} & \multicolumn{1}{c|}{\multirow{-2}{*}{\cellcolor[HTML]{FFFFFF}1538 B}} & \multicolumn{1}{c|}{\multirow{-2}{*}{\cellcolor[HTML]{FFFFFF}1538 B}} & \multicolumn{1}{c|}{\cellcolor[HTML]{FFFFFF}}                                          & \multicolumn{1}{c|}{\cellcolor[HTML]{FFFFFF}4 Mbps}       & \multicolumn{1}{c|}{\cellcolor[HTML]{FFFFFF}100 Mbps}     & \multicolumn{1}{c|}{\cellcolor[HTML]{FFFFFF}50 KB}        & \multicolumn{1}{c|}{\cellcolor[HTML]{FFFFFF}1538 B}        & \multicolumn{1}{c|}{\cellcolor[HTML]{FFFFFF}}                        & \multicolumn{1}{c|}{\cellcolor[HTML]{FFFFFF}}                         & \multicolumn{1}{c|}{\cellcolor[HTML]{FFFFFF}}                                                                                           \\ \cline{1-9} \cline{11-14}
\rowcolor[HTML]{FFFFFF} 
\multicolumn{1}{|c|}{\cellcolor[HTML]{FFFFFF}VC}                                       & \multicolumn{1}{c|}{\cellcolor[HTML]{FFFFFF}52.8 Mbps}    & \multicolumn{1}{c|}{\cellcolor[HTML]{FFFFFF}100 Mbps}     & \multicolumn{1}{c|}{\cellcolor[HTML]{FFFFFF}50 KB}        & \multicolumn{1}{c|}{\cellcolor[HTML]{FFFFFF}50 KB}        & \multicolumn{1}{c|}{\cellcolor[HTML]{FFFFFF}}                            & \multicolumn{1}{c|}{\cellcolor[HTML]{FFFFFF}}                           & \multicolumn{1}{c|}{\cellcolor[HTML]{FFFFFF}}                      & \multicolumn{1}{c|}{\cellcolor[HTML]{FFFFFF}}                      & \multicolumn{1}{c|}{\cellcolor[HTML]{FFFFFF}}                                          & \multicolumn{1}{c|}{\cellcolor[HTML]{FFFFFF}52.8 Mbps}    & \multicolumn{1}{c|}{\cellcolor[HTML]{FFFFFF}100 Mbps}     & \multicolumn{1}{c|}{\cellcolor[HTML]{FFFFFF}50 KB}        & \multicolumn{1}{c|}{\cellcolor[HTML]{FFFFFF}1538 B}        & \multicolumn{1}{c|}{\multirow{-2}{*}{\cellcolor[HTML]{FFFFFF}1538 B}} & \multicolumn{1}{c|}{\multirow{-2}{*}{\cellcolor[HTML]{FFFFFF}10766 B}} & \multicolumn{1}{c|}{\cellcolor[HTML]{FFFFFF}}                                                                                           \\ \cline{1-5} \cline{11-16}
\rowcolor[HTML]{FFFFFF} 
\multicolumn{1}{|c|}{\cellcolor[HTML]{FFFFFF}BE}                                       & \multicolumn{1}{c|}{\cellcolor[HTML]{FFFFFF}N/A}          & \multicolumn{1}{c|}{\cellcolor[HTML]{FFFFFF}N/A}          & \multicolumn{1}{c|}{\cellcolor[HTML]{FFFFFF}N/A}          & \multicolumn{1}{c|}{\cellcolor[HTML]{FFFFFF}N/A}          & \multicolumn{1}{c|}{\multirow{-2}{*}{\cellcolor[HTML]{FFFFFF}52.8 Mbps}} & \multicolumn{1}{c|}{\multirow{-2}{*}{\cellcolor[HTML]{FFFFFF}100 Mbps}} & \multicolumn{1}{c|}{\multirow{-2}{*}{\cellcolor[HTML]{FFFFFF}1538 B}} & \multicolumn{1}{c|}{\multirow{-2}{*}{\cellcolor[HTML]{FFFFFF}1538 B}} & \multicolumn{1}{c|}{\multirow{-5}{*}{\cellcolor[HTML]{FFFFFF}100 Mbps}}                & \multicolumn{1}{c|}{\cellcolor[HTML]{FFFFFF}N/A}          & \multicolumn{1}{c|}{\cellcolor[HTML]{FFFFFF}100 Mbps}     & \multicolumn{1}{c|}{\cellcolor[HTML]{FFFFFF}N/A}        & \multicolumn{1}{c|}{\cellcolor[HTML]{FFFFFF}1538 B}        & \multicolumn{1}{c|}{\cellcolor[HTML]{FFFFFF}1538 B}                   & \multicolumn{1}{c|}{\cellcolor[HTML]{FFFFFF}1538 B}                    & \multicolumn{1}{c|}{\multirow{-6}{*}{\cellcolor[HTML]{FFFFFF}\begin{tabular}[c]{@{}c@{}}A: HPQ \\ B: LPQ \end{tabular}}} \\ \hline
\multicolumn{1}{l}{}                                                                   & \multicolumn{1}{l}{}                                      & \multicolumn{1}{l}{}                                      & \multicolumn{1}{l}{}                                      & \multicolumn{1}{l}{}                                      & \multicolumn{1}{l}{}                                                     & \multicolumn{1}{l}{}                                                    & \multicolumn{1}{l}{}                                               & \multicolumn{1}{l}{}                                               & \multicolumn{1}{l}{}                                                                   & \multicolumn{1}{l}{}                                      & \multicolumn{1}{l}{}                                      & \multicolumn{1}{l}{}                                      & \multicolumn{1}{l}{}                                      & \multicolumn{1}{l}{}                                                 & \multicolumn{1}{l}{}                                                  & \multicolumn{1}{l}{}                                                                                                                    \\
\multicolumn{1}{l}{}                                                                   & \multicolumn{1}{l}{}                                      & \multicolumn{1}{l}{}                                      & \multicolumn{1}{l}{}                                      & \multicolumn{1}{l}{}                                      & \multicolumn{1}{l}{}                                                     & \multicolumn{1}{l}{}                                                    & \multicolumn{1}{l}{}                                               & \multicolumn{1}{l}{}                                               & \multicolumn{1}{l}{}                                                                   & \multicolumn{1}{l}{}                                      & \multicolumn{1}{l}{}                                      & \multicolumn{1}{l}{}                                      & \multicolumn{1}{l}{}                                      & \multicolumn{1}{l}{}                                                 & \multicolumn{1}{l}{}                                                  & \multicolumn{1}{l}{}                                                                                                                    \\
\multicolumn{1}{l}{}                                                                   & \multicolumn{1}{l}{}                                      & \multicolumn{1}{l}{}                                      & \multicolumn{1}{l}{}                                      & \multicolumn{1}{l}{}                                      & \multicolumn{1}{l}{}                                                     & \multicolumn{1}{l}{}                                                    & \multicolumn{1}{l}{}                                               & \multicolumn{1}{l}{}                                               & \multicolumn{1}{l}{}                                                                   & \multicolumn{1}{l}{}                                      & \multicolumn{1}{l}{}                                      & \multicolumn{1}{l}{}                                      & \multicolumn{1}{l}{}                                      & \multicolumn{1}{l}{}                                                 & \multicolumn{1}{l}{}                                                  & \multicolumn{1}{l}{}                                                                                                                   
\end{tabular}%
}
\vspace*{-0.45cm}
\caption{Experiment C System Parameter Configuration}
\label{tab:expc-params}
\end{table*}   

This experiment lasts 60 seconds and is divided into the three following time intervals: 

\begin{enumerate}
    \item Interval 1 (from \( t = 0 \, \text{s} \) to \( t = 10 \, \text{s} \)): All traffic classes are active, but there are no traffic bursts during this interval. 
    \item Interval 2 (from \( t = 10 \, \text{s} \) to \( t = 45 \, \text{s} \)): All traffic classes are active and transmitting at a constant rate. However, traffic bursts are introduced for all classes: \gls{urllc} generates a burst every second, the \gls{tod} video every two seconds, and both \gls{tod} telemetry and \gls{embb} VC every five seconds.
    \item Interval 3 (from \( t = 45 \, \text{s} \) to \( t = 60 \, \text{s} \)): the same conditions as Interval 2, except that the traffic burst from \gls{urllc} is no longer present.
\end{enumerate}

\rev{All flows have a background constant rate that is significantly below their guaranteed rate, except for BE traffic, which is transmitted at 100\,Mbps. This approach ensures that the buckets have time to refill, allowing multiple bursts per class to be observed rather than just one. Moreover, throughout the experiment, all generated bursts have a uniform size of 100\,KB. The traffic generated for this experiment is shown in Fig. \ref{expc-conf}a.}

The results are presented in Figs.~\ref{expc-conf}b, \ref{expc-conf}c, \ref{expc-conf}d and \ref{expc-conf}e, illustrating  the \gls{qos} achieved by the traffic flows in terms of both bandwidth and latency with both models. The \textit{Conf1} configuration shown in Table~\ref{tab:expc-params} for the coarse-grained resource control has been used in the \gls{hctns} model to obtain these results.

\afterpage{
\begin{figure*}[!h]
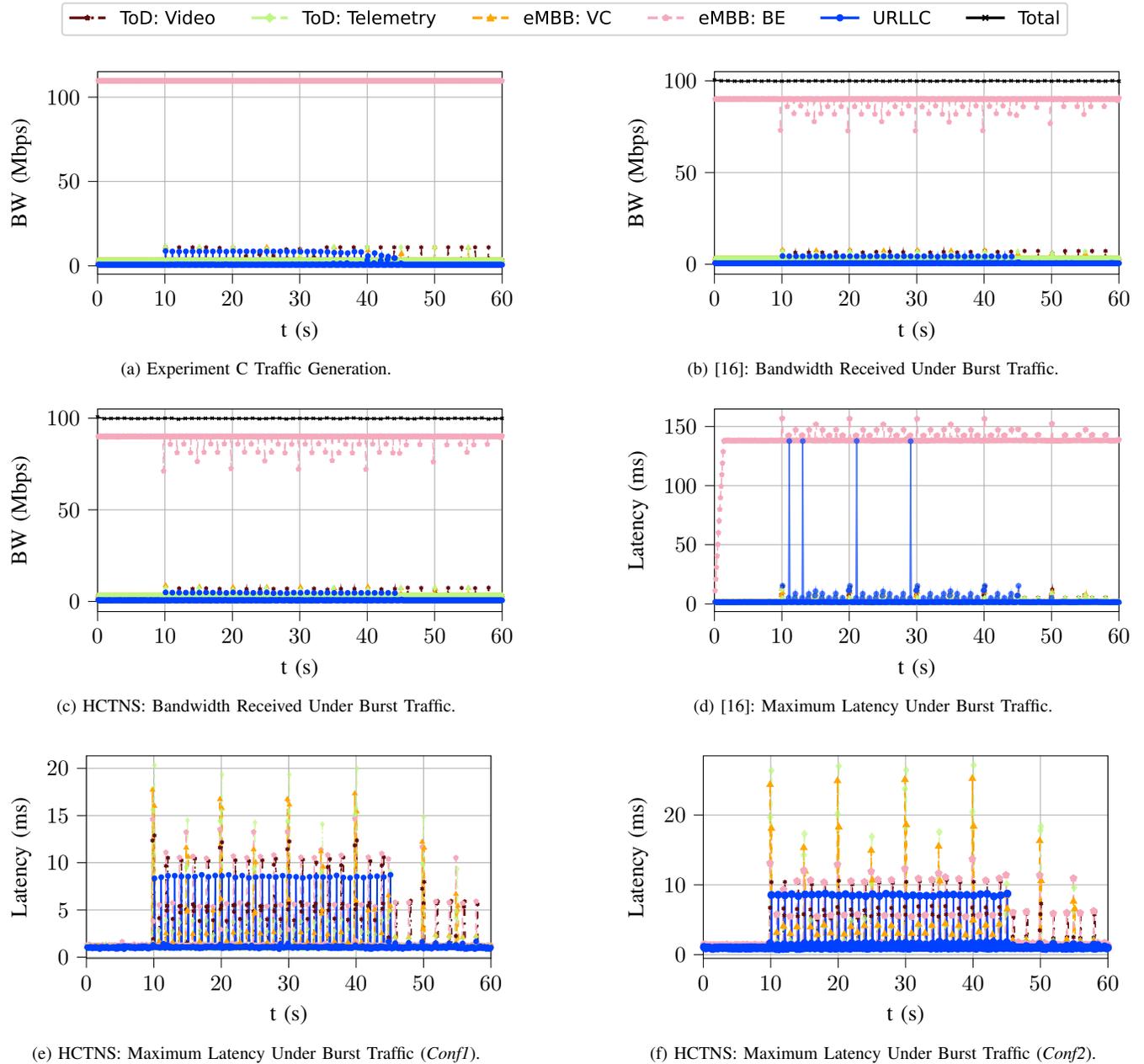

    \vspace*{-2em}
    \centering
    \includegraphics[width=1\textwidth, trim=50 10 50 10, clip]{figures/results/ExperimentA/legend_5g.pdf}
    \begin{minipage}[t]{0.45\textwidth} 
        \centering
        \input{figures/results/ExperimentC/Proposed/bandwidth_sent}
        \centering \small (a) Experiment C Traffic Generation.
    \end{minipage}
    \vspace*{0.4cm}
    \hfill
    \begin{minipage}[t]{0.45\textwidth} 
        \centering
        \input{figures/results/ExperimentC/SoA/bandwidth_received}
        \centering \small (b) \cite{Lin2021}: Bandwidth Received Under Burst Traffic.
    \end{minipage}
    \vspace*{0.4cm}
    \begin{minipage}[t]{0.45\textwidth} 
        \centering
        \input{figures/results/ExperimentC/Proposed/bandwidth_received}
        \centering \small (c) \gls{hctns}: Bandwidth Received Under Burst Traffic.
    \end{minipage}
    \vspace*{0.2cm}
    \hfill
    \begin{minipage}[t]{0.45\textwidth} 
        \centering
        \input{figures/results/ExperimentC/SoA/max_lat}
        \centering \small (d) \cite{Lin2021}: Maximum Latency Under Burst Traffic.
    \end{minipage}
    \vspace*{0.2cm}
    \begin{minipage}[t]{0.45\textwidth}
        \centering
        \input{figures/results/ExperimentC/Proposed/max_lat}
        \centering \small (e) \gls{hctns}: Maximum Latency Under Burst Traffic (\textit{Conf1}).
    \end{minipage}
    \hfill
    \begin{minipage}[t]{0.45\textwidth}
        \centering
        \input{figures/results/ExperimentC/Proposed/max_lat_conf2}
        \centering \small (f) \gls{hctns}: Maximum Latency Under Burst Traffic (\textit{Conf2}).
    \end{minipage}
    \caption{Experiment C: \gls{hctns} and \cite{Lin2021} model performance under burst traffic.}
    \label{expc-conf}
\end{figure*}
}

\rev{During \textit{Interval 1}, traffic classes transmit at a constant rate without generating traffic bursts. As depicted in Figs.~\ref{expc-conf}b and \ref{expc-conf}c, both models show similar bandwidth behavior, with BE traffic using all the remaining bandwidth not consumed by other traffic classes. However, the latency behavior (Figs.~\ref{expc-conf}d and~\ref{expc-conf}e) for the BE traffic differs, since the~\cite{Lin2021} model, in addition to accepting background traffic from other flows, admits 100\,Mbps of BE traffic. As a result, the low priority queue is congested and BE packets experiment a high latency. In contrast, since \gls{hctns} incorporates a global policer that limits the traffic admitted at the transport network's ingress, it ensures that the accepted traffic does not exceed the queues' capacity. This results in low latency values, similar to the rest of the flows.}

During \textit{Interval 2}, traffic bursts are generated for all defined traffic classes. \rev{In the~\cite{Lin2021} model, the bursts accepted from all traffics are classified as ``green", being forwarded to the highest priority queue. As the priority queue is not saturated, these packets are transmitted. When bursts are accepted, the highest priority queue uses a higher transmission rate, which reduces the transmission rate of the lower priority queue, causing the downward spikes in bandwidth in Fig.~\ref{expc-conf}b.} In \gls{hctns}, a very close bandwidth behavior (Fig.~\ref{expc-conf}c) is observed, accepting a similar amount of burst packers. In this case, the configured \gls{cbs} and \gls{pbs} values allow the defined amount of bytes to ingress the \gls{tn}, while the excess traffic is queued in the \gls{htb} ingress queue. If this queue becomes full, any additional packets are discarded. As explained in Section~\ref{sec:proposal}.\ref{subsec:proposal-finegraine}, when traffic burst control is enabled and bursts are accepted by the class policers (or the slice policer for \gls{urllc} traffic), both the slice and global policers may temporally have ``negative'' tokens. Consequently, BE traffic is not admitted by the ingress policer until the slice policer's bucket recovers the tokens it owed restoring a positive balance. Until the global policer's bucket is not fully recovered, BE traffic cannot borrow bandwidth beyond the \gls{embb} slice's \gls{cir}. As a result, during burst periods, the bandwidth available to the BE traffic shows also temporary downward spikes. This mechanism ensures that, while packets arriving at rates exceeding the \gls{cir} defined in the global policer may be temporally accepted, the average bandwidth will remain within the defined limits. 

\rev{As shown in Fig.~\ref{expc-conf}d, traffic bursts impact the latency of all flows when using the~\cite{Lin2021} model. This is because the accepted burst packets (``green") are forwarded to the highest priority queue, regardless of the traffic flow to which they belong. As a result, all burst traffic from the flows undergoes similar treatment. For example, all  traffic flows experience a similar maximum latency,  around 15\,ms. Packets marked as ``yellow" are forwarded to the lowest priority queue, which is congested. Since BE traffic continuously enters the lowest priority queue at 100\,Mbps and the lowest priority queue cannot transmit a packet if there are packets in the highest priority queue, most ``yellow" packets from the burst will be discarded due to queue overflow. For the same reason, ``yellow" packets, such as those from \gls{urllc} traffic, experience latencies exceeding 130 ms, which is unacceptable.}

In \gls{hctns}, as illustrated in Fig.~\ref{expc-conf}e, traffic bursts also have a significant impact on the latency. While the average rate at which packets arrive at the \gls{pe}1 input port does not exceed the output port capacity, bursts temporarily surpass this capacity. This causes packets to be enqueued at the \gls{pe}1 output port, resulting in delays. However, since the average accepted rate remains below the global policer's \gls{cir}, which is adjusted to the available network capacity, packets do not accumulate over time in these queues. Therefore, it is crucial to properly dimension the queue sizes to handle bursty traffic effectively, preventing packet loss. 

Since \gls{urllc} traffic is assigned to \gls{tn} \gls{qos} class A, which has a priority queue, its latency remains unaffected  by traffic bursts from the other \gls{tn} \gls{qos} classes. However, \gls{urllc} bursts impact the delay in the \gls{drr} queues, as these queues must wait for the priority queues to empty before transmitting packets. Additionally, the latency experienced by traffic classes in \gls{tn} \gls{qos} classes with \gls{drr} queues is also influenced by the bursts in other \gls{drr} queues. For example, when bursts from the telemetry and VC traffic classes are transmitted, the latency experienced by BE and \gls{tod} video traffic increases. As will be explained below, different coarse-grained resource control configurations can be applied to manage and control this latency. \rev{Unlike \cite{Lin2021}, all traffic accepted by the policers from the same \gls{tn} \gls{qos} class receives the same treatment, achieving \gls{urllc}, video, telemetry, and VC latencies of under 9\,ms, 13\,ms, 21\,ms, and 21\,ms, respectively. The BE queue is not saturated, so latency values remain within these ranges.}

\rev{In \textit{Interval 3}, when there are no \gls{urllc} traffic bursts, the latency perceived by the other traffic classes decreases in both models. In the~\cite{Lin2021} model, fewer packets are queued in the highest priority queue, which reduces the waiting time for packets arriving at this queue. As a result, the maximum latency values achieved are around 12\,ms. Additionally, although some packets are marked as ``yellow'', the saturation of the lowest priority queue prevents any ``yellow" packets from entering the buffer. This explains why the latency spikes have disappeared. However, as the lowest priority queue remains saturated, the latency perceived by BE traffic does not change.}

\rev{In \gls{hctns}, the latency experienced by traffic classes assigned to \gls{tn} \gls{qos} classes with an allocated \gls{drr} queue is still influenced by bursts from other \gls{drr} queues. However, the latency values decrease to under 8\,ms for video and under 15\,ms for telemetry and VC. Since the BE queue is not saturated, its latency values also remain within these ranges.}

In \gls{hctns}, as expected, all the traffic classes sharing a \gls{tn} \gls{qos} class experiment similar latency values, due to the \gls{phb} applied at the output port through the coarse-grained resource control configuration. 

Fig.~\ref{expc-conf}f shows how different coarse-grained resource control configurations (quantum values)  affect latencies experienced by the traffic classes. The \textit{Conf2} configuration, as shown in Table~\ref{tab:expc-params}, was used to obtain the following results. 

This new configuration of the coarse-grained resource control mechanism allows the \gls{tn} \gls{qos} class B (which carries \gls{tod} video) to have a higher transmission rate, compared to the previous configuration. As a result, its queued packets are transmitted earlier, reducing the latency experienced. For example, a maximum latency of 12.89\,ms was measured in the experiment shown in Fig.~\ref{expc-conf}e, compared to 10.69\,ms in the experiment shown in Fig.~\ref{expc-conf}f. The other \gls{tn} \gls{qos} classes now have a lower transmission rate than in the previous coarse-grained resource control configuration, which leads to increased latency for those traffic classes.

\rev{Despite both models achieving similar bandwidth results, in \gls{hctns}, by adjusting the quantum values in the coarse-grained resource control configuration, the latency experienced by the traffic of a \gls{tn} \gls{qos} class can be controlled. As observed throughout all the experiments, when using \gls{hctns}, packets experience queuing delays under burst traffic conditions. However, even in these cases, unlike with the~\cite{Lin2021} model, this latency can be controlled by \gls{hctns}. In \gls{hctns}, the latency of \gls{urllc} traffic is solely influenced by the traffic bursts of this \gls{tn} \gls{qos} class, while the latency in the remaining traffic classes depends not only on the accepted \gls{urllc} traffic bursts, but also on the bursts accepted by other \gls{tn} \gls{qos} classes. In contrast, in the~\cite{Lin2021} model, the latency of each traffic flow is influenced by the bursts of other traffic flows, including BE traffic, when some of its packets are classified as ``yellow". This causes the~\cite{Lin2021} model to perform worse than \gls{hctns}.}

\section{Conclusions}
\label{sec:conclusion}

In this paper, we have proposed \gls{hctns}, a slicing model  that employs a three-level hierarchical \gls{htb} to implement a fine-grained resource control mechanism at the ingress node of the transport network. This mechanism is combined with coarse-grained resource control in the transport network nodes to allow meeting the bandwidth and latency requirements of services defined within the slices, even in worst-case scenarios and with bursty traffic, while achieving a high level of network bandwidth sharing both within and across slices. One of the strengths of \gls{hctns} is that all the traffic admitted to be processed in the transport network is treated with \gls{qos} guarantees, not only the traffic under the guaranteed bitrate but also traffic exceeding it (up to certain limit), and even for limited traffic bursts.

\rev{In real deployments, each network segment operates independently as an administrative domain, meaning that end-to-end performance depends on the specific traffic management solutions implemented in each segment. HCTNS ensures that the latency, bandwidth, and packet loss requirements defined in the SLAs are met within the transport network, contributing to their fulfillment end-to-end. While HCTNS does not impact the performance of the RAN or core network, it is essential for maintaining service guarantees in the transport network and, by extension, across the entire system.}

\rev{An experimental platform combining virtual and physical network elements has allowed us to verify that \gls{hctns} outperforms not only the model being standardized by the \gls{ietf}, but also a state-of-the-art reference approach, in terms of bandwidth utilization, latency and traffic burst control.}

As future steps for this research, we believe that centralized policies are needed to guarantee end-to-end delay bounds in the transport network. In addition, models are required to translate these policies into configurations for data plane functions in the transport network equipment. This includes the establishment and configuration of IPv6 tunnels based on \gls{sr} with routing algorithms driven by \gls{qos} metrics.

\bibliography{ojcoms}

\begin{IEEEbiography}[{\includegraphics[width=1in,height=1.25in,clip,keepaspectratio]{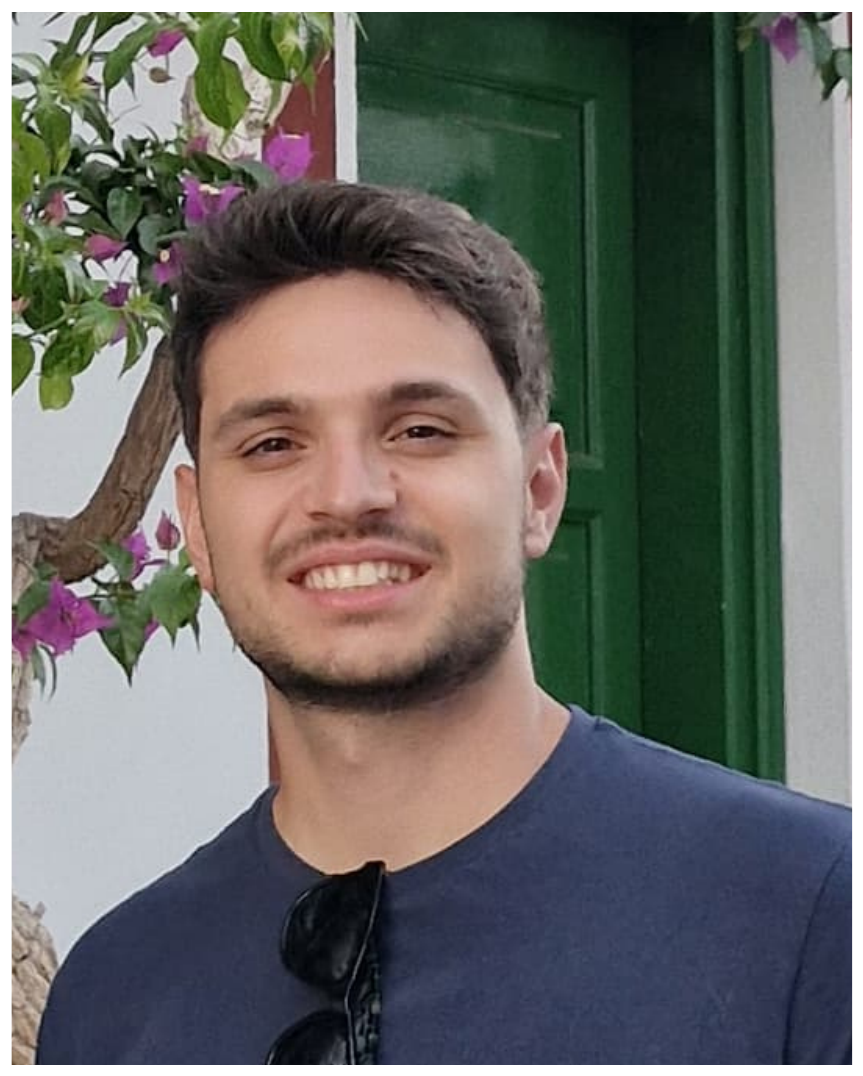}}]{AITOR ENCINAS-ALONSO}{\,} received his B.S. degree in technologies and telecommunication services in 2021, and his M.S. degree in telecommunications engineering in 2023, both from the Universidad Politécnica de Madrid (UPM). He is currently a Ph.D. candidate at UPM. His research interests include network slicing, QoS in 5G networks, software-defined networks and remote driving services.
\end{IEEEbiography}
\begin{IEEEbiography}[{\includegraphics[width=1in,height=1.25in,clip,keepaspectratio]{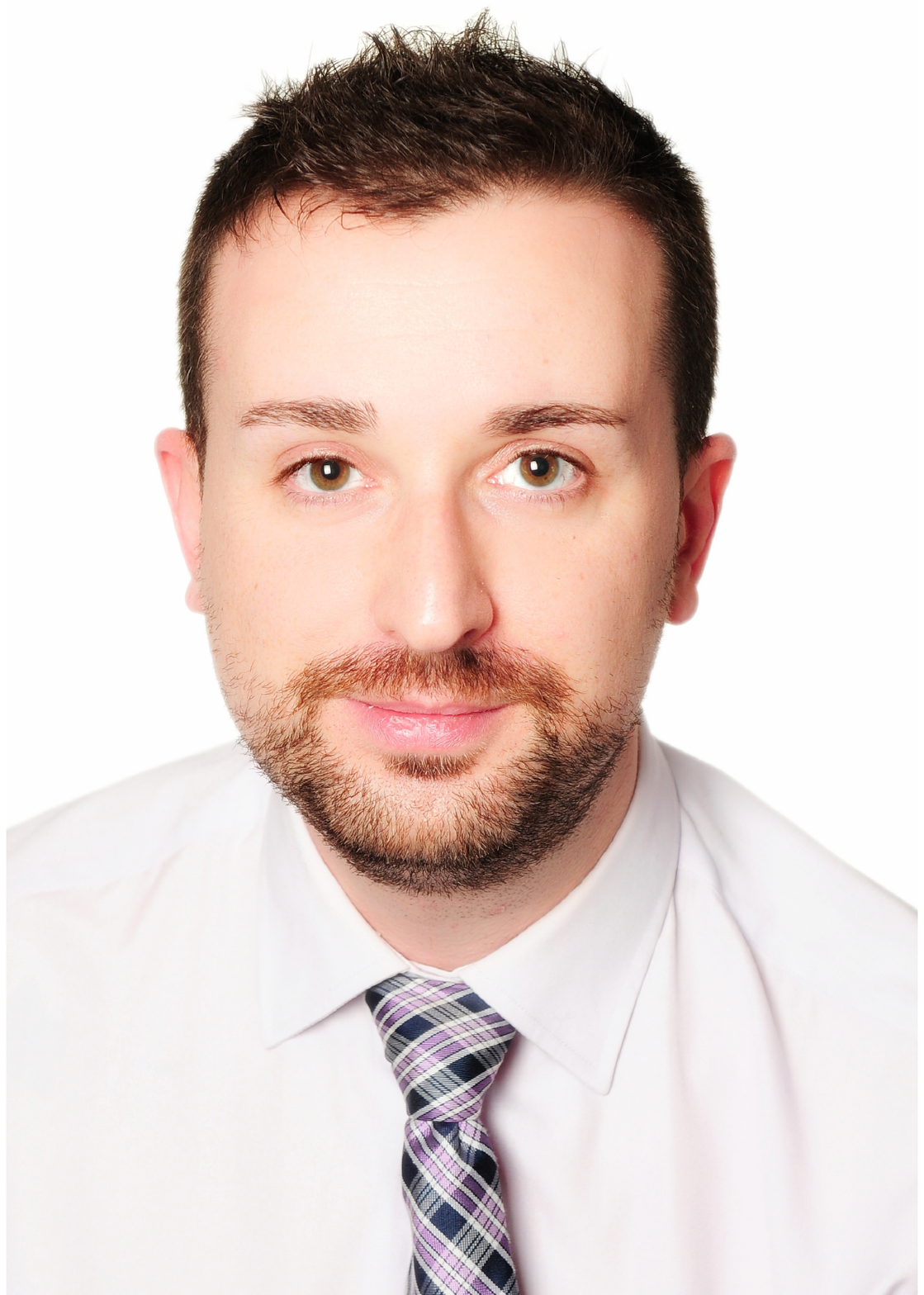}}]{CARLOS M. LENTISCO}{\,} received the M.S. and Ph.D. degrees in telecommunications engineering from the Universidad Politécnica de Madrid (UPM), Madrid,
Spain, in 2014 and 2019, respectively. He is currently
an Associate Professor with UPM, specializing in the
fields of computer networking, multimedia services,
and Internet technologies. His current research
interests include software networks, mobile communications, and vehicular networking. 
\end{IEEEbiography}
\begin{IEEEbiography}[{\includegraphics[width=1in,height=1.25in,clip,keepaspectratio]{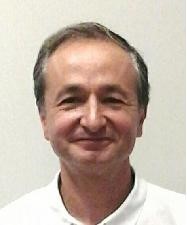}}]{IGNACIO SOTO}{\,} received a telecommunication engineering degree in 1993, and a Ph.D. in telecommunications in 2000, both from the Universidad de Vigo, Spain. He was a Research and Teaching Assistant in telematics engineering at the University of Valladolid from 1993 to 1999. In 1999, he joined Universidad Carlos III de Madrid, where he was an Associate Professor from 2002 to 2021. In 2021, he joined Universidad Politécnica de Madrid where he currently works as Professor in telematics engineering. His research activities focus on vehicular networks, future transportation systems, mobility support in packet networks, software defined networks and network virtualization. 
\end{IEEEbiography}
\begin{IEEEbiography}[{\includegraphics[width=1in,height=1.25in,clip,keepaspectratio]{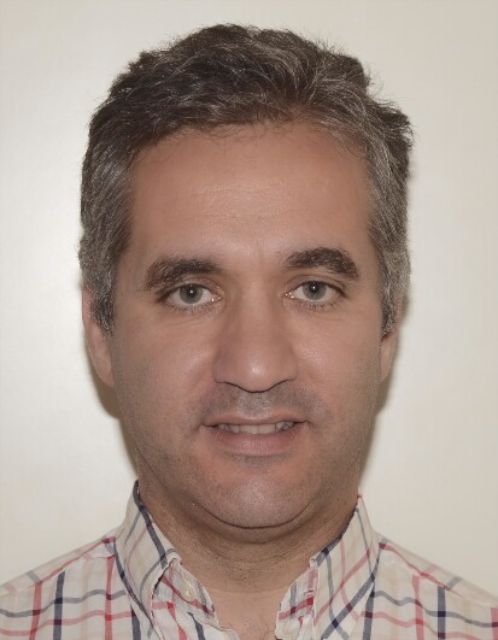}}]{LUIS BELLIDO}{\,} received the M.S. and Ph.D. degrees in telecommunications engineering from the Universidad Politécnica de Madrid (UPM), Madrid, Spain, in 1994 and 2004, respectively. He is currently an Associate Professor with UPM, specializing in computer networking, internet technologies, and quality of service. His research interests include mobile networks and services, software-defined networks, network virtualization, and data-driven network management.
\end{IEEEbiography}
\begin{IEEEbiography}[{\includegraphics[width=1in,height=1.25in,clip,keepaspectratio]{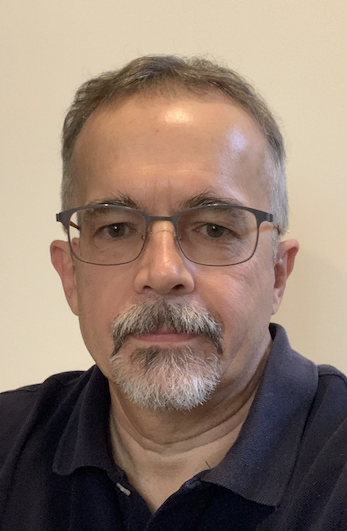}}]{DAVID FERNANDEZ}{\,} received the M.S. degree in telecommunications engineering and the Ph.D. degree in telematics engineering from Universidad Politécnica de Madrid (UPM), Spain, in 1988 and 1993, respectively. Since 1995, he has been an Associate Professor with the Department of Telematics Systems Engineering (DIT), UPM. His current research interests include software-defined networks, network virtualization, cloud computing data center technologies, and network security.
\end{IEEEbiography}

\end{document}